\documentclass[fleqn,usenatbib]{mnras}

\usepackage{newtxtext,newtxmath}
\usepackage[T1]{fontenc}

\DeclareRobustCommand{\VAN}[3]{#2}
\let\VANthebibliography\thebibliography
\def\thebibliography{\DeclareRobustCommand{\VAN}[3]{##3}\VANthebibliography}

\usepackage{graphicx}	% Including figure files
\usepackage{amsmath}	% Advanced maths commands
\usepackage{tablefootnote}
\usepackage{pdflscape}
\usepackage{orcidlink}
\usepackage{enumerate}
\usepackage{comment}

\title[Resolution criteria to avoid artificial clumping]{Resolution criteria to avoid artificial clumping in Lagrangian hydrodynamic simulations with a multi-phase interstellar medium}

\author[S. Ploeckinger et al.]{
Sylvia Ploeckinger,\orcidlink{0000-0002-1965-1650}$^{1,2}$\thanks{E-mail: sylvia.ploeckinger@univie.ac.at}
Folkert S.J. Nobels,$^{3}$
Matthieu Schaller,\orcidlink{0000-0002-2395-4902}$^{2,3}$
Joop Schaye\orcidlink{0000-0002-0668-5560}$^{3}$
\\
$^{1}$Department of Astrophysics, University of Vienna, T\"urkenschanzstrasse 17, 1180 Vienna, Austria\\
$^{2}$Lorentz Institute for Theoretical Physics, Leiden University, PO Box 9506, 2300 RA Leiden, the Netherlands\\
$^{3}$Leiden Observatory, Leiden University, PO Box 9513, NL-2300 RA Leiden, The Netherlands
}

\date{Accepted XXX. Received YYY; in original form ZZZ}

\pubyear{2023}

\begin{document}
\label{firstpage}
\pagerange{\pageref{firstpage}--\pageref{lastpage}}
\maketitle

\begin{abstract}
Large-scale cosmological galaxy formation simulations typically prevent gas in the interstellar medium (ISM) from cooling below $\approx 10^4\,\mathrm{K}$. This has been motivated by the inability to resolve the Jeans mass in molecular gas ($\ll 10^5\,\mathrm{M}_{\odot}$) which would result in undesired artificial clumping. We show that the classical Jeans criteria derived for Newtonian gravity are not applicable in the simulated ISM if the spacing of resolution elements representing the dense ISM is below the gravitational force softening length and gravity is therefore softened and not Newtonian. We re-derive the Jeans criteria for softened gravity in Lagrangian codes and use them to analyse gravitational instabilities at and below the hydrodynamical resolution limit for simulations with adaptive and constant gravitational softening lengths. In addition, we define criteria for which a numerical runaway collapse of dense gas clumps can occur caused by over-smoothing of the hydrodynamical properties relative to the gravitational force resolution. This effect is illustrated using simulations of isolated disk galaxies with the smoothed particle hydrodynamics code \textsc{Swift}. We also demonstrate how to avoid the formation of artificial clumps in gas and stars by adjusting the gravitational and hydrodynamical force resolutions. 
\end{abstract}

\begin{keywords}
hydrodynamics -- instabilities -- methods: numerical -- galaxies: ISM -- galaxies: general -- galaxies: formation
\end{keywords}

%%%%%%%%%%%%%%%%%%%%%%%%%%%%%%%%%%%%%%%%%%%%%%%%%%

%%%%%%%%%%%%%%%%% BODY OF PAPER %%%%%%%%%%%%%%%%%%

\section{Introduction}\label{sec:intro}

Numerical simulations of cosmological structure formation are an invaluable tool to test theories of galaxy formation, galaxy evolution and cosmology. These simulations have become increasingly complex in the last decades - from pure N-body simulations (e.g. \citealp{Millennium}) to simulations that model the gas and stars in and around galaxies in great detail (e.g. Horizon-AGN: \citealp{HorizonAGN}, Illustris: \citealp{Illustris}, \textsc{eagle}: \citealp{EAGLE}, IllustrisTNG: \citealp{IllustrisTNGPillepich}, Simba: \citealp{Simba}, FIREbox: \citealp{FIREbox}, see also the review from \citealp{Crain2023Review}). Gravity is a key ingredient in the Universe on a wide range of scales, from the formation of individual stars to cosmological structure formation, and modelling gravity adequately is critical for a multitude of simulations. 

An important approach to validate numerical methods is comparing the gravitational collapse of various objects in simulations to analytical expectations. For gaseous, self-gravitating structures the analytical framework for stability conditions is largely based on \citet{Jeans1902}. 
His analysis of the stability of spherical nebulae to ``vibrations'' led to various derivations of ``Jeans lengths'' which generally refer to a critical length scale above which a spherical, self-gravitating gas cloud of a given density and temperature collapses under its own gravity. 
This scale can be derived by (i) solving the linearized fluid equations and defining a wavenumber $k_{\mathrm{J}}$ as the limiting wavenumber between oscillatory solutions ($k>k_{\mathrm{J}}$, short wavelengths) and exponentially growing solutions ($k<k_{\mathrm{J}}$, long wavelengths); (ii) defining the Jeans length as the size of the gas cloud above which the free-fall time is smaller than the sound crossing time; or (iii) comparing the gravitational potential energy $W$ to the internal energy $U$ of a spherical gas cloud. Perturbations can grow for scales where $W+U<0$ (for derivations and discussion see e.g. ~\citealp{BinneyTremaineBook}).

\noindent 
Each of these derivations leads to a length scale $\lambda_{\mathrm{J}}$ above which density perturbations (or: gas clouds) of size $r$, sound speed $c_{\mathrm{s}}$, and (unperturbed) density $\rho$ are gravitationally unstable:

\begin{equation}
    \lambda_{\mathrm{J}} = A \left ( \frac{c_{\mathrm{s}}^2}{G\rho}\right)^{1/2} \; ,
\end{equation}

\noindent
where $G$ is the gravitational constant and $A$ is a dimensionless pre-factor of order unity which depends on the exact derivation. Because the assumptions of perfect spherical symmetry and initial zero velocity are rarely fulfilled in real world applications, the Jeans criterion is an approximation and the pre-factors of order unity may vary.  

Unlike in Eulerian (i.e. grid-based) codes, in Lagrangian (i.e. particle-based) codes, resolution elements can have arbitrarily small separations. If the resolution elements represent an underlying smooth density distribution, gravity is softened below a given length scale in order to avoid close 2-body interactions. A classical prescription for softened gravity is the \citet{Plummer1911} potential $\phi_{\mathrm{P}} = -GM (r^2 + \epsilon^2)^{-1/2}$ for a point mass, $M$, and the gravitational softening length, $\epsilon$. The optimal choice of $\epsilon$ is already non-trivial for a collisionless N-body system (e.g.~\citealp{merritt1996,romeo1998,athanassoula2000, Dehnen2001,power2003,Rodionov2005,Romeo2008,Ludlow2019darkmatter}) and is further complicated in hydrodynamical simulations (e.g.~\citealp{PriceMonaghan2007,Ludlow2020hydro}). The softening length, $\epsilon$, is a measure for the gravitational force resolution of a simulation, because gravity is non-Newtonian (i.e. softened) for $r\lesssim\epsilon$. 

A common technique to solve the equations of motions for a collisional fluid is Smoothed Particle Hydrodynamics (SPH, originally from \citealp{Lucy1977, GingoldMonaghan1977} but see e.g.~\citealp{Price2012, Hopkins2013,Hu2014,Beck2016, sphenix} for modern examples) where gas properties, such as the gas density and energy, are smoothed over a kernel which consist of multiple particles, so-called neighbours. The softening length is related to the gravitational force resolution while the kernel size, with the smoothing length, $h$, as the characteristic length scale, is related to the hydrodynamical force resolution.

The smoothing length, $h$, generally decreases with increasing gas density, $h\propto \rho^{-1/3}$, for a density-independent number of neighbours per kernel.
Depending on the code, the gravitational and hydrodynamical force resolutions can be coupled and $\epsilon \propto h$ (''adaptive softening") or $\epsilon$ can be set to a constant value for all densities. 
Adaptive softening is implemented e.g. in the SPH codes \textsc{Gadget4} \citep{Gadget4}, \textsc{Phantom} \citep{Phantom2018}, and the SPH and meshless finite mass / volume code \textsc{Gizmo} \citep{Gizmo}. Their adaptive softening lengths are defined such that they are tied to the kernel length ($\approx h$, the smoothing length of gas particles) or cell size and the hydrodynamic force resolution equals the gravitational force resolution. 

A constant parameter to define the Plummer-equivalent softening length, $\epsilon$, in SPH is used for example in the \textsc{eagle} (\citealp{EAGLE}, code: \textsc{Gadget3}, based on \citealp{gadget2}), the \textsc{Astrid} (\citealp{Astrid2022}, code: \textsc{Mp-Gadget}, \citealp{MpGadget2018}) and the \textsc{Flamingo} (\citealp{Flamingo2023}, code: \textsc{Swift} \citealp{swift2016, swift2018, swift2023}) simulations. 

In the moving mesh code \textsc{Arepo} \citep{arepo}, the softening length is adaptive and follows $\epsilon_{\mathrm{Arepo}} = f_{\mathrm{h}} (3V/(4\pi) )^{1/3}$, where $f_{\mathrm{h}}$ is an input parameter which defines the softening length relative to the cell radius, and $V$ is the volume of the Voronoi cell. In the \textsc{IllustrisTNG} project \citep{IllustrisTNGWeinberger, IllustrisTNGPillepich}, which uses \textsc{Arepo}, the softening length, $\epsilon$ is set to a minimum value $\epsilon_{\mathrm{min}}$ much larger than their minimum cell size (TNG50: $\epsilon_{\mathrm{min}} = 72\,\mathrm{pc}$, minimum cell size: 6.5~pc, \citealp{TNG50Annalisa}), an example for a combination of adaptive (for low densities) and constant ($\epsilon = \epsilon_{\mathrm{min}}$, for high densities) softening lengths. 
In grid-based codes, such as \textsc{Ramses} \citep{ramses}, gas resolution elements cannot have separations smaller than the minimum cell size and an additional gravitational softening parameter is unnecessary.

Whether gravitational instabilities are modelled correctly (i.e. follow the Jeans conditions for gravitational instabilities) has been studied mostly for adaptive softening. \citet{BateBurkert1997} have defined resolution criteria based on the Jeans length for SPH codes. Their main test case is the isothermal collapse of a one solar mass gas cloud with solid body rotation and they follow both its collapse and fragmentation. \citet{BateBurkert1997} advocate using an adaptive softening length equal to the kernel smoothing length ($\epsilon = h$) and ensuring that both resolution parameters can get small enough to resolve the smallest Jeans mass cloud ($M_{\mathrm{J}} = 4\pi  \lambda_{\mathrm{J}}^3 \rho/3$) in the simulation by $2N_{\mathrm{neigh}}$ particles (later reduced to $1.5 N_{\mathrm{neigh}}$ by \citealp{Bate2003}), where $N_{\mathrm{neigh}}$ is the number of neighbours in the SPH kernel. If these conditions are not fulfilled, the collapse or lack thereof of self-gravitating structures is determined by the resolution parameters (collapse / fragmentation artificially induced for $\epsilon < h$ and artificially inhibited for $\epsilon > h$) and not by the physical conditions. 

\citet{Hubber2006} introduced the ``Jeans test'', a numerical test of a plane-wave perturbation within a homogeneous medium, and found that if the resolution criterion from \citet{BateBurkert1997} is violated, the collapse of marginally unstable modes is suppressed but artificial fragmentation does not occur in SPH for their setup with an adaptive softening length $\epsilon = h$, confirming the analytical results from \citet{Whitworth1998}. \citet{Yamamoto2021} repeated the Jeans test with various hydrodynamic schemes within the Gizmo code (meshless finite mass, meshless finite volume, density-based SPH, and pressure-based SPH, see \citealp{Gizmo} for details). They confirm the findings of \citet{Hubber2006} for SPH and extend them to other meshless methods, all for adaptive softening with $\epsilon = h$. They found that for none of the hydrodynamic solvers artificial fragmentation is induced, for all solvers the collapse is slowed down if the resolution is lower than required by \citet{BateBurkert1997}, and the collapse is slowed down more for the meshless methods compared to the SPH methods. 

Large-scale simulations of cosmological volumes cannot follow the collapse of individual gas clouds within the interstellar medium (ISM) down to the formation of individual stars and doing so will remain computationally too expensive for the foreseeable future. Utilizing appropriate subgrid prescriptions for star formation and stellar feedback nevertheless allows for realistic galaxy populations (see e.g. the review from \citealp{VogelsbergerReview}). As an example, the \textsc{eagle} simulation project \citep{EAGLE} used an SPH code and a constant gravitational softening length for all baryon particles of $\epsilon = 700\,\mathrm{pc}$ for their $(100\,\mathrm{Mpc})^3$ simulation with an initial baryon particle mass of $1.81 \times 10^6\,\mathrm{M}_{\odot}$. This marginally fulfills the Jeans mass criterion set by \citet{BateBurkert1997} (but not their recommendation to set $\epsilon = h$) for the warm neutral medium (WNM) with a typical Jeans mass of a few times $10^7\,\mathrm{M}_{\odot}$ but would violate it drastically for the cold gas phase with typical Jeans masses below $10^4\,\mathrm{M}_{\odot}$. An effective pressure floor, sometimes referred to as a polytropic equation of state (EOS), $P_{\mathrm{eff}} \propto n_{\mathrm{H}}^{\gamma_{\mathrm{eff}}}$ for gas with a density of $n_{\mathrm{H}} > 0.1\,\mathrm{cm}^{-3}$ formally solves this issue. For an effective polytropic index of $\gamma_{\mathrm{eff}}=4/3$, the Jeans mass is constant for gas that is limited by the effective pressure floor (see the discussion in \citealp{SchayeDallaVecchia2008}). In \textsc{eagle} the normalization for the effective pressure, $P_{\mathrm{eff}}$, is chosen to be at an effective temperature $T_{\mathrm{eff}} = 8000\,\mathrm{K}$ for $n_{\mathrm{H}} = 0.1\,\mathrm{cm}^{-3}$. These are typical conditions for the WNM and since the Jeans mass is (marginally) resolved for the WNM, it remains resolved for arbitrarily high densities at their effective pressure. 

Therefore, if it is too computationally expensive to resolve the ``real'' Jeans mass, the ``numerical'' Jeans mass can be artificially increased. A polytropic EOS was already used for this purpose in the simulations by \citet{BateBurkert1997}. \citet{RichingsSchaye2016} studied the effect of various pressure floor normalisations in simulations of a dwarf galaxy with a particle mass of $750\,\mathrm{M}_{\odot}$ and showed that the lowest mass gaseous clumps that formed in their galaxy disks are less compact and less gravitationally bound with a higher artificial pressure floor. An artificially increased Jeans mass in the form of a polytropic EOS, a pressure or entropy floor, or the \citet{SpringelHernquist} subgrid model\footnote{Hydrodynamically, the \citet{SpringelHernquist} model acts as a polytropic EOS with a density-dependent $\gamma_{\mathrm{eff}}$, see their figure 1.} is implemented in almost\footnote{One recent exception is the FIREbox project \citep{FIREbox}.} all simulations of cosmologically representative volumes that reach $z=0$.

\begin{table*}
    \centering
    \caption{Overview of equations for Newtonian gravity, a Plummer softened gravitational potential, and a Wendland C2 softened potential.}
    \begin{tabular}{l|lcl}
        \hline
        \multicolumn{4}{l}{Newtonian (no softening)}  \\
        \hline
        Potential &   $\varphi_{\mathrm{N}}(r)$ &=& $- GMr^{-1}$ \\
        Acceleration & $|a_{\mathrm{N}}(r)| $ &=& $ GMr^{-2}$\\
        Free-fall time & $t_{\mathrm{ff,N}}$ &=& $\left(\frac{3\pi}{32 G \rho}\right)^{1/2} = 4.4\, \mathrm{Myr} \left ( \frac{n_{\mathrm{H}}}{100\,\mathrm{cm}^{-3}}\right)^{-1/2}$  \\
        Jeans length & $\lambda_{\mathrm{J,N}}$ &=& $\left ( \frac{3\pi\gamma X_{\mathrm{H}}k_{\mathrm{B}}T}{32 G m_{\mathrm{H}}^2 n_{\mathrm{H}}}\right )^{1/2} = 1.5\,\mathrm{pc} \left (\frac{T}{10\,\mathrm{K}}\right)^{1/2} \left (\frac{n_{\mathrm{H}}}{100 \,\mathrm{cm}^{-3}}\right)^{-1/2} $\\  
        Jeans mass & $M_{\mathrm{J,N}}$ &=& $\frac{4\pi\rho}{3}  \lambda_{\mathrm{J,N}}^3 =  46 \,\mathrm{M}_{\odot} \,\left( \frac{T}{10\,\mathrm{K}} \right)^{3/2} \left ( \frac{n_{\mathrm{H}}}{100\,\mathrm{cm}^{-3}}\right )^{-1/2} $ \\
        \hline
        \multicolumn{4}{l}{Plummer softening with softening scale $\epsilon = \epsilon_{\mathrm{Plummer}}$}  \\
        \hline        
        Potential &  $\varphi_{\mathrm{P}}(r, \epsilon) $ &=& $ - GM\left(r^2 + \epsilon^2 \right)^{-1/2}$  \\
        Acceleration & $|a_{\mathrm{P}}(r,\epsilon)| $ &=& $ GMr \left(r^2 + \epsilon^2\right )^{-3/2}$ \\
        Free-fall time & $t_{\mathrm{ff,P,fit}}$ &=&$t_{\mathrm{ff,N}} \left ( 1 + 2^{2/3} \left (\frac{\epsilon}{R} \right )^{2} \right )^{3/4}$ \\
        Jeans length   & $\lambda_{\mathrm{J,P,fit}}$ &=& $\lambda_{\mathrm{J,N}} \left ( 1 + 1.42 \left (\frac{\epsilon}{\lambda_{\mathrm{J,N}}}\right)^{3/2} \right)^{2/5}$\\
        Jeans mass     & $M_{\mathrm{J,P,fit}}$ &=& $M_{\mathrm{J,N}} \left ( 1 + 1.42 \left ( \frac{\epsilon}{\lambda_{\mathrm{J,N}}}\right )^{3/2}\right )^{6/5} $\\
        \hline
        \multicolumn{4}{l}{Wendland C2 / \textsc{Swift} softening with softening scale $H = 3\epsilon=\epsilon_{\mathrm{Plummer}}$ and $u = r/H$}  \\
        \hline       
        Potential &  $\varphi_{\mathrm{W}}(r<H, H) $ &=& $ -GMH^{-1} W(u)$  \\
                  & $\quad\quad$ with $W(u)$ &=& $\left( -3u^7 + 15u^6 - 28u^5 +21u^4 - 7u^2 +3\right)$\\
                  &  $\varphi_{\mathrm{W}}(r \ge H) $ &=& $ - GMr^{-1}$ \\
        Acceleration & $|a_{\mathrm{W}}(r<H,H)| $ &=& $ GMr H^{-3} V(u)$\\
                  & $\quad\quad$ with $V(u)$ &=&$-W'(u)/u$ = $\left (21u^5 - 90u^4 + 140u^3 - 84u^2 + 14 \right )$\\
                     & $|a_{\mathrm{W}}(r\ge H)| $ &=& $ GM r^{-2}$ \\
        Free-fall time & $t_{\mathrm{ff,W,fit}}$ &=&$t_{\mathrm{ff,N}} \left ( 1 + \frac{1}{7} \left (\frac{H}{R} \right )^{3} \right )^{1/2}$  = $t_{\mathrm{ff,N}} \left ( 1 + \frac{27}{7} \left (\frac{\epsilon}{R} \right )^{3} \right )^{1/2}$\\
        Jeans length  & $\lambda_{\mathrm{J,W,fit}}$ &=& $\lambda_{\mathrm{J,N}} \left ( 1 + 0.27 \left (\frac{H}{\lambda_{\mathrm{J,N}}}\right)^2 \right)^{3/10}$ = $\lambda_{\mathrm{J,N}} \left ( 1 + 2.43 \left (\frac{\epsilon}{\lambda_{\mathrm{J,N}}}\right)^2 \right)^{3/10}$\\
        Jeans mass & $M_{\mathrm{J,W,fit}}$ &=& $M_{\mathrm{J,N}} \left ( 1 + 0.27 \left ( \frac{H}{\lambda_{\mathrm{J,N}}}\right )^2\right )^{9/10}$ = $M_{\mathrm{J,N}} \left ( 1 + 2.43 \left ( \frac{\epsilon}{\lambda_{\mathrm{J,N}}}\right )^2\right )^{9/10}$\\
        \hline
    \end{tabular}
    \label{tab:overview}
\end{table*}

The resolution of the gaseous component in a simulation is consequently not only defined by one mass and two spatial (gravity and hydrodynamic) resolutions per particle type (i.e. dark matter, gas, and stars) but also affected by a polytropic EOS. For example, a simulation with a theoretically infinite mass and spatial resolution would still not capture the dynamics of the cold gas phase correctly in the presence of a polytropic EOS. An obvious step forward is to remove any artificial pressure floor and allow for a multi-phase ISM (i.e. hot ionized, warm ionized / neutral, cold neutral, as well as molecular gas) to form. Based on the Jeans mass arguments above, this would only be possible for simulations with a baryon mass resolution of $\ll 10^4\,\mathrm{M}_{\odot}$. We argue here that the Newtonian Jeans mass does not need to be resolved to avoid numerical clumping and given the success of modern cosmological simulations in reproducing general galaxy properties, it can be questioned how important it really is - in the context of the objectives of large-scale simulations - to model the collapse of individual gas clouds within the ISM correctly. 

The aim of this work is to define criteria that help to avoid artificial fragmentation and collapse for simulations with both adaptive and constant values for the gravitational force softening in Lagrangian codes. This is particularly relevant for simulations that model the multi-phase ISM without an artificial effective pressure floor. 

This paper is organized as follows. In section~\ref{sec:Jeans} we introduce the ``softened'' Jeans criteria (softened Jeans length and softened Jeans mass) and explain why they are more appropriate for describing the conditions of gravitational instabilities in a simulation with softened gravity than the physical Jeans criteria that assume Newtonian gravity. We discuss in section~\ref{sec:application} for which densities and temperatures gravitational instabilities are modelled physically correctly, or are numerically induced or suppressed, for both constant and adaptive softening. This section includes an independent criterion on the minimum smoothing length, $h_{\mathrm{min}}$ for which a too large value can lead to a numerical runaway collapse (section~\ref{sec:runawaycollapse}).  
The impact of different choices for the force resolution parameters $\epsilon$ and $h_{\mathrm{min}}$ is illustrated using simulations of isolated disk galaxies with the modern hydrodynamics code \textsc{Swift}\footnote{\textsc{Swift} is publicly available at: \url{www.swiftsim.com}} \citep{swift2018,swift2023} in section~\ref{sec:simulations}.
The implications for (cosmological) galaxy formation simulations are discussed in section~\ref{sec:discussion} and we summarize the results in section~\ref{sec:summary}.

In the literature, the terms ``smoothing'' and ``softening'' are each sometimes used for either the hydrodynamical or gravitational force calculations involving a ``smoothing'' / ``softening'' kernel. Throughout this work we strictly use ``smoothing'' when referring to the calculation of hydrodynamical quantities and ``softening'' for calculations of gravitational forces. We use $\log$ as $\log_{\mathrm{10}}$ throughout this work.

\section{Softened Jeans criteria for a constant softening length}\label{sec:Jeans}

In galaxy formation simulations, mass distributions are discretized by resolution elements such as particles, static grid cells, or moving mesh cells. In order to reduce the artificial 2-body scattering from individual particles that represent a continuous distribution, the gravitational forces are softened for close interactions at a distance $r$ smaller than the softening length $\epsilon$ in Lagrangian codes. The Newtonian gravitational acceleration, $a_{\mathrm{N}} \propto r^{-2}$ is replaced by the softened gravitational acceleration, for example $a_{\mathrm{P}}\propto r (r^2 + \epsilon^2)^{-3/2}$ for a \citet{Plummer1911} potential (see Table~\ref{tab:overview} for an overview). In the limit of large separations, $r \gg \epsilon$, the softened acceleration equals the Newtonian acceleration ($a_{\mathrm{P}} \rightarrow a_{\mathrm{N}} \propto r^{-2}$) but for $r \ll \epsilon$, the softened acceleration approaches $a_{\mathrm{P}}\propto r$ and therefore diverges from the Newtonian gravity as $r \rightarrow 0$. 

In simulations with a particle mass of $m_{\mathrm{B}} \gtrsim 10^5\,\mathrm{M}_{\odot}$, gravitational instabilities for densities typical for the cold neutral and molecular gas phases in the ISM are unresolved (examples will be shown and discussed in section~\ref{sec:application} and in particular in Fig.~\ref{fig:zoneexamples}) and the classical (Newtonian) Jeans criteria provide an inaccurate estimate of the conditions for gravitational instability.

We re-derive the classical Jeans criteria in order to obtain a better description of gravitational instabilities that occur in numerical simulations with softened gravity\footnote{For the derivation of the softened Jeans criteria we assume that the hydrodynamic forces are calculated accurately. The impact of unresolved hydrodynamic forces on numerical instabilities is discussed in detail in section~\ref{sec:application}.}. We show the derivation for two different softened gravitational potentials: the classical \citet{Plummer1911} potential as well as the modern \citet{Wendland1995} C2 potential (as implemented in \textsc{Swift}) in appendix~\ref{sec:derivation}. We do not repeat the derivations for other potential shapes, such as the cubic spline kernel (as e.g. used in \textsc{Gadget4}, \citealp{Gadget4}) but the difference between the different softening potentials is negligible compared to the difference between Newtonian and softened Jeans criteria. 

Table~\ref{tab:overview} summarizes the equations for free-fall times, Jeans lengths and Jeans masses for Newtonian and softened (Plummer and Wendland C2) gravity. The free-fall time in softened gravity (here: $t_{\mathrm{ff,s}} = t_{\mathrm{ff,W,fit}}$ from equation~\ref{eq:tffC2fit}) 

\begin{equation}\label{eq:tffNintext}
    t_{\mathrm{ff,s}} = t_{\mathrm{ff,N}}\left (1 + \frac{27}{7} \left ( \frac{\epsilon}{R} \right )^3 \right )^{1/2}
\end{equation}

\noindent
is longer than the free-fall time in Newtonian gravity for an object of radius $R$ and is proportional to

\begin{equation}\label{eq:tffsintext}
    t_{\mathrm{ff,s}} \propto t_{\mathrm{ff,N}} \left (\frac{\epsilon}{R} \right )^{3/2}
\end{equation}

\noindent
for $\epsilon \gg R$. As example, doubling the softening length therefore slows down the gravitational free-fall in the simulations by a factor of 2.8 for constant initial $R$.

\begin{figure}
    \includegraphics[width=\columnwidth]{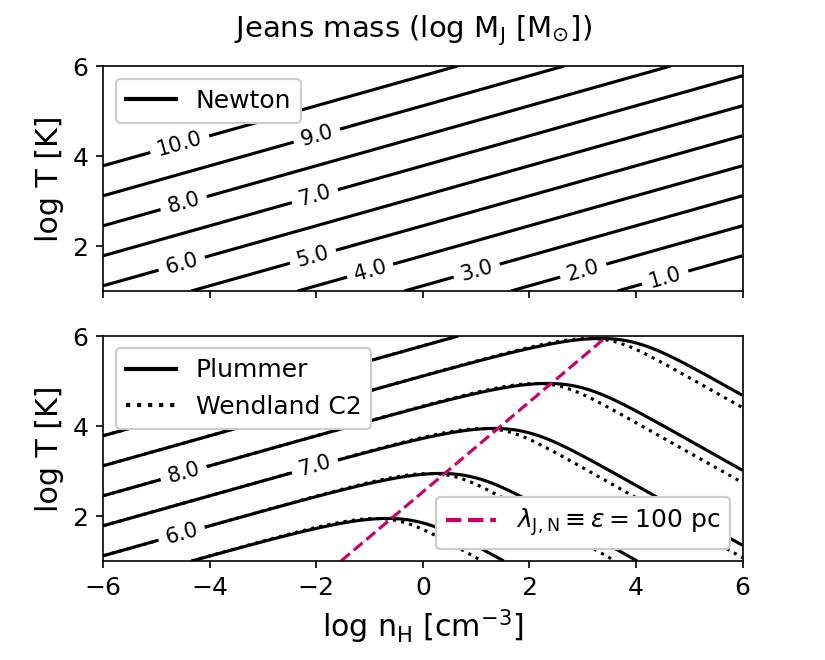}
    \caption{The contours show the Jeans mass in units of $\log M_{\mathrm{J}} [\mathrm{M}_{\odot}]$ for a Newtonian gravitational potential (top panel), a  Plummer softened (bottom panel, solid lines) and a  Wendland C2 softened (bottom panel, dotted lines) potential. The red dashed line in the bottom panel indicates where the Newtonian Jeans mass equals the Plummer softening scale $\epsilon$ (here: $\epsilon = 100\,\mathrm{pc}$).}
    \label{fig:MJ100pc}
\end{figure}

\begin{figure*}
    \centering
    \includegraphics[width=0.19\linewidth]{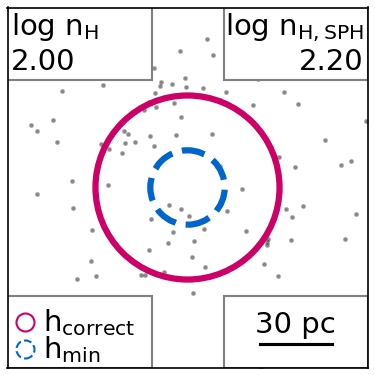}
    \includegraphics[width=0.19\linewidth]{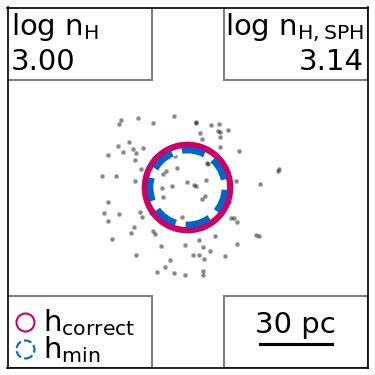}
    \includegraphics[width=0.19\linewidth]{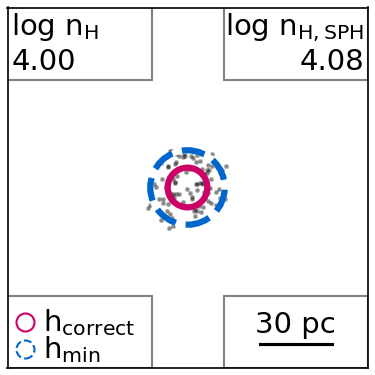}
    \includegraphics[width=0.19\linewidth]{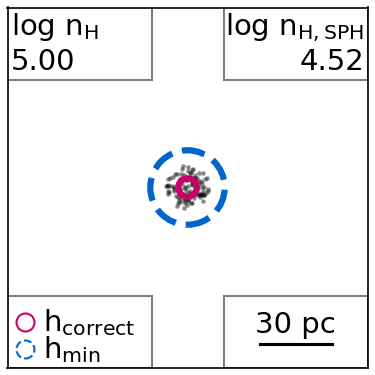}
    \includegraphics[width=0.19\linewidth]{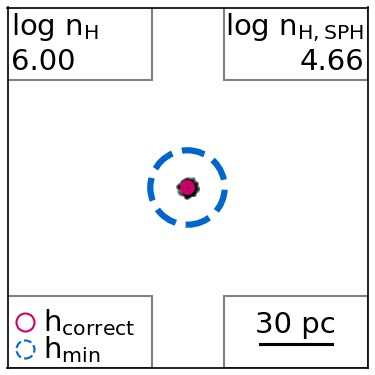}
    \includegraphics[width=0.19\linewidth]{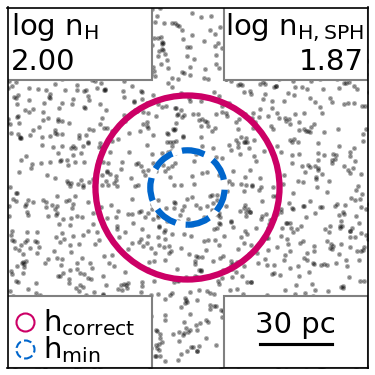}
    \includegraphics[width=0.19\linewidth]{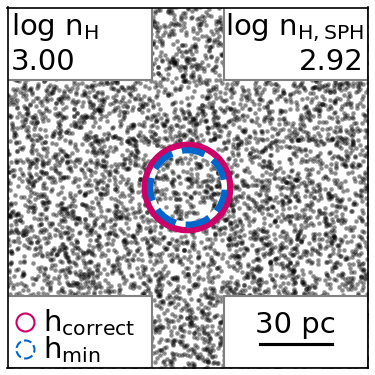}
    \includegraphics[width=0.19\linewidth]{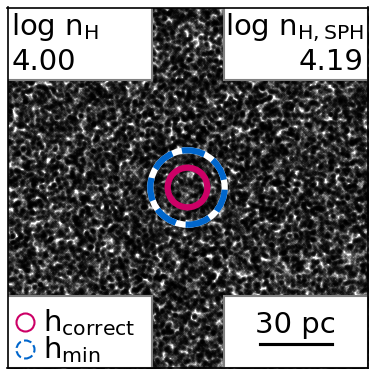}
    \includegraphics[width=0.19\linewidth]{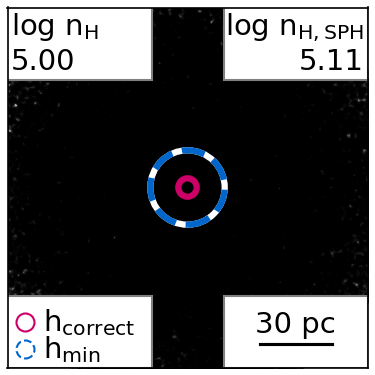}
    \includegraphics[width=0.19\linewidth]{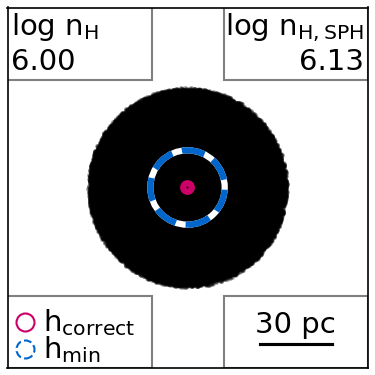}    
    \caption{Random distributions of particles (particle mass $m_{\mathrm{B}} = 10^5\,\mathrm{M}_{\odot}$) in 3D that represent input densities from  $\log n_{\mathrm{H}} [\mathrm{cm}^{-3}] =2$ (leftmost panels) to $\log n_{\mathrm{H}} [\mathrm{cm}^{-3}] =6$ (rightmost panels). The density at the centre of each distribution is calculated with a Wendland C2 kernel and a smoothing length which is the maximum of $h_{\mathrm{correct}}$ (red solid circle) and $h_{\mathrm{min}}$ (blue dashed circle) and indicated in the top right of each panel ($n_{\mathrm{H,SPH}}$). In the top panels the dense gas clump (black circles) consists of 100 particles ($\approx$ number of neighbours in the kernel) and the bottom panels represent a more homogeneous medium without clumping on scales below $h_{\mathrm{min}}$. If the particles clump on scales below $h_{\mathrm{min}}$, then the density that the SPH solver calculates is too low (see top right panel) but if the gas distribution is smooth on scales smaller than $h_{\mathrm{min}}$, then the SPH density estimate is still accurate, even if $h_{\mathrm{correct}} \ll h_{\mathrm{min}}$ (see bottom right panel).}
    \label{fig:hminillustration}
\end{figure*}

Fig.~\ref{fig:MJ100pc} shows an overview of the Jeans masses from equations~(\ref{eq:MJ0}), (\ref{eq:MJP}), and (\ref{eq:MJC2}) for the Newtonian (top panel), Plummer softened (solid line, bottom panel), and the Wendland C2 softened (dotted line, bottom panel) potential, respectively, for a Plummer equivalent softening length of $\epsilon = 100\,\mathrm{pc}$. At densities above (or at temperatures below) the dashed red line in the bottom panel, the Newtonian Jeans length is smaller than the softening length $\epsilon$, and the softened Jeans masses (bottom panel) exceed the Newtonian Jeans mass (top panel). This means that the limit between growing and decaying density perturbations is described by the softened Jeans mass (length) rather than the Newtonian Jeans mass (length). In addition to the different fragmentation scale, the gravitational collapse follows the longer softened free-fall time, rather than the Newtonian free-fall time.

The difference between the Plummer (solid lines) and the Wendland C2 (dotted lines) softened potentials is small considering the approximate nature of the Jeans criterion. In contrast, the softened Jeans mass can be several orders of magnitude larger than the Newtonian Jeans mass for gas densities and temperatures that are typical for the cold ISM. For example, at a gas temperature of a few tens of K and a gas density of $100\,\mathrm{cm}^{-3}$, the Newtonian Jeans mass is $M_{\mathrm{J,N}}\approx 100\,\mathrm{M}_{\odot}$ while the softened Jeans masses for $\epsilon = 100\,\mathrm{pc}$ is  $M_{\mathrm{J,s}}\approx 10^5\,\mathrm{M}_{\odot}$ (and for $\epsilon = 1000\,\mathrm{pc}$, $M_{\mathrm{J,s}}\approx 10^7\,\mathrm{M}_{\odot}$) for both $M_{\mathrm{J,s}} = M_{\mathrm{J,P,fit}}$ (Plummer) and $M_{\mathrm{J,s}} = M_{\mathrm{J,W,fit}}$ (Wendland C2).

Throughout this work we consider for simplicity only thermal pressure as a stabilizing force against gravitational collapse in the derivation of the Newtonian Jeans length and mass. For applications where (unresolved) turbulent pressure or magnetic pressure is important, their contributions can be added to the Newtonian Jeans criteria by substituting the sound crossing time in equation~(\ref{eq:tsc}) with a more general form of a signal crossing time, $t_{\mathrm{sig}}$, that can also include a turbulent or a magnetic component (see \citealp{Folkert2023arXiv230913750N} for an example of using the turbulent velocity dispersion, $\sigma_{\mathrm{turb}}$, and the velocity dispersion of ions, $\sigma_{\mathrm{Alfven}}$, as additional terms, when deriving the Jeans length). The softened Jeans mass and length are defined relative to the Newtonian Jeans mass and length. For example, 

\begin{align}
    \lambda_{\mathrm{J,s}} &= \lambda_{\mathrm{J,N}} \left ( 1 + 2.43 \left (\frac{\epsilon}{\lambda_{\mathrm{J,N}}}\right)^2 \right)^{3/10} \\
    M_{\mathrm{J,s}} &= M_{\mathrm{J,N}} \left ( 1 + 2.43 \left ( \frac{\epsilon}{\lambda_{\mathrm{J,N}}}\right )^2\right )^{9/10} \\
\end{align}

\noindent
for the Wendland C2 kernel (Table~\ref{tab:overview}).
The definitions of the softened Jeans mass and length in Table~\ref{tab:overview} are therefore unaffected by these additional components.

The softened Jeans masses derived in appendix~\ref{sec:derivation} and listed in Table~\ref{tab:overview} differ from the Newtonian Jeans mass whenever the particle separations are below the softening length. This might be most prominent in simulations with a (large) constant value for $\epsilon$ but is also applicable for simulations with adaptive softening when $\epsilon$ is limited by a minimum value $\epsilon_{\mathrm{min}}$.

\section{Numerical instabilities related to limited resolution}\label{sec:application}

In large-scale simulations, for example of cosmological volumes, gas can reach gas densities and temperatures that are not formally resolved. Either because gravity or hydrodynamic forces are softened or smoothed on scales larger than the Jeans length or because realistic fragmentation masses are not resolved by enough resolution elements. This is often acceptable, for example if the averaged properties of the ISM within a galaxy are more of interest than correctly following the collapse of individual gas clouds, especially if the collapse of molecular clouds is approximated by subgrid prescriptions, e.g. for star formation. For code stability it is typically preferable to numerically suppress the collapse of gas clouds than to numerically induce collapse because the latter can lead to prohibitively small timesteps, and artificial clumps in the stellar distribution. 

In this section we discuss two distinct instabilities that can occur in Lagrangian simulations and how they depend on the gravitational softening length (i.e. the gravitational force resolution), the hydrodynamic smoothing length (i.e. the hydrodynamical force resolution) and the mass resolution. In subsection~\ref{sec:runawaycollapse} we identify the problematic behaviour of SPH-like simulation codes when the hydrodynamical smoothing length $h$ is limited by a too large minimum value, $h_{\mathrm{min}}$. The resulting instability is caused by an underestimate of the gas density and leads to artificial clumps of closely spaced particles. Setting $h_{\mathrm{min}}$ to a very small non-zero value resolves this issue and we cannot identify any disadvantages of letting $h_{\mathrm{min}} \rightarrow 0$ (to be discussed in detail in section~\ref{sec:simulations}).

In subsection~\ref{sec:adaptivesoft} we analyse gravitational instabilities in simulations with either constant or adaptive softening lengths at and below the hydrodynamical resolution limit, i.e. the size of an individual smoothing kernel.

\subsection{Numerical instability caused by imposing a minimum hydrodynamical smoothing length}\label{sec:runawaycollapse}

We have shown that depending on the gravitational softening scale, the softened Jeans mass can be orders of magnitude larger than the Newtonian Jeans mass and increases with density at constant temperature for $\lambda_{\mathrm{J,N}} < \epsilon$ (Fig.~\ref{fig:MJ100pc}). Gas clumps would therefore be increasingly stabilised against gravitational collapse as their density gets larger. However, this assumes that the hydrodynamic forces are modelled accurately. For simulations with a minimum smoothing length, $h_{\mathrm{min}}\ne 0$, the hydrodynamic forces can be over-smoothed for high gas densities and we explain below how this can cause a numerical runaway collapse. 

In this section, the smoothing length, $h$, and its lower limit, $h_{\mathrm{min}}$, are defined as in the hydrodynamics solver \textsc{Sphenix} \citep{sphenix}, implemented in the \textsc{Swift} code\footnote{We discuss in section~\ref{sec:adaptivesoft} how this can be adapted to other simulations codes by defining a general smoothing length scale, $l_{\mathrm{smooth}}$.}. In \textsc{Swift}, the smoothing length is based on the kernel standard deviation as in \citet{DehnenAly2012} and is calculated as

\begin{align}
    h &= \eta_{\mathrm{res}} \left (\frac{X_{\mathrm{H}} m_{\mathrm{B}}}{m_{\mathrm{H}} n_{\mathrm{H}}} \right )^{1/3} \nonumber \\
                         & = 38.3\,\mathrm{pc} \left ( \frac{m_{\mathrm{B}}}{10^5\,\mathrm{M}_{\odot}} \right )^{1/3} \left ( \frac{n_{\mathrm{H}}}{100\,\mathrm{cm}^{-3}} \right )^{-1/3} \label{eq:hcorrect}
\end{align}\label{eq:smoothinglength}

\noindent
where $\eta_{\mathrm{res}}$ is a constant related to the number of neighbours in the kernel, $X_{\mathrm{H}}$ is the hydrogen mass fraction, $m_{\mathrm{B}}$ the baryon particle mass, $m_{\mathrm{H}}$ the hydrogen particle mass, $n_{\mathrm{H}}$ and the hydrogen number density. We use the typical values $\eta_{\mathrm{res}} = 1.2348$ (this corresponds to 58 neighbours with a Wendland C2 kernel) and $X_{\mathrm{H}} = 0.74$.

The minimum smoothing length in \textsc{Swift} is set by the dimensionless parameter $h_{\mathrm{min,ratio}}$ which is defined as the ratio between the kernel support, $h \gamma_{\mathrm{k}}$, and the length scale above which gravity is Newtonian, $H=3\epsilon$. For the Wendland C2 kernel $\gamma_{\mathrm{k}} = 1.936492$ \citep{DehnenAly2012} and the minimum smoothing length is therefore $h_{\mathrm{min}} = 1.55 \, \epsilon \, h_{\mathrm{min,ratio}}$. The density above which the hydrodynamic forces are over-smoothed is

\begin{align}\label{eq:nHhmin}
    n_{\mathrm{H,hmin}} & = \frac{X_{\mathrm{H}}}{m_{\mathrm{H}}} \eta_{\mathrm{res}}^3   \left ( \frac{m_{\mathrm{B}}}{h_{\mathrm{min}}^{3}} \right ) \nonumber \\
                        & = 5.6\,\mathrm{cm}^{-3} \left ( \frac{m_{\mathrm{B}}}{10^5\,\mathrm{M}_{\odot}} \right ) \left ( \frac{h_{\mathrm{min}}}{100\,\mathrm{pc}}\right )^{-3} \;.
\end{align}

\noindent
Note that the critical density $n_{\mathrm{H,hmin}}$ can remain constant for simulations with different mass resolutions $m_{\mathrm{B}}$ if $h_{\mathrm{min}}$ is adjusted accordingly.

The softened Jeans criteria derived in appendix~\ref{sec:derivation} are therefore only valid for $n_{\mathrm{H}} \le n_{\mathrm{H,hmin}}$. For higher densities the hydrodynamic forces can become inaccurate due to over-smoothing and this inaccuracy depends on the density contrast within $h_{\mathrm{min}}$. The top panels of Fig.~\ref{fig:hminillustration} demonstrate this for an extreme case with one clump of dense gas (represented by 100 resolution elements: circles in Fig.~\ref{fig:hminillustration}). The 3D particle positions are random but restricted to a volume that corresponds to a set input density $n_{\mathrm{H}}$ which increases from $\log n_{\mathrm{H}} [\mathrm{cm}^{-3}] =2$ (leftmost panel) to $\log n_{\mathrm{H}} [\mathrm{cm}^{-3}] =6$ (rightmost panel). The red solid circle represents the smoothing length for this gas density and particle mass (equation~\ref{eq:hcorrect}) and the blue dashed circle represents the minimum smoothing length which is set to $h_{\mathrm{min}} = 15.5\,\mathrm{pc}$ (corresponding to e.g. $\epsilon = 100\,\mathrm{pc}$ and $h_{\mathrm{min,ratio}} = 0.1$) in this example. 

Each panel in Fig.~\ref{fig:hminillustration} lists the input density as well as the density $n_{\mathrm{H,SPH}}$ that the SPH solver calculates at the centre of each box using a Wendland C2 kernel. If $h$ becomes smaller than $h_{\mathrm{min}}$ and the gas is clumpy on scales smaller than $h_{\mathrm{min}}$, then the estimated density (and pressure) is increasingly underestimated. This can lead to a runaway collapse if the underestimated pressure does not balance the (softened) gravitational forces. The artificial, unresolved collapse is difficult to spot in the gas properties because the SPH gas density does not correctly represent the particle distribution, but it can lead to the formation of clusters of stellar particles that are denser than expected based on the resolution of the simulation and the gas densities.
The bottom panels of Fig.~\ref{fig:hminillustration} show a similar situation, but here the dense gas is homogeneous on scales below $h_{\mathrm{min}}$, in which case the output densities $n_{\mathrm{H,SPH}}$ remain accurate, even for $h \ll h_{\mathrm{min}}$ (see also figure~1 in \citealp{Borrow2021} for an illustration). 

\begin{figure}
    \centering
    \includegraphics[width=0.9\linewidth]{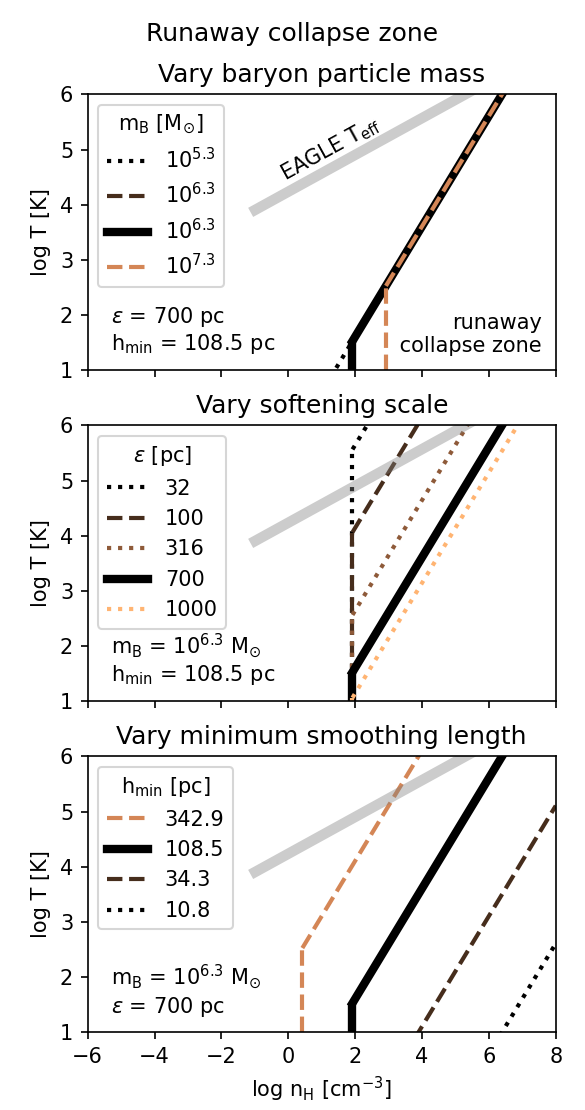}
    \caption{Lines indicate the borders of the runaway collapse zone (see text) for various combinations of the baryon particle mass $m_{\mathrm{B}}$, the gravitational softening length $\epsilon$, and the minimum smoothing length $h_{\mathrm{min}}$. At densities above the indicated limits (both the vertical and the inclined part), the SPH density estimate may be incorrect and a runaway collapse can occur. Each panel varies one resolution parameter (top: $m_{\mathrm{B}}$, middle: $\epsilon$, bottom: $h_{\mathrm{min}}$) while keeping the remaining two constant at the fiducial parameters of the \textsc{eagle} $(100\,\mathrm{Mpc})^3$ simulation (black solid line, see labels). The thick grey line represents the effective temperature used in \textsc{eagle} for gas with densities $>0.1\,\mathrm{cm}^{-3}$.}
    \label{fig:zone}
\end{figure}

To avoid a potential run-away collapse, which is desirable for both numerical and physical reasons, we can define a problematic region in the density-temperature phase-space based on the conditions discussed above: 

\begin{enumerate}[i]
  \item $h < h_{\mathrm{min}}$ (or: $n_{\mathrm{H}} > n_{\mathrm{H,hmin}}$)
  \item fragmentation on scales below $h_{\mathrm{min}}$
\end{enumerate}

\noindent
 For an estimate of the clumping that is expected in the simulation, we use the softened Jeans length and (ii) then translates to $\lambda_{\mathrm{J,s}} < h_{\mathrm{min}}$.

Assuming that $h_{\mathrm{min}} < \epsilon$ (to avoid inducing numerical collapse, see e.g.~\citealp{BateBurkert1997}), the condition $\lambda_{\mathrm{J,s}} < h_{\mathrm{min}}$ falls into the softened part of the phase-space where $H \equiv 3\epsilon \gg \lambda_{\mathrm{J,N}}$ and we therefore approximate the softened Jeans mass (here for the Wendland C2 kernel, equation~\ref{eq:lJfitWendland}) with $\lambda_{\mathrm{J,s}}\approx 0.27^{0.3} \, \lambda_{\mathrm{J,N}}^{0.4} \, H^{0.6}$ and the conditions (i) and (ii) for a potential run-away collapse correspond to the following regions in the phase-space diagram as

\begin{align}
    n_{\mathrm{H}} & > 5.6\,\mathrm{cm}^{-3} \left ( \frac{m_{\mathrm{B}}}{10^5\,\mathrm{M}_{\odot}}\right ) \left ( \frac{h_{\mathrm{min}}}{100\,\mathrm{pc}}\right )^{-3} \; ,\label{eq:zonedens}\\
   T &< 27\,\mathrm{K} \left (\frac{h_{\mathrm{min}}}{100\,\mathrm{pc}}\right)^5 \left (\frac{\epsilon}{700\,\mathrm{pc}}\right)^{-3} \left ( \frac{n_{\mathrm{H}}}{100\,\mathrm{cm}^{-3}}\right) \;. 
   \label{eq:zonetemp}
\end{align}

This ``runaway collapse zone'' is indicated in Fig.~\ref{fig:zone} for various combinations of the particle mass $m_{\mathrm{B}}$, Plummer equivalent softening length $\epsilon$, and the minimum smoothing length $h_{\mathrm{min}}$. The default values (thick, black lines) with $\log m_{\mathrm{B}} [\mathrm{M}_{\odot}] =6.3$, $\epsilon = 700\,\mathrm{pc}$, and $h_{\mathrm{min}} = 108.5\,\mathrm{pc}$ (i.e. $h_{\mathrm{min,ratio}} = 0.1$) are those from the \textsc{eagle} $(100\,\mathrm{Mpc})^3$ simulation and two of these parameters are constant in each panel while the third parameter is varied (top panel: $m_{\mathrm{B}}$, middle panel: $\epsilon$, bottom panel: $h_{\mathrm{min}}$).

Each zone is limited through a combination of a vertical line, condition (i): $n_{\mathrm{H}} = n_{\mathrm{H,hmin}}$ (equation~\ref{eq:zonedens}) and a line of $\log T \propto \log n_{\mathrm{H}}$, i.e. for a constant softened Jeans length of $\lambda_{\mathrm{J,s}} = h_{\mathrm{min}}$ (condition ii, equation~\ref{eq:zonetemp}). Densities with higher values than the indicated lines are reported incorrectly by the SPH kernel density estimator if the gas is clumpy on scales below $h_{\mathrm{min}}$ and can lead to a runaway collapse.

Varying the baryon particle mass $m_{\mathrm{B}}$ at constant $\epsilon$ and $h_{\mathrm{min,ratio}}$ (Fig.~\ref{fig:zone}, top panel) only changes $n_{\mathrm{H,hmin}}$ but condition (ii) remains unaffected. A better mass resolution is even slightly counter-productive in resolving higher densities because $n_{\mathrm{H,hmin}}\propto m_{\mathrm{B}}$ for constant $h_{\mathrm{min}}$ (equation~\ref{eq:nHhmin}). Similarly, reducing the gravitational softening length decreases the softened Jeans length and the runaway collapse zone includes higher temperatures (middle panel of Fig.~\ref{fig:zone}). Decreasing the minimum smoothing length at constant $m_{\mathrm{B}}$ and constant $\epsilon$ on the other hand pushes the runaway collapse zone to significantly higher densities (from $10^2\,\mathrm{cm}^{-3}$ to $>10^6\,\mathrm{cm}^{-3}$ when reducing $h_{\mathrm{min}}$ by a factor of 10, bottom panel of Fig.~\ref{fig:zone}). 

If the highest (expected) density in a simulation is $n_{\mathrm{sim,max}}$, the minimum smoothing length should be set so that $h(n_{\mathrm{sim,max}}) \ge h_{\mathrm{min}}$. With $h$ from equation~(\ref{eq:hcorrect}), this means that the minimum smoothing length should follow

\begin{equation}\label{eq:hminmax}
    h_{\mathrm{min}} < 8.3\,\mathrm{pc} \left ( \frac{m_{\mathrm{B}}}{10^5\,\mathrm{M}_{\odot}} \right )^{1/3} \left ( \frac{n_{\mathrm{sim,max}}}{10^4\,\mathrm{cm}^{-3}} \right )^{-1/3} \; ,
\end{equation}

\noindent
for a criterion based only on the vertical lines in Fig.~\ref{fig:zone}. As we will discuss in section~\ref{sec:discussion}, we cannot identify any benefit for larger values for $h_{\mathrm{min}}$ and therefore recommend to use a very small but non-zero\footnote{A non-zero value for $l_{\mathrm{smooth,min}}$ ensures that the simulation does not break down in the highly unlikely, but not impossible, case of a vanishing smoothing length for individual particles.} value for $h_{\mathrm{min}}$ that generously fulfills the conservative criterion from equation~(\ref{eq:hminmax}). 

Equation~\ref{eq:hminmax} does not depend on $\epsilon$ and applies to both simulations with adaptive and constant softening lengths.

\subsubsection{Examples from the literature}\label{sec:examplesliterature}

\begin{figure}
    \centering
    \includegraphics[width=0.85\linewidth]{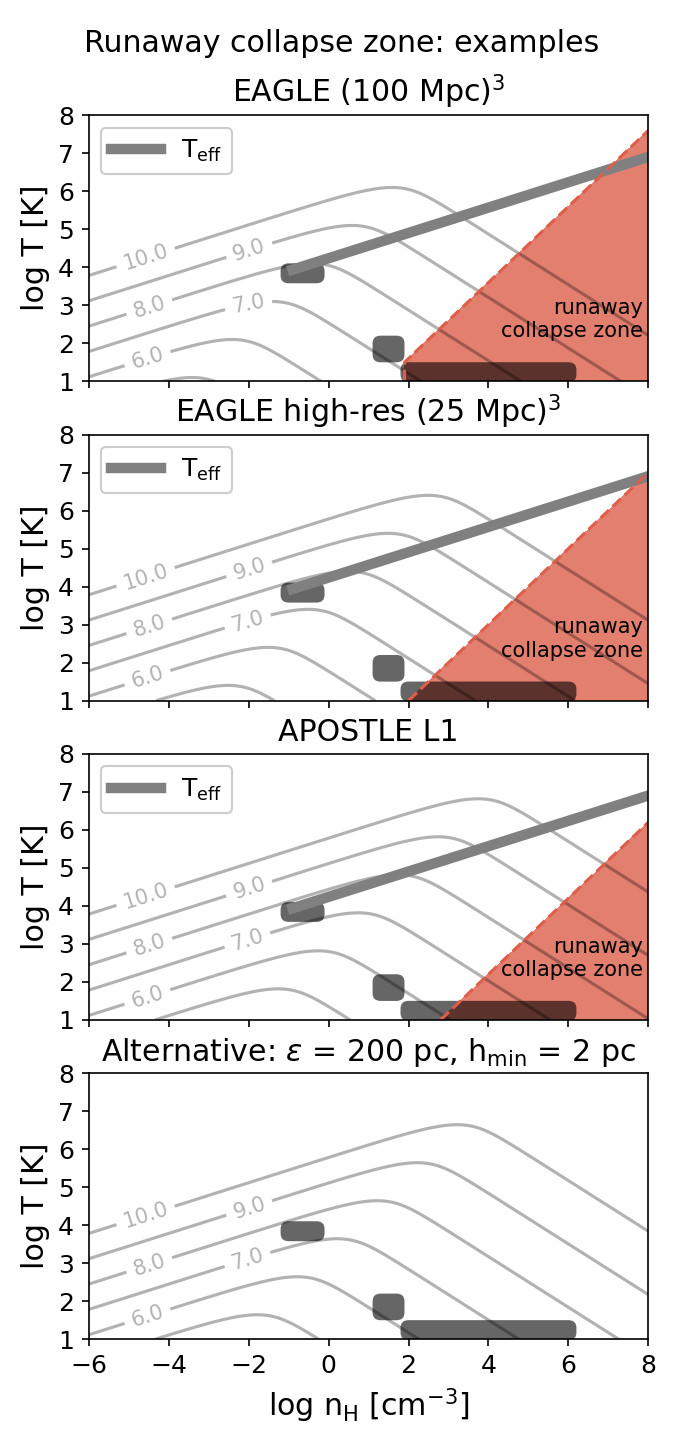}
    \caption{Overview of the density-temperature phase-space for the \textsc{eagle}, \textsc{eagle-high-res}, and \textsc{apostle-L1} resolution parameters, as well as for an alternative set (from top to bottom, see also Table~\ref{tab:simulationsliteratures}). The contours indicate the softened Jeans mass in units of $\log M_{\mathrm{J,s}} [\mathrm{M}_{\odot}]$. The red shaded region at high densities is the runaway collapse zone and SPH-estimated gas densities within this zone may be underestimated (see section~\ref{sec:runawaycollapse}). The grey thick line is the effective temperature floor used in \textsc{eagle}, \textsc{eagle-high-res}, and \textsc{Apostle-L1}. Typical densities and temperatures for the warm neutral medium, the cold neutral medium and molecular clouds (from low to high densities) are indicated with black patches for reference. An alternative set of parameters is shown in the bottom panel: for a mass resolution of order $10^5$ to $10^6\,\mathrm{M}_{\odot}$, a softening scale of 200~pc and a minimum smoothing length of 2~pc ($h_{\mathrm{min,ratio}} = 0.006$) push the runaway collapse zone to densities  $\gtrsim 10^8\,\mathrm{cm}^{-3}$). All phases in the ISM can therefore be modelled without triggering a numerical runaway collapse.}
    \label{fig:zoneexamples}
\end{figure}

The runaway collapse zone is indicated in Fig.~\ref{fig:zoneexamples} for three simulation projects that use the same code (a modified version of \textsc{gadget3}, \citealp{gadget2}) but span more than 2 orders of magnitude in mass resolution with baryon particle masses of $m_{\mathrm{B}} = 1.81\times10^6\,\mathrm{M}_{\odot}$ (\textsc{eagle}), $m_{\mathrm{B}} = 2.26\times10^5\,\mathrm{M}_{\odot}$ (\textsc{eagle-high-res}), and $m_{\mathrm{B}} = 10^4\,\mathrm{M}_{\odot}$ (\textsc{apostle}).

The resolution parameters (see Table~\ref{tab:simulationsliteratures}) used in both \textsc{eagle} flagship runs (\textsc{eagle}, \textsc{eagle-high-res}, \citealp{EAGLE}) and the highest resolution zoom-in simulations from the \textsc{apostle} simulation suite \citep{Apostle2016Fattahi, Apostly2016Sawala} violate the condition from equation~(\ref{eq:hminmax}) for densities of $n_{\mathrm{sim,max}}\gtrsim 100\,\mathrm{cm}^{-3}$. However, for these simulations, the numerically induced runaway collapse was avoided because gas with densities above $0.1\,\mathrm{cm}^{-3}$ follows a polytropic equation of state where the effective temperature\footnote{The polytropic equation of state is in practice implemented as an entropy or internal energy floor. The exact effective temperature therefore depends on the mean particle mass, $\mu$, but this is irrelevant here as we show $T_{\mathrm{eff}}$ only for reference.} is 

\begin{equation}
    T_{\mathrm{eff}} \approx 8000\,\mathrm{K} \left ( \frac{n_{\mathrm{H}}}{0.1\,\mathrm{cm}^{-3}} \right )^{\gamma_{\mathrm{eff}}-1}
\end{equation}

\noindent
with the polytropic index $\gamma_{\mathrm{eff}} = 4/3$. This effective temperature is indicated for reference as thick grey line in Figs.~\ref{fig:zone} and \ref{fig:zoneexamples} and is identical for \textsc{eagle} and \textsc{apostle}.

The \textsc{apostle} zoom-in simulations have a particle mass of $m_{\mathrm{B}} = 10^4\,\mathrm{M}_{\odot}$ but the runaway collapse zone is barely affected by the higher mass resolution, compared to e.g. \textsc{eagle}. The effective temperature floor was therefore necessary for their values for $h_{\mathrm{min}}$, despite the low particle mass. 

%Fig.~\ref{fig:zoneexamples} illustrates why a pressure floor was necessary for a selection of cosmological simulations with constant softening and which minor changes in the resolution parameters, $\epsilon$ and $h_{\mathrm{min}}$, avoid numerical runaway collapse in simulations with the same particle mass but without a polytropic equation of state (bottom panel). The first and second panels shows the runaway collapse zones ((equations~\ref{eq:zonedens} and \ref{eq:zonetemp}, red shaded area) for the resolution parameters of the two flaghship simulations within the \textsc{eagle} project (``\textsc{eagle}'', ``\textsc{eagle-high-res}''). The third panel uses the resolution parameters from the highest resolution (``L1'') simulations within the \textsc{apostle} project \citep{Apostle2016Fattahi,Apostly2016Sawala}.

%The resolution parameters for all simulations are listed in Table~\ref{tab:simulationsliteratures}.

\begin{table}
    \caption{Values for the gas particle mass ($m_{\mathrm{B}}$), the gravitational force softening length ($\epsilon$), and the minimum hydrodynamical smoothing length ($h_{\mathrm{min}}$) for a few simulations with constant softening from the literature and an alternative set of resolution parameter proposed in this work as shown in Fig.~\ref{fig:zoneexamples}.}
    \begin{tabular}{llrrl}
        \hline
        Simulation & $m_{\mathrm{B}} [\mathrm{M}_{\odot}]$ & $\epsilon$ [pc] & $h_{\mathrm{min}}$ [pc] & Reference\\
        \hline
        \textsc{eagle}          & $1.81\times10^6$ & 700 & 108.50 & (1) \\
        \textsc{eagle-high-res} & $2.26\times10^5$ & 350 & 54.25 & (1) \\
        \textsc{apostle-L1}     & $1.00\times10^4$ & 135 & 20.77 &  (2,3)  \\
        Alternative    & $\approx 10^5$ to $10^6$ & 200 & 2.00 & -\\
        \hline
    \end{tabular}
    \footnotesize{(1) \citet{EAGLE}; (2) \citet{Apostle2016Fattahi}; (3) \citet{Apostly2016Sawala}}
    \label{tab:simulationsliteratures}
\end{table}

%The runaway collapse zone (equations~\ref{eq:zonedens} and \ref{eq:zonetemp}) is indicated as a red shaded area in each panel. Because the ISM is described by a single phase medium that follows an effective temperature (thick grey line), the gas can enter the runaway collapse zone only at very high densities ($n_{\mathrm{H}} > 10^6\,\mathrm{cm}^{-3}$) and the \textsc{eagle} and \textsc{apostle} simulations do not suffer from artificial runaway collapse and incorrect gas density estimates. 

%Without this effective temperature floor, these simulations would however have been affected by a numerical runaway collapse at densities $n_{\mathrm{H}}\gtrsim 100\,\mathrm{cm}^{-3}$.  
%The adoption of a polytropic equation of state with a polytropic index of $\gamma_{\mathrm{eff}} = 4/3$ in \textsc{eagle} was motivated by the numerical advantage of a constant Jeans mass for the ISM and a constant ratio of the Jeans length to the smoothing length \citep{SchayeDallaVecchia2008}. Based on \citet{BateBurkert1997}, \citet{SchayeDallaVecchia2008} argue that a constant Jeans mass prevents artificial fragmentation because the (Newtonian) Jeans mass is always (marginally) resolved by the same number of particles. 

The softened Jeans mass, as derived in appendix~\ref{sec:derivation} for a Wendland C2 kernel and for the \textsc{eagle} and \textsc{eagle-high-res} softening parameters, is overlaid  as contours in units of $\log M_{\mathrm{J,s}} \,[\mathrm{M}_{\odot}]$ in Fig.~\ref{fig:zoneexamples}. Artificial fragmentation that is caused by not resolving the Jeans mass with enough resolution elements (as in \citealp{BateBurkert1997}) would not be expected as the softened Jeans mass increases with density (at a constant temperature) as soon as gravity is no longer Newtonian. While we argue that numerical runaway collapse would occur in these simulations without an effective temperature floor, the reason for this is not related to the Newtonian Jeans mass but to the adoption of a too large minimum smoothing length as discussed in section~\ref{sec:runawaycollapse} and therefore not very sensitive to the particle mass.

\begin{figure*}
    \centering
  \includegraphics[width=1.0\linewidth]{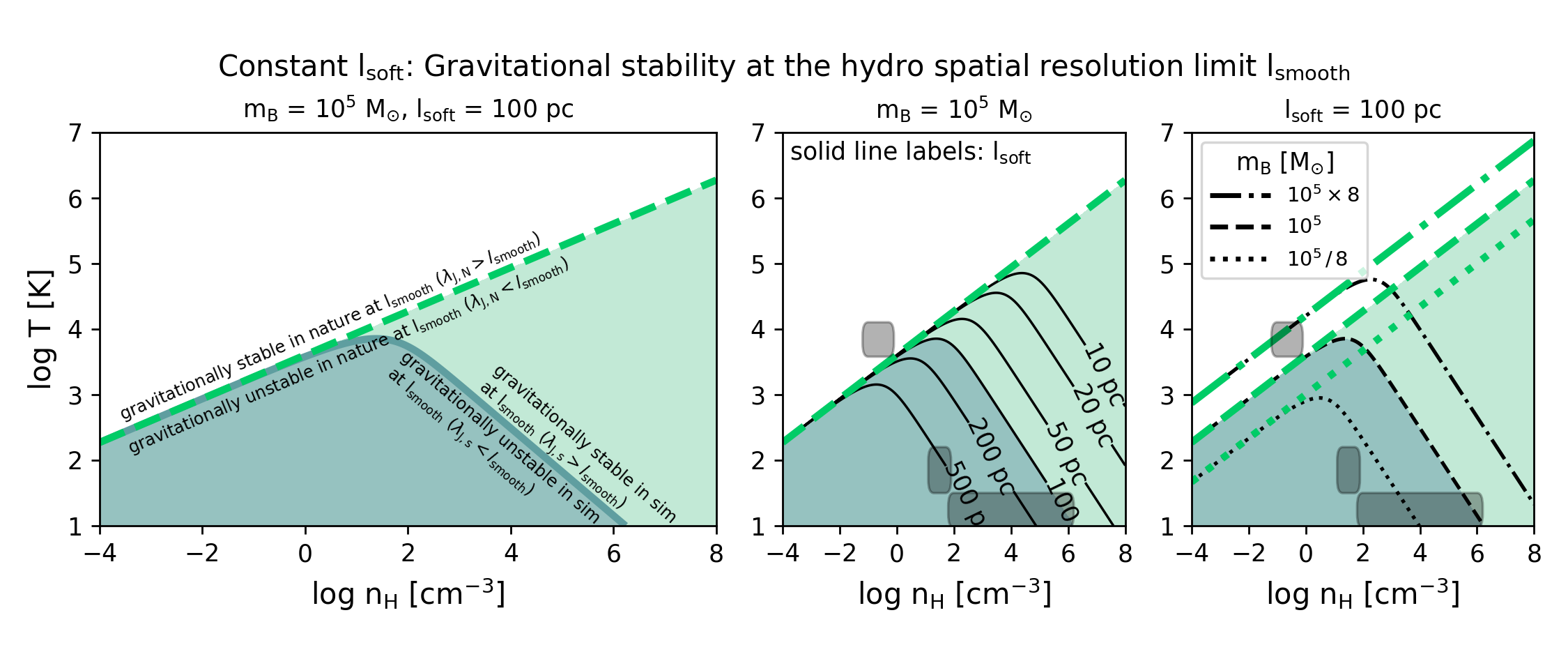}
    \caption{Gravitational stability at the hydrodynamical resolution limit, $l_{\mathrm{smooth}}$, for simulations with a constant softening length $l_{\mathrm{soft}}$. Perturbations at length scale, $l_{\mathrm{smooth}}$, are expected to grow in Newtonian gravity for $\lambda_{\mathrm{J,N}} < l_{\mathrm{smooth}}$ (``gravitationally unstable in nature'') and to decay for $\lambda_{\mathrm{J,N}} > l_{\mathrm{smooth}}$ (``gravitationally stable in nature''), separated by the thick dashed line with a slope of 1/3 ($T\propto n_{\mathrm{H}}^{1/3}$) in each panel for  $l_{\mathrm{smooth,min}}=0$. Here, the gas density, $n_{\mathrm{H}}$, refers to the SPH-estimated density. In simulations with softened gravity, such perturbations are expected to grow for $\lambda_{\mathrm{J,s}} < l_{\mathrm{smooth}}$ (``gravitationally unstable in sim'') and to decay for $\lambda_{\mathrm{J,s}} > l_{\mathrm{smooth}}$ (``gravitationally stable in sim''), separated by the solid line which diverges from the dashed line and follows a slope of -2/3 ($T\propto n_{\mathrm{H}}^{-2/3}$) at high densities. The left panel shows the boundary $\lambda_{\mathrm{J,s}} = l_{\mathrm{smooth}}$ for $m_{\mathrm{B}}=10^5\,\mathrm{M}_{\odot}$ and a constant softening length $l_{\mathrm{soft}} = 100\,\mathrm{pc}$ (thick solid line). In the middle panel the solid lines represent $\lambda_{\mathrm{J,s}} = l_{\mathrm{smooth}}$ for different values for $l_{\mathrm{soft}}$ (see contour labels) and $m_{\mathrm{B}}=10^5\,\mathrm{M}_{\odot}$. The right panel shows the dependence of both $\lambda_{\mathrm{J,N}} = l_{\mathrm{smooth}}$ (thick green lines) and $\lambda_{\mathrm{J,s}} = l_{\mathrm{smooth}}$ (thin black lines) on the particle mass $m_{\mathrm{B}}$ for $l_{\mathrm{soft}} = 100\,\mathrm{pc}$ and $m_{\mathrm{B}} = 10^5/8\,, 10^5, \,\mathrm{and}\, 10^5\times 8\,\mathrm{M}_{\odot}$ (dotted, dashed, dash-dotted lines, respectively). Typical densities and temperatures for the WNM, CNM, and MCs are indicated with dark patches, as in Fig.~\ref{fig:zoneexamples}.}
    \label{fig:zonesatlsmooth}
\end{figure*}

The bottom panel of Fig.~\ref{fig:zoneexamples} illustrates an alternative set of resolution parameters for any mass resolution between those of \textsc{eagle} and \textsc{apostle-L1} (the dependence on $m_{\mathrm{B}}$ is very weak, see top panel of Fig.~\ref{fig:zone}). A Plummer equivalent softening of $\epsilon=200\,\mathrm{pc}$ is small enough to model gravitational instabilities correctly in the WNM but still large enough for a minimum softened Jeans mass of $\approx 10^{6}\,\mathrm{M}_{\odot}$ in all neutral phases of the ISM. Counter-intuitively, a \emph{smaller} softening length would decrease the softened Jeans mass in the cold neutral medium (CNM) and molecular clouds (MCs), which means that the softened Jeans mass would be resolved by fewer particles in these phases. With a minimum smoothing length of $h_{\mathrm{min}} = 2\,\mathrm{pc}$ (instead of $54.25\,\mathrm{pc}$ in \textsc{eagle-high-res}), the runaway collapse zone only covers $n_\mathrm{H}> 10^8\,\mathrm{cm}^{-3}$ and the molecular phase can be directly modelled without suffering from the numerical issues described above. Simulations of isolated galaxies at comparable mass resolutions and without an entropy floor are shown below in section~\ref{sec:simulations}.

Future simulations that include a multi-phase ISM need to carefully select the combination of mass (baryon particle mass), gravity (softening scale), and hydrodynamic resolutions (minimum smoothing length) to avoid unwanted numerical artefacts. 
%If the ISM in the \textsc{eagle} simulations had contained the CNM or the molecular cloud phase, i.e. if gas had not been limited by the effective temperature floor, the choices for the gravitational softening and the minimum smoothing length would have led to issues already for gas densities of $n_{\mathrm{H}}>  100 \,\mathrm{cm}^{-3}$ (see top and middle panels of Fig.~\ref{fig:zoneexamples}), even for the high-resolution simulation.  
%For this case, the following options, or combinations thereof, are suggested to ensure numerical stability by keeping gas particles out of the runaway collapse zone: (i) \emph{increasing} the gravitational softening length, (ii) decreasing the values of the minimum smoothing length; moderate changes already have a big impact (see bottom panel in Fig.~\ref{fig:zone}) or (iii) choosing a star formation condition that converts gas particles to star particles on a very short timescale when they enter the runaway collapse zone.

\subsection{Gravitational instabilities at and below the resolution limits}\label{sec:adaptivesoft}

In subsection~\ref{sec:runawaycollapse} we defined a numerical instability that is caused by an incorrect density estimate if the smoothing length is limited by a constant minimum value. In this subsection, we focus on the formation of gas clumps through gravitational fragmentation and collapse, considering both codes with adaptive and constant constant values for the gravitational softening length. 

For a code-independent discussion of the conditions under which gravitational instabilities develop in simulations, we define the code-agnostic length scales  $l_{\mathrm{soft}}$, over which gravitational forces are softened, and $l_{\mathrm{smooth}}$ (with a minimum of $l_{\mathrm{smooth,min}}$), over which the hydrodynamical forces are smoothed. We explain briefly how they connect to the smoothing lengths $h$ and the softening lengths $\epsilon$ in some codes as examples:

We use the kernel support as measure of $l_{\mathrm{smooth}}$. In \textsc{Swift}, $h$ is defined following \citet{DehnenAly2012} and based on the kernel standard deviation which is directly related to the numerical resolution of sound waves (equation~\ref{eq:hcorrect}). For the Wendland C2 kernel and the \textsc{Swift} definition of $h$, $l_{\mathrm{smooth}} =\gamma_{\mathrm{k}}\,h = 1.936492\,h$. In practice, the definition of the smoothing length varies for each code. The SPH code \textsc{Gadget4} \citep{Gadget4} defines $h$ as the finite support of the kernel and therefore $l_{\mathrm{smooth}}$ would be equal to $h_{\mathrm{Gadget4}}$.

The softening length is typically given as the Plummer-equivalent softening length, $\epsilon$. Depending on the softening potential, the gravitational force is exactly Newtonian for particle separations of $\ge 2.8\epsilon$ (cubic spline kernel as in \textsc{Gadget4}, \citealp{Gadget4}) or $\ge 3\epsilon$ (Wendland C2 kernel as in \textsc{Swift}). Because the transition to softened gravity is very gradual, the deviations from Newtonian gravity are only significant for separations $\lesssim 1 - 1.5\epsilon$ (see e.g. figure 1 in \citealp{Gadget4}). We therefore use $l_{\mathrm{soft}} = 1.5\epsilon$ as an estimate for the softening length scale. 

For the following discussion, using a different measure for either $l_{\mathrm{soft}}$ or $l_{\mathrm{smooth}}$ would shift the lines slightly but does not change the general interpretation of the individual zones in temperature-density space. After concluding in section~\ref{sec:runawaycollapse} that the minimum smoothing length should be very small to avoid a runaway collapse, we assume in this subsection $l_{\mathrm{smooth,min}}\rightarrow 0$, if not explicitly mentioned otherwise.

\subsubsection{Gravitational instabilities at the hydro resolution limit $l_{\mathrm{smooth}}$}\label{sec:gravinstatlsmooth}

Gravitational instabilities cannot be modelled accurately in the simulation if the Newtonian Jeans length, $\lambda_{\mathrm{J,N}}$, is smaller than the size of an individual smoothing kernel, $l_{\mathrm{smooth}}$. The hydrodynamical forces are inaccurate on scales below $l_{\mathrm{smooth}}$ and fragmentation on scales of $\lambda_{\mathrm{J,N}}<l_{\mathrm{smooth}}$ is unresolved. For the Newtonian Jeans length, the condition $\lambda_{\mathrm{J,N}} = l_{\mathrm{smooth}}$ depends on the gas density and temperature as well as on the particle mass, $m_{\mathrm{B}}$ ($l_{\mathrm{smooth}}\propto m_{\mathrm{B}}^{1/3} n_{\mathrm{H}}^{-1/3}$), but it does not depend on the gravitational softening length, $l_{\mathrm{soft}}$. The condition $\lambda_{\mathrm{J,N}} = l_{\mathrm{smooth}}$ therefore applies in the same way for simulations with constant and adaptive softening lengths. In each case, gravitational instabilities are only resolved (i.e. modelled accurately) for $\lambda_{\mathrm{J,N}} > l_{\mathrm{smooth}}$.

For gravitational instabilities at the hydrodynamic resolution limit, $l_{\mathrm{smooth}}$, differences between the formation and further evolution of gravitational instabilities emerge that depend on the assumed gravitational softening lengths. In this subsection we discuss the formation of gravitational instabilities at the hydrodynamical resolutions limit, $l_{\mathrm{smooth}}$, for simulations with constant and adaptive softening lengths. 

\begin{figure*}
    \centering
    \includegraphics[width=0.9\linewidth]{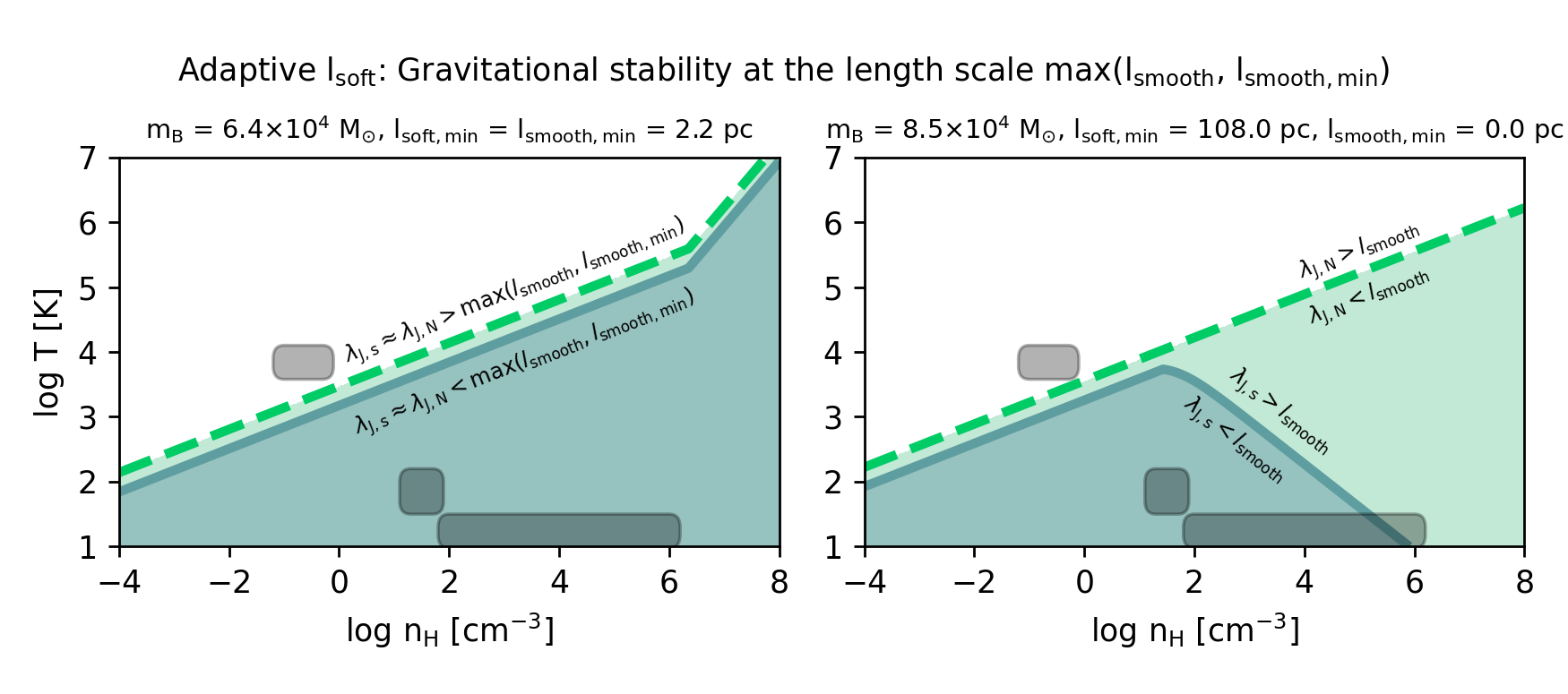}
    \caption{Gravitational stability at the hydrodynamical resolution limit, $\mathrm{max}(l_{\mathrm{smooth}},l_{\mathrm{smooth,min}})$, for simulations with an adaptive softening length, $l_{\mathrm{soft}}$. Line styles as in the left panel of Fig.~\ref{fig:zonesatlsmooth} but for an adaptive softening length, $l_{\mathrm{soft}}$. In the left panel, the minimum value for the gravitational softening length equals the minimum value for the smoothing length. Here, the slope of both lines (dashed line: $\lambda_{\mathrm{J,N}} = \mathrm{max}(l_{\mathrm{smooth}}, l_{\mathrm{smooth,min}}$), solid line: $\lambda_{\mathrm{J,s}} = \mathrm{max}(l_{\mathrm{smooth}}, l_{\mathrm{smooth,min}}$)) changes when $l_{\mathrm{smooth}}$ is limited by $l_{\mathrm{smooth,min}}$. In the right panel, the softening length is adaptive down to a minimum value $l_{\mathrm{soft,min}}$, for which the softening length is effectively constant. In this panel, the slope of the solid line, $\lambda_{\mathrm{J,s}} = \mathrm{max}(l_{\mathrm{smooth}}, l_{\mathrm{smooth,min}}$), changes for densities above which the softening length, $l_{\mathrm{soft}}$, is limited by a minimum value, $l_{\mathrm{soft,min}}$. Values for $m_{\mathrm{B}}$, $l_{\mathrm{soft,min}}$, and $l_{\mathrm{smooth,min}}$ were selected to represent \textsc{FIREbox} (left panel) and \textsc{TNG50} (right panel, see text for details). As in Fig.~\ref{fig:zonesatlsmooth}, the gas density, $n_{\mathrm{H}}$, refers to the SPH-estimated density. Typical densities and temperatures for the WNM, CNM, and MCs are indicated with dark patches, as in Fig.~\ref{fig:zoneexamples}.}
    \label{fig:zonesatlsmoothadap}
\end{figure*}

\paragraph{Constant softening length:} 

In nature (or in simulations with $m_{\mathrm{B}}\rightarrow 0$, $l_{\mathrm{soft}} \rightarrow 0$, and $l_{\mathrm{smooth,min}}\rightarrow0$), density perturbations with a length scale of $l_{\mathrm{smooth}}$ grow if $\lambda_{\mathrm{J,N}} < l_{\mathrm{smooth}}$, and decay until they reach a new equilibrium for $\lambda_{\mathrm{J,N}} > l_{\mathrm{smooth}}$. The border\footnote{The condition $\lambda_{\mathrm{J,N}} = l_{\mathrm{smooth}}$ can also be interpreted as the condition that the Jeans mass, $M_{\mathrm{J,N}} = 4\pi\lambda_{\mathrm{J,N}}^3 \rho /3$, is equal to the mass in the kernel $N_{\mathrm{neigh}} m_{\mathrm{B}} \approx l_{\mathrm{smooth}}^3 \rho$ (with a constant pre-factor of order unity). However, the number of neighbours, $N_{\mathrm{neigh}}$, is not well defined and we therefore use the better defined condition $\lambda_{\mathrm{J,N}} = l_{\mathrm{smooth}}$ instead.} (i.e. case $\lambda_{\mathrm{J,N}} = l_{\mathrm{smooth}}$) is indicated as a thick dashed line in all panels in Fig.~\ref{fig:zonesatlsmooth}. As discussed above, gravitational instabilities are only modelled accurately (i.e. are resolved) if $\lambda_{\mathrm{J,N}} > l_{\mathrm{smooth}}$ (unshaded area).  

For softened gravity with a constant softening length (left and right panel: $l_{\mathrm{soft}} = 100\,\mathrm{pc}$, middle panel: various values for $l_{\mathrm{soft}}$, see the labels), the solid lines ($\lambda_{\mathrm{J,s}} = l_{\mathrm{smooth}}$) separate perturbations with length scales $l_{\mathrm{smooth}}$ that are gravitationally (un)stable in softened gravity, i.e. in the simulation. Densities and temperatures for which gas is gravitationally stable in nature for perturbations at the resolution limit $l_{\mathrm{smooth}}$ are also gravitationally stable in simulations with softened gravity (white area in the left panel of Fig.~\ref{fig:zonesatlsmooth} for a simulation with $m_{\mathrm{B}} = 10^5\,\mathrm{M}_{\odot}$, $l_{\mathrm{soft}} = 100\,\mathrm{pc}$, and $l_{\mathrm{smooth,min}} \rightarrow 0\,\mathrm{pc}$). 

Gas with densities and temperatures within the shaded areas ($\lambda_{\mathrm{J,N}}<l_{\mathrm{smooth}}$), is gravitationally unstable for perturbations at the size of a smoothing kernel in Newtonian gravity. In softened gravity, gravitational instabilities in part of this area are suppressed at scales of $l_{\mathrm{smooth}}$ because the softened Jeans length exceeds the kernel size ($\lambda_{\mathrm{J,s}} > l_{\mathrm{smooth}}$) and therefore perturbations on scales of $l_{\mathrm{smooth}}$ decay. The middle panel shows that the boundary $\lambda_{\mathrm{J,s}} = l_{\mathrm{smooth}}$ depends on the value for the constant softening length, $l_{\mathrm{soft}}$. The right panel shows that increasing (dash-dotted lines) or decreasing (dotted lines) the particle mass by a factor of 8 from the fiducial value of $m_{\mathrm{B}} = 10^5\,\mathrm{M}_{\odot}$ (dashed lines), affects both the thick green lines for $\lambda_{\mathrm{J,N}} = l_{\mathrm{smooth}}$ as well as the thin black lines for the condition $\lambda_{\mathrm{J,s}} = l_{\mathrm{smooth}}$, because $l_{\mathrm{smooth}}\propto m_{\mathrm{B}}^{1/3}$ (right panel of Fig.~\ref{fig:zonesatlsmooth} for $l_{\mathrm{soft}} = 100\,\mathrm{pc}$). 

\paragraph{Adaptive softening length:} 

For adaptive softening, typically $l_{\mathrm{soft}} = l_{\mathrm{smooth}}$ and therefore $\lambda_{\mathrm{J,s}} \approx \lambda_{\mathrm{J,N}}$. The instability criteria for perturbations at the resolution limit, i.e. at the kernel size $l_{\mathrm{smooth}}$, therefore follow the instability criteria for Newtonian gravity. The left panel of Fig.~\ref{fig:zonesatlsmoothadap} (linestyles as in Fig.~\ref{fig:zonesatlsmooth}) illustrates this for parameters representative for the \textsc{FIREbox} \citep{FIREbox} simulation project with $m_{\mathrm{B}} = 6\times10^4\,\mathrm{M}_{\odot}$. Their adaptive softening length is equal to the average gas particle separation ($l_{\mathrm{soft}} = 1.5 \epsilon (m_{\mathrm{B}}/ \rho)^{1/3}$, for gas density $\rho$), down to a minimum value of $l_{\mathrm{soft,min}} = l_{\mathrm{smooth,min}} = 2.25\,\mathrm{pc}$ ($\epsilon_{\mathrm{min}} = 1.5\,\mathrm{pc}$). Density perturbations of length-scale $l_{\mathrm{smooth}}$ therefore follow the Newtonian Jeans criteria. The small offset between the dashed and solid lines is from the order of unity prefactors related to the kernel shapes and exact definitions of $l_{\mathrm{smooth}}$ and $l_{\mathrm{soft}}$. For simplicity we use the same kernel shapes and pre-factors as in Fig.~\ref{fig:zonesatlsmooth}. 

The slope change of the $\lambda_{\mathrm{J,N}} = l_{\mathrm{smooth}}$ (dashed) and $\lambda_{\mathrm{J,s}} = l_{\mathrm{smooth}}$ (solid) lines is caused by the minimum smoothing length $l_{\mathrm{smooth,min}}$. The critical density $n_{\mathrm{H,hmin}}$ above which densities are over-smoothed, and hence $n_{\mathrm{H,SPH}} < n_{\mathrm{H}}$, is $\approx 2\times10^6\,\mathrm{cm}^{-3}$ (equation~\ref{eq:nHhmin} for $m_{\mathrm{B}} = 6\times10^4\,\mathrm{M}_{\odot}$ and $h_{\mathrm{min}}\approx 1.2\,\mathrm{pc}$). The small values for $l_{\mathrm{smooth,min}}$ and $l_{\mathrm{soft,min}}$ therefore prevent the instability described in section~\ref{sec:runawaycollapse} for densities below $\approx 2\times10^6\,\mathrm{cm}^{-3}$.

In the \textsc{IllustrisTNG} project \citep{TNG2018Pillepich, TNG2018Nelson} which uses the moving mesh code \textsc{Arepo} \citep{arepo}, the softening length is related to the adaptive sizes of the gaseous cells but here the minimum softening length (\textsc{TNG50}: $l_{\mathrm{soft,min}} = 1.5 \epsilon_{\mathrm{gas,min}} = 108\,\mathrm{pc}$, \textsc{TNG100}: $l_{\mathrm{soft,min}} = 1.5 \epsilon_{\mathrm{gas,min}} = 285\,\mathrm{pc}$, \textsc{TNG300}: $l_{\mathrm{soft,min}} = 1.5\epsilon_{\mathrm{gas,min}} = 555\,\mathrm{pc}$, see table~1 in \citealp{TNG50Annalisa}) differs from the minimum smoothing length (the minimum cell size in \textsc{TNG50} is reported as $6.5\,\mathrm{pc}$, \citealp{TNG50Annalisa}). 

Gravitational instabilities in gas cells for which the gravitational softening is limited by a minimum value ($\epsilon=\epsilon_{\mathrm{gas,min}}$) follow the softened Jeans criteria outlined in section~\ref{sec:Jeans}. 
The right panel of Fig.~\ref{fig:zonesatlsmoothadap} shows the $\lambda_{\mathrm{J,N}} = l_{\mathrm{smooth}}$ (dashed) and $\lambda_{\mathrm{J,s}} = l_{\mathrm{smooth}}$ (solid) lines for values representative for \textsc{TNG50}: $m_{\mathrm{B}} = 8.5\times 10^4\,\mathrm{M}_{\odot}$, $l_{\mathrm{soft,min}} = 108\,\mathrm{pc}$ and an assumed negligible value for $l_{\mathrm{smooth,min}}$.

Comparing the right panel of Fig.~\ref{fig:zonesatlsmoothadap} with the left panel of Fig.~\ref{fig:zonesatlsmooth}, we see that gravitational instabilities on scales $<l_{\mathrm{soft,min}}$ in simulations with adaptive softening lengths effectively behave as in simulations with constant softening lengths.

\subsubsection{Gravitational instabilities below the hydro resolution limit $l_{\mathrm{smooth}}$}\label{sec:gravinstbelowlsmooth}

Density perturbations grow in nature on length-scales smaller than $l_{\mathrm{smooth}}$ in gas with temperatures and densities for which $\lambda_{\mathrm{J,N}}<l_{\mathrm{smooth}}$ (shaded regions in Figs.~\ref{fig:zonesatlsmooth} and \ref{fig:zonesatlsmoothadap}). The gravitational collapse of these perturbations within an individual smoothing kernel is not resolved and therefore not expected to be modelled accurately with any simulation method.  

We can see from Figs.~\ref{fig:zonesatlsmooth} and \ref{fig:zonesatlsmoothadap} that a large fraction of the neutral ISM in simulations might form sub-kernel instabilities ($\lambda_{\mathrm{J,s}} < l_{\mathrm{smooth}}$, dark shaded regions). While the resulting clumpy particle configurations within a kernel might have a limited effect on the simulation because the cooling rates, star formation rates and other density- or pressure-dependent sub-grid models use the smoother SPH estimates, we briefly discuss the differences in the treatment of sub-kernel perturbations between codes with constant and adaptive softening lengths. 

\begin{figure*}
    \centering
  \includegraphics[width=0.9\linewidth]{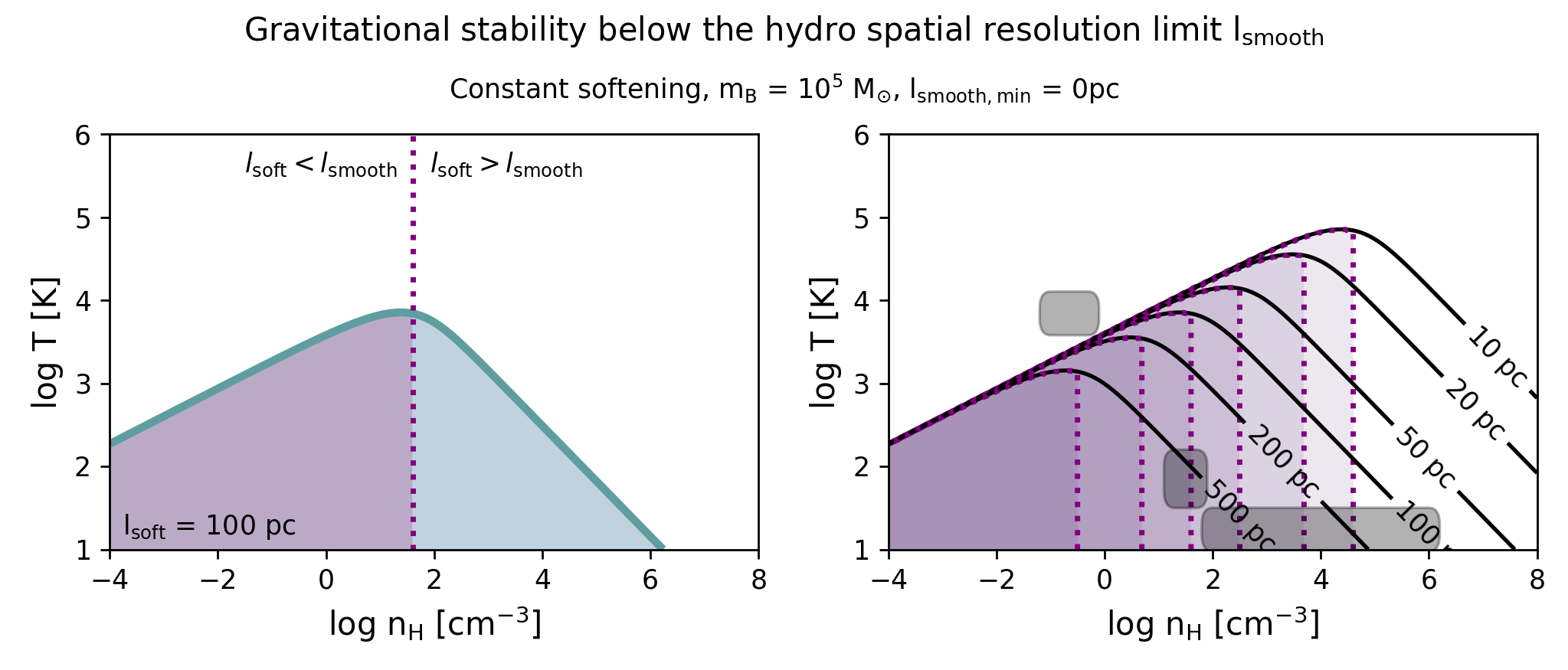}
    \caption{As Fig.~\ref{fig:zonesatlsmooth} but focusing on densities and temperatures for which gravitation instabilities on sub-kernel scales are expected ($\lambda_{\mathrm{J,s}} < l_{\mathrm{smooth}}$). Left panel: The dotted vertical line indicates the boundary $l_{\mathrm{soft}} = l_{\mathrm{smooth}}$ for $m_{\mathrm{B}} = 10^5\,\mathrm{M}_{\odot}$ and a constant softening length of $l_{\mathrm{soft}} = 100\,\mathrm{pc}$. If $l_{\mathrm{soft}} < l_{\mathrm{smooth}}$ gas clumps within a smoothing kernel can fragment further, but this process is suppressed at higher densities where $l_{\mathrm{soft}} > l_{\mathrm{smooth}}$ (see text for details). In the right panel all lines are repeated for different values of $l_{\mathrm{soft}}$ (see contour labels). As in Fig.~\ref{fig:zonesatlsmooth}, the gas density, $n_{\mathrm{H}}$, refers to the SPH-estimated density. Typical densities and temperatures for the WNM, CNM, and MCs are indicated with dark patches, as in Fig.~\ref{fig:zoneexamples}.}
    \label{fig:zonesbelowlsmooth}
\end{figure*}

\paragraph{Constant softening length:} 

Gas with $\lambda_{\mathrm{J,s}} < l_{\mathrm{smooth}}$ is expected to be gravitationally unstable when exposed to fluctuations on length scales between $\lambda_{\mathrm{J,s}}$ and $l_{\mathrm{smooth}}$. In the derivation of the softened Jeans length, $\lambda_{\mathrm{J,s}}$, we assume an accurate gas density and pressure estimate. If the gas pressure is underestimated, gravitational instabilities may form from perturbations on length scales below $\lambda_{\mathrm{J,s}}$.
Therefore, the softened Jeans criteria serve as an upper limit to the expected instabilities in the simulation because the hydrodynamic forces might be underestimated due to the smoothing over the full kernel (an extreme case is discussed in subsection~\ref{sec:runawaycollapse}). If $l_{\mathrm{smooth}} < l_{\mathrm{soft}}$, the gravitational forces are softened on larger scales than a smoothing kernel and further gravitational collapse and fragmentation within a kernel is suppressed. On the other hand, $l_{\mathrm{soft}}$ can be (much) smaller than the size of a smoothing kernel, for a constant value of $l_{\mathrm{soft}}$. In this case, dense particle configurations with sizes possibly even smaller than $\lambda_{\mathrm{J,N}}$ can form because the density and pressure estimates of sub-kernel clumps can be inaccurate. 

Fig.~\ref{fig:zonesbelowlsmooth} shows an overview of the behaviour of dense particle configurations within a smoothing kernel. Gas that is gravitationally unstable at the scale of an individual smoothing kernel ($\lambda_{\mathrm{J,s}} < l_{\mathrm{smooth}}$, as in Fig.~\ref{fig:zonesatlsmooth}) is further split into densities for which further fragmentation to even smaller scales is suppressed ($l_{\mathrm{soft}} > l_{\mathrm{smooth}}$) or potentially induced ($l_{\mathrm{soft}} < l_{\mathrm{smooth}}$), with the boundary, $l_{\mathrm{soft}} = l_{\mathrm{smooth}}$, indicated as vertical, dotted lines (left panel: $l_{\mathrm{soft}} = 100\,\mathrm{pc}$, right panel: various values for $l_{\mathrm{soft}}$, see contour labels). 

We conclude that for a constant softening length, gas in simulations with $\lambda_{\mathrm{J,s}} < l_{\mathrm{smooth}}$ and $l_{\mathrm{soft}} < l_{\mathrm{smooth}}$ (purple areas in Fig.~\ref{fig:zonesbelowlsmooth}) might be affected by sub-kernel clumping. For example: within a kernel of 100 particles, a subset of 10 particles has particle separations that are smaller than average for this kernel and hence smaller than the smoothing length, $l_{\mathrm{smooth}}$. In contrast to simulations with adaptive softening, the constant value for $l_{\mathrm{soft}}$ can be smaller than the particle separation of these 10 particles. The gravitational forces between this subset of particles are therefore Newtonian, while the pressure is smoothed over the full kernel with 100 particles. An initial perturbation within the kernel may grow artificially or at a rate that is artificially high until $l_{\mathrm{smooth}}=l_{\mathrm{soft}}$ (vertical lines in Fig.~\ref{fig:zonesbelowlsmooth}). For higher gas densities (i.e. $l_{\mathrm{smooth}}<l_{\mathrm{soft}}$), the further collapse is suppressed. The detailed impact of sub-kernel clumping on galaxy properties in large-scale simulations is beyond the scope of this work.

If sub-kernel clumping (``subk") is undesired in simulations with constant softening, the conditions $\lambda_{\mathrm{J,s}} < l_{\mathrm{smooth}}$ and $l_{\mathrm{soft}} < l_{\mathrm{smooth}}$ (purple areas in Fig.~\ref{fig:zonesbelowlsmooth}) should not cover regions in density-temperature space that are populated by many gas particles in the simulation. If we want to confine the area in density temperature space for which $l_{\mathrm{soft}} < l_{\mathrm{smooth}}$ to densities below $n_{\mathrm{max,subk}}$, this condition translates into a minimum constant softening length 

\begin{align}
    l_{\mathrm{min,subk}} = 1.5 \epsilon_{\mathrm{min,subk}} & > 1.5 \frac{\gamma_{\mathrm{k}} \eta_{\mathrm{res}}}{1.5} \left ( \frac{X_{\mathrm{H}} m_{\mathrm{B}}}{m_{\mathrm{H}} n_{\mathrm{max,subk}}}\right)^{1/3} \;\mathrm{or}\nonumber\\
    l_{\mathrm{min,subk}} = 1.5 \epsilon_{\mathrm{min,subk}} & > 160\,\mathrm{pc} \left ( \frac{n_{\mathrm{max,subk}}}{10\,\mathrm{cm}^{-3}}\right )^{-1/3}  \left ( \frac{m_{\mathrm{B}}}{10^5\,\mathrm{M}_{\odot}} \right )^{1/3}  \label{eq:epsilonD}
\end{align}

\noindent
for the smoothing length and softening length definition as above.

Unlike in simulations with adaptive softening, unresolved gravitational instabilities are not suppressed by design in simulations with constant softening lengths. Fulfilling equation (\ref{eq:epsilonD}) avoids potentially undesired, sub-kernel gravitational instabilities in large parts of the cold ISM, but at the expense of suppressing physical instabilities at the kernel scale, $l_{\mathrm{smooth}}$ (see middle panel of Fig.~\ref{fig:zonesatlsmooth}).

We focus in this work on simulations with $m_{\mathrm{B}}\gtrsim 10^5\,\mathrm{M}_{\odot}$ but the analysis presented here is relevant for simulations of any mass resolution. In Appendix~\ref{sec:highres} (Fig.~\ref{fig:highres}), the right panel of Fig.~\ref{fig:zonesbelowlsmooth} is repeated for simulations with a particle mass of $m_{\mathrm{B}} = 4\,\mathrm{M}_{\odot}$, for which a constant softening length of at least $\epsilon = 2\,\mathrm{pc}$ is needed to avoid sub-kernel clumping.

\paragraph{Adaptive softening length:} 

In simulations with an adaptive softening length and $l_{\mathrm{soft}} \ge l_{\mathrm{smooth}}$ any further fragmentation within a smoothing kernel is generally suppressed because gravitational forces are softened on scales larger than or equal to the smoothing kernel. Physical gravitational instabilities on scales $\lambda_{\mathrm{J,N}}<l_{\mathrm{smooth}}$ are therefore artificially suppressed. 

In order to avoid the runaway collapse described in section~\ref{sec:runawaycollapse}, the minimum smoothing length, $l_{\mathrm{smooth,min}}$, needs to be small (see equation~\ref{eq:hminmax}). In simulations, such as \textsc{FIREbox}, where $l_{\mathrm{soft,min}} = l_{\mathrm{smooth,min}}$, the minimum softening length needs to have the same small value as the minimum smoothing length.

\begin{figure*}
    \centering
    \includegraphics[width=\linewidth]{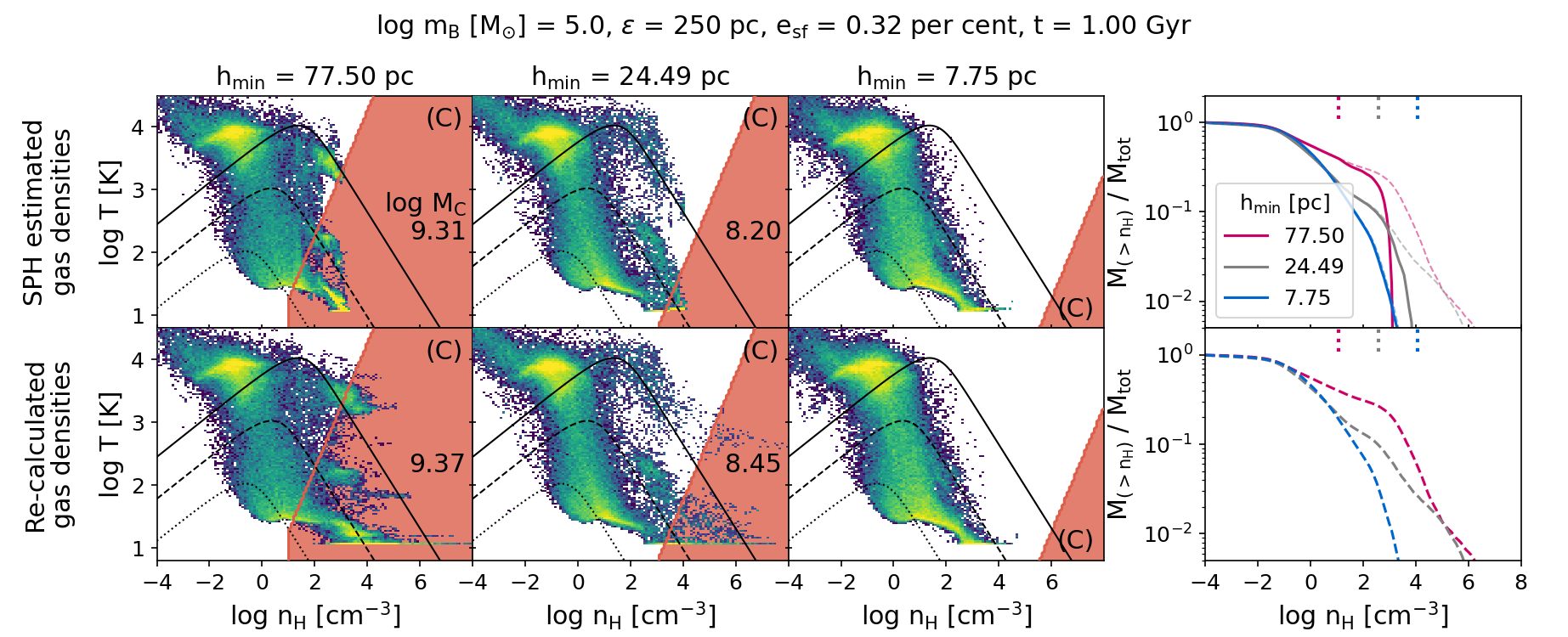}
    \caption{Illustration of the runaway collapse described in section~\ref{sec:runawaycollapse}. Temperature-density histograms at time $t=1\,\mathrm{Gyr}$ for 3 simulations (columns 1 to 3) with identical initial conditions. The densities (x-axis) in the top row are the SPH gas density estimates, $n_{\mathrm{H,SPH}}$, as used in the simulations and stored in the snapshots, while the bottom row shows the re-calculated densities for $h_{\mathrm{min}}\rightarrow 0$, based on the same particle positions. The colour of the temperature-density histograms is proportional to the log of gas mass per pixel and the minimum and maximum values for the colour maps are constant across all plots. From left to right the minimum smoothing length, $h_{\mathrm{min}}$, decreases from $77.5\,\mathrm{pc}$ ($h_{\mathrm{min,ratio}} = 0.2$; first column) and $h_{\mathrm{min}} = 24.5\,\mathrm{pc}$ (second column) to $h_{\mathrm{min}} = 7.75\,\mathrm{pc}$ (third column). The red shaded area is the runaway collapse zone as defined in section~\ref{sec:runawaycollapse} and each panel includes a number for the total mass of particles within this zone as $\log M_{\mathrm{zoneC}} [\mathrm{M}_{\odot}]$ if $M_{\mathrm{zoneC}}>0$. For reference, the black lines indicate where the softened Jeans mass is resolved by 1 (dotted), 10 (dashed), and 100 (solid) particles. The panels in the rightmost column show the cumulative mass fraction for particles above a given gas density $n_{\mathrm{H}}$. The solid lines in the top panel show the density distributions for the phase-space diagrams in the top row and the dashed lines in the bottom panel for the density distribution from the bottom row (dashed lines are repeated for reference in the top panel). The short vertical dotted lines at the top of the figures in the rightmost row indicate the densities $n_{\mathrm{H,hmin}}$ (equation~\ref{eq:nHhmin}) above which the smoothing length is limited by $h_{\mathrm{min}}$.}
    \label{fig:GasPhaseOverview}
\end{figure*}

\section{Galaxy simulations}\label{sec:simulations}

In section~\ref{sec:adaptivesoft}, we used the smoothing and softening lengths $l_{\mathrm{smooth}}$ and $l_{\mathrm{soft}}$ for a code independent discussion on the individual zones. In this section we use simulations of isolated galaxies with the public \textsc{Swift} code\footnote{The simulations use version 0.9.0 and in particular revision v0.9.0-1182-g423e9dd8.} (\citealp{swift2016,swift2018,swift2023}, \href{https://swift.dur.ac.uk/}{www.swiftsim.com}). The resolution parameters set by the user are $\epsilon$ and $h_{\mathrm{min}}$ which relate to $l_{\mathrm{soft}} = 1.5 \epsilon$ and $l_{\mathrm{smooth,min}} = 1.94 h_{\mathrm{min}}$, respectively (see text in section~\ref{sec:adaptivesoft} for details). We will demonstrate the numerical issues that can occur for different choices of $\epsilon$ and $h_{\mathrm{min}}$, focusing on simulations of isolated galaxies with a constant softening length. 
The individual simulations are based on the ``IsolatedGalaxy-feedback'' example in \textsc{Swift} and use the modern and open source SPH scheme SPHENIX \citep{sphenix}, implemented as the default hydrodynamic solver in \textsc{Swift}. Gravity is softened with a Wendland C2 kernel and a constant softening length, defined as the Plummer equivalent softening length, $\epsilon$. 

The initial conditions were created with \textsc{MakeNewDisk} which is based on the code used in \citet{MakeDiskCode} but modified to use the exact definition of the analytic dark matter halo mass $M_{\mathrm{200}}$ (i.e. mass within $R_{\mathrm{200}}$, the radius within which the average density is 200 times the critical density of the Universe) instead of the halo mass integrated to infinity (see \citealp{Folkert2023arXiv230913750N} for details). 
The isolated galaxy initially has a gas mass of $1.644\times 10^{10}\,\mathrm{M}_{\odot}$ with solar metallicity ($Z_{\odot} = 0.0134$, \citealp{Asplund2009}) and an exponential stellar disk with a radial scale length of $4.3\,\mathrm{kpc}$ and a mass of $3.836\times 10^{10}\,\mathrm{M}_{\odot}$ The initial disk gas fraction is 30~per~cent and the gas is initialized with a temperature of $10^4\,\mathrm{K}$. The baryonic disk is in equilibrium with an analytic dark matter halo with a \citet{Hernquist1990} profile of $M_{\mathrm{200}} = 1.37 \times 10^{12}\,\mathrm{M}_{\odot}$ and a scale radius such that the central density profile matches that of a \citet{NFW} profile with a concentration of $c=9$. The initial conditions are available within the ``IsolatedGalaxy'' example in \textsc{Swift}. 

We use the fiducial cooling tables from \citet{PS20} which include the effects of self-shielding, dust, cosmic rays, an interstellar radiation field and a UV background. In Appendix~\ref{sec:ISRFvariation} we demonstrate that our conclusions remain unchanged if we instead use cooling tables appropriate for a weaker and stronger radiation field. No artificial pressure or entropy floor is included. A supernova energy of $10^{51}\,\mathrm{erg}$ per SN is injected stochastically as thermal energy, following \citet{DallaVecchiaSchaye2012}, with a heating temperature of $10^{7.5}\,\mathrm{K}$. The high heating temperature of the stochastic feedback model increases the efficiency of thermal feedback because the cooling time of gas with $10^{7.5}\,\mathrm{K}$ is long and less energy is lost radiatively lost compared to using many smaller thermal energy injections that would heat up the gas to e.g. $10^5\,\mathrm{K}$, the peak of the cooling curve (see \citealp{DallaVecchiaSchaye2012} for a detailed discussion). We inject the stellar feedback energy into the gas particle closest to the star, which has been shown to further increase the efficiency of supernova feedback, compared to selecting a random gas particle in the star's kernel (see ``Min distance'' model in \citealp{Evgenii22}). Additional simulations with 2 and 4 times higher supernova energies are presented in Appendix~\ref{sec:ESNvariation}. 

Star formation is limited to densities of $n_{\mathrm{H}} > 0.1\,\mathrm{cm}^{-3}$ and cold gas (temperatures of $T<1000\,\mathrm{K}$) and the star formation rate for each gas particle with mass $m_{\mathrm{gas}}$ is given by the \citet{SchmidtLaw} relation

\begin{align}\label{eq:sfr}
    \dot{m}_{\star} &= \frac{e_{\mathrm{sf}} }{t_{\mathrm{ff,N}}} m_{\mathrm{gas}}
\\
&=7.2 \times 10^{-5} \,\mathrm{M}_{\odot}\,\mathrm{yr}^{-1} \left ( \frac{e_{\mathrm{sf}}}{0.32\%} \right ) \left (\frac{n_{\mathrm{H}}}{100\,\mathrm{cm}^{-3}}\right )^{1/2} \left ( \frac{m_{\mathrm{gas}}}{10^5\,\mathrm{M}_{\odot}}\right) \nonumber
\end{align}

\noindent
with the star formation efficiency $e_{\mathrm{sf}}$ and the Newtonian free-fall time $t_{\mathrm{ff,N}}$ (equation~\ref{eq:tffNewton}). A gas particle is converted into a star particle stochastically. The simulations analysed in this work vary the resolution parameters: gravitational softening $\epsilon = [250, 500]\,\mathrm{pc}$, minimum SPH smoothing length $h_{\mathrm{min}} = [7.75, 24.5, 77.5]\,\mathrm{pc}$ ($h_{\mathrm{min,ratio}} = [0.02, 0.063, 0.2]$ for $\epsilon = 250\,\mathrm{pc}$), and baryon particle mass

\begin{landscape}
\begin{figure}
    \centering
    \includegraphics[height=9.5cm]{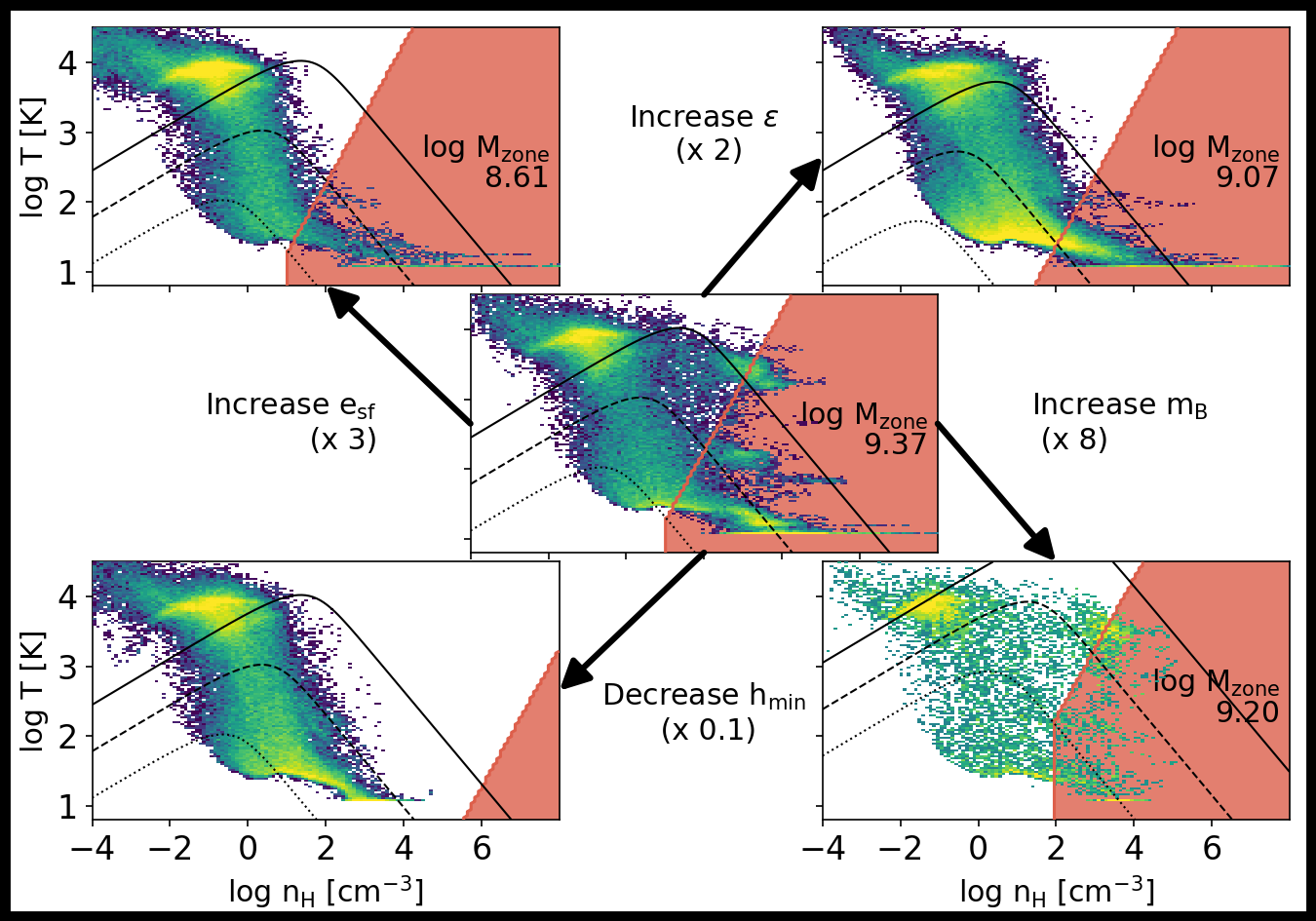}
    \includegraphics[height=9.5cm]{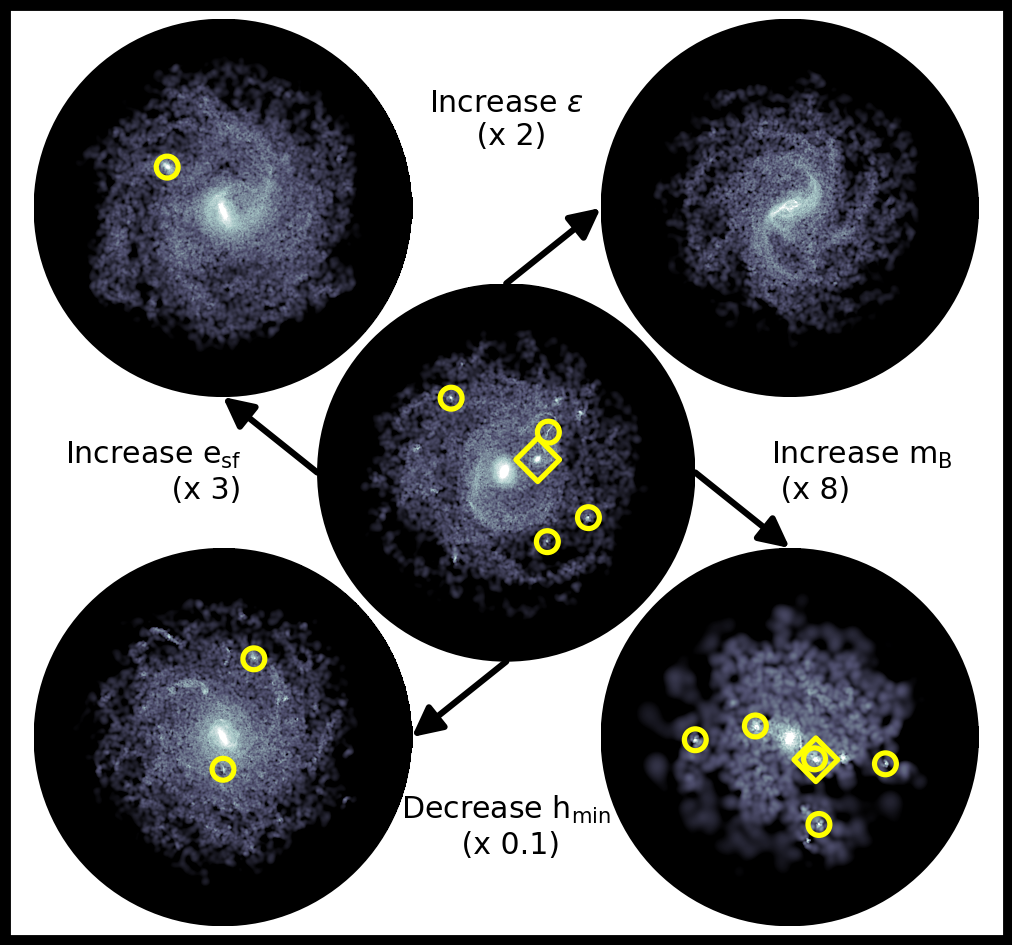}
    \caption{The distribution of gas (left) is represented as a (recalculated) density - temperature histogram (as in the bottom row of Fig.~\ref{fig:GasPhaseOverview}) and the distribution of stars (right) as a stellar mass surface density map for simulations with various resolution and star formation efficiency parameters at $t = 1\,\mathrm{Gyr}$. The central panel in each figure shows the results for the fiducial parameters $m_{\mathrm{B}} = 10^5\,\mathrm{M}_{\odot}$, $\epsilon = 250\,\mathrm{pc}$, $h_{\mathrm{min}} = 77.5\,\mathrm{pc}$ ($h_{\mathrm{min,ratio}} = 0.2$), and $e_{\mathrm{sf}} = 0.32\,\mathrm{per\,cent}$ and the panels in each corner vary one parameter while keeping the others constant: top right: $\epsilon$ is increased by a factor of 2, bottom right: $m_{\mathrm{B}}$ is increased by a factor of 8, bottom left: $h_{\mathrm{min}}$ is decreased by a factor of 10, top left: $e_{\mathrm{sf}}$ is increased by a factor of 3. The phase-space plots on the left hand side show the zones where the numerical instability as described in section~\ref{sec:runawaycollapse} can form as red-shaded areas. The total gas mass in this zone is shown as $\log M_{\mathrm{zone}} [\mathrm{M}_{\odot}]$ in each panel, if $M_{\mathrm{zone}}>0$.
    The right hand side consists of images of the stellar surface mass density within $r = 25\,\mathrm{kpc}$ (excluding stars already present in the initial conditions). The clumps identified by the friends-of-friends algorithm are highlighted with circles (diamonds) if their mass is below (above) $10^8\,\mathrm{M}_{\odot}$. Both colour maps are logarithmic and span the same range across all simulations. }
    \label{fig:gallerygasstars}
\end{figure}
\end{landscape}

\noindent $m_{\mathrm{B}}= [10^5, 8\times10^5]\,\mathrm{M}_{\odot}$; as well as the star formation efficiency $e_{\mathrm{sf}} = [0.32, 1]\,\mathrm{per\,cent}$ in order to change the amount of high-density gas in a controlled way.

All simulations run until $t_{\mathrm{end}}=1.2\,\mathrm{Gyr}$. We use \textsc{Swift} to re-calculate the gas densities based on the gas particle positions for $h_{\mathrm{min}}\rightarrow 0$ by restarting the original simulations at $t = 1\,\mathrm{Gyr}$ for one very small timestep ($\Delta t_{\mathrm{max}} = 10\,\mathrm{yr}$) with a very small minimum smoothing length\footnote{We use a non-zero value for $h_{\mathrm{min,ratio}}$ but the selected value is small enough that the re-calculated smoothing lengths are larger than $h_{\mathrm{min}}$ for all gas particles.} ($h_{\mathrm{min,ratio}} = 10^{-4}$). The snapshot file that is produced after this timestep contains the correct (i.e. not limited by a minimum smoothing length) gas densities based on the gas particle positions.

Fig.~\ref{fig:GasPhaseOverview} shows the difference between the original SPH densities (top row) and the recalculated densities for $h_{\mathrm{min}}\rightarrow 0$ (bottom row) for 3 simulations (first three columns; all with $m_{\mathrm{B}}=10^5\,\mathrm{M}_{\odot}$, $\epsilon = 250\,\mathrm{pc}$, $e_{\mathrm{sf}}=0.32\,\mathrm{per\,cent}$) with decreasing values for $h_{\mathrm{min}}$ (from left to right). The runaway collapse described in section~\ref{sec:runawaycollapse} is obvious in the simulation with the largest value for $h_{\mathrm{min}}$ (1st column). The distribution of recalculated, ``real'' densities (bottom row) extends to much higher values than the densities used during the simulation (top row) for many particles as they approach the runaway collapse zone (red shaded area). For the simulation with the smallest value of $h_{\mathrm{min}}$ (3rd column), the two density-temperature diagrams look identical. 

The panels in the rightmost column show the cumulative mass fraction of particles above density $n_{\mathrm{H}}$ (x-axis) for the SPH densities (solid lines, top panel) and the recalculated densities (dashed lines, bottom panel; repeated for reference in the top panel). 
As the densities exceed $n_{\mathrm{H,hmin}}$ (equation~\ref{eq:nHhmin}; indicated as short vertical dotted lines) the SPH densities and the recalculated densities diverge. In this case, the formally lowest resolution simulation ($h_{\mathrm{min}} = 77.5\,\mathrm{pc}$) has much higher ``real'' (but unresolved) densities, up to $n_{\mathrm{H}}>10^6\,\mathrm{cm}^{-3}$, than the formally highest resolution simulation ($h_{\mathrm{min}} = 7.75\,\mathrm{pc}$). If these dense clumps would appear through physical processes such as instabilities in the galaxy disk, they would persist in the simulation with better resolved hydrodynamic forces. We therefore conclude that the very high densities seen in the particle distributions are indeed artificial as outlined in section~\ref{sec:runawaycollapse}.

While the dense gas clumps form, they may become eligible to star formation and if their densities are underestimated, their star formation rates are underestimated as well ($\dot{m}_{\star}\propto n_{\mathrm{H}}^{1/2}$, equation~\ref{eq:sfr}). As result, the gas clumps in the runaway collapse zone turn into star particles slower than expected from the value for the star formation efficiency and the particle positions (i.e.~``real densities"). 

After star particles form within dense gas clumps, stellar feedback is modelled by injecting thermal energy. \citet{DallaVecchiaSchaye2012} derived a maximal gas density below which thermal feedback is efficient, which for the simulations used in this work means that thermal feedback is inefficient for densities above $n_{\mathrm{fb}} \approx 10\,\mathrm{cm}^{-3}$. Their derivation assumes $h>h_{\mathrm{min}}$ which is fulfilled at densities of $\approx 10\,\mathrm{cm}^{-3}$ in all simulations in this work.
Comparing the values of $n_{\mathrm{fb}}$ to the particle densities in Figs.~\ref{fig:GasPhaseOverview} and \ref{fig:gallerygasstars} reveals that the artificially dense gas clumps are unlikely to be destroyed by stellar feedback and may thus result in artificially dense clusters of star particles. 

In Fig.~\ref{fig:gallerygasstars} we compare images of the stellar mass surface density of five simulations and their respective gas distributions in temperature - recalculated density phase-space (all at $t=1\,\mathrm{Gyr}$) The lines and labels are as in Fig.~\ref{fig:GasPhaseOverview}. The central images for both gas and stars are for the fiducial simulation with resolution parameters $m_{\mathrm{B}}=10^5\,\mathrm{M}_{\odot}$, $\epsilon = 250\,\mathrm{pc}$, and $h_{\mathrm{min}} = 77.5\,\mathrm{pc}$ and a star formation efficiency of $e_{\mathrm{sf}}=0.32\,\mathrm{per\,cent}$. The four simulations in each corner vary one of these parameters at a time, keeping the other parameters at their fiducial values (see labels). 

Some stellar mass surface density images (right side of Fig.~\ref{fig:gallerygasstars}) show dense star clumps. We identify the clump properties by running the \textsc{Swift} friends-of-friends algorithm as a stand-alone routine on the star particles within a snapshot output. We use a small linking length of 25~pc and a minimum number of particles of 50 for simulations with $m_{\mathrm{B}}=10^5\,\mathrm{M}_{\odot}$ and 18 for simulations with $m_{\mathrm{B}}=8 \times 10^5\,\mathrm{M}_{\odot}$. The minimum clump mass is a factor of $\approx 2.9$ times larger for the low mass-resolution simulation compared to the simulation with the fiducial particle mass. This is a compromise between choosing the same number of particles and the same mass in a clump when comparing simulations with different mass resolutions. Slightly different parameters in the clump finding algorithm might affect the detailed analysis on the exact properties of the individual clumps but the qualitative conclusions that we focus on are insensitive to these choices. The positions of all clumps identified outside of the innermost 3 kpc are indicated by circles and the most massive clumps ($M_{\mathrm{clump}} > 10^8\,\mathrm{M}_{\odot}$) are highlighted by diamonds. 

The galaxy with the fiducial resolution parameters (centre) has several stellar clumps due to the large amount of mass ($\log M\mathrm{(zoneC)} [\mathrm{M}_{\odot}] = 9.37$) within the runaway collapse zone (zone~C in section~\ref{sec:adaptivesoft}) in the gas phase. These clumps are mostly artificial as they largely disappear when increasing the hydrodynamic force resolution (shown in Fig.~\ref{fig:GasPhaseOverview} and when comparing centre and lower left panels in Fig.~\ref{fig:gallerygasstars}) and therefore moving the runaway collapse zone to higher gas densities. 

\begin{figure*}
    \centering
    \includegraphics[width=0.45\linewidth]{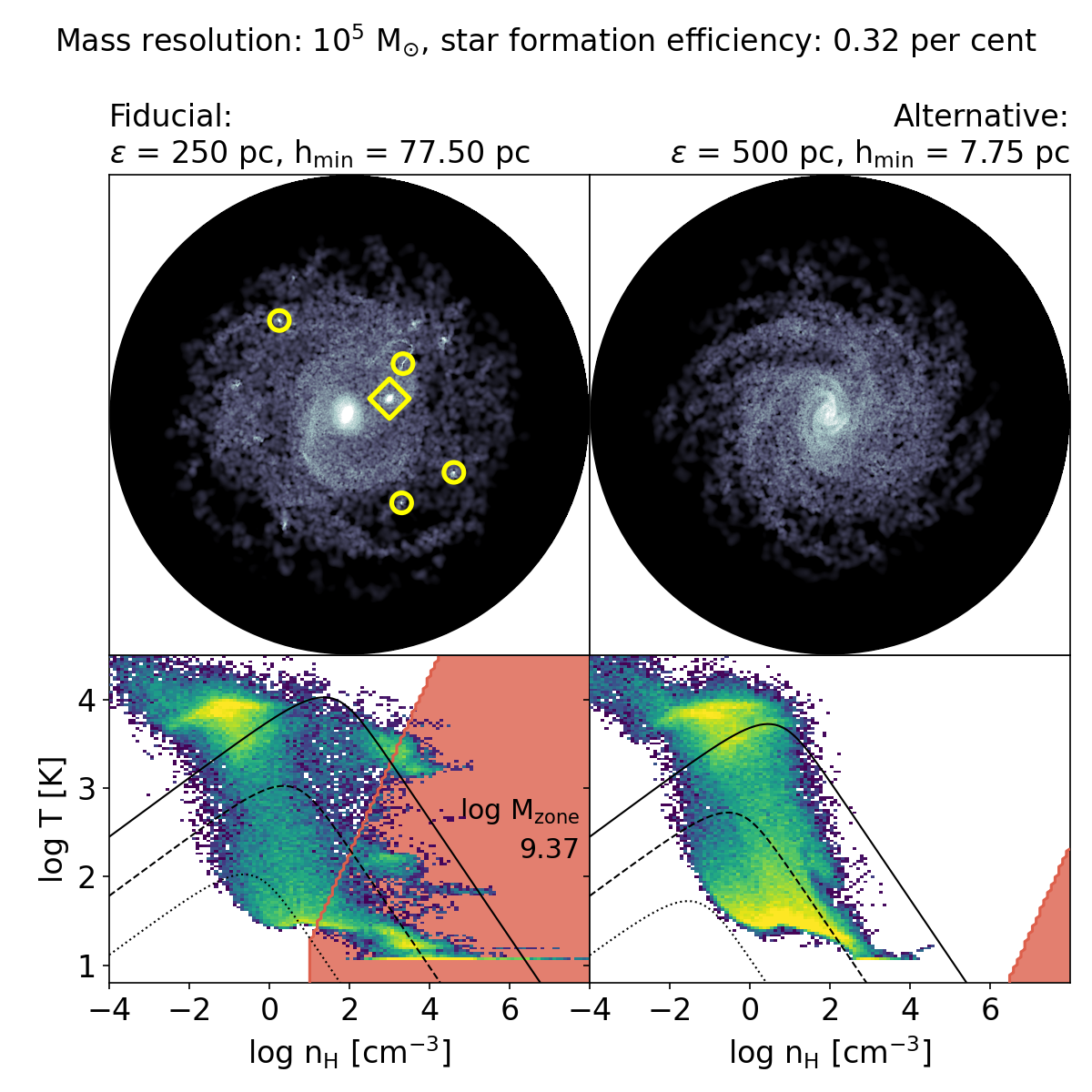} 
    \includegraphics[width=0.45\linewidth]{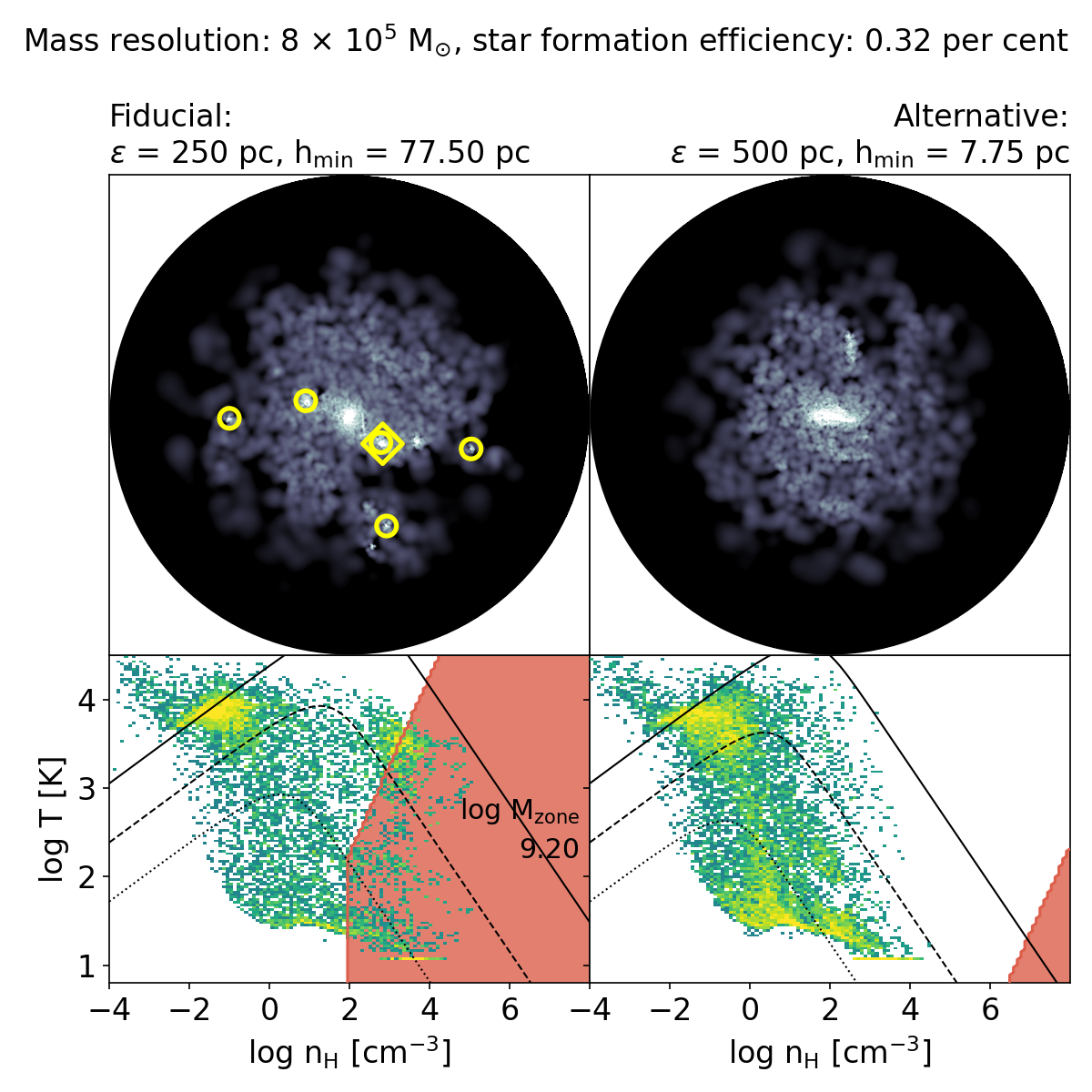}     
    \caption{Stellar mass surface density (top row) and gas phase-space distribution (bottom row) as in Fig.~\ref{fig:gallerygasstars} for the fiducial simulation parameters (first and third columns) and a set of alternative simulation parameters (as in Table~\ref{tab:simulationsliteratures} and Fig.~\ref{fig:zoneexamples}; second and fourth columns) for a particle mass of $m_{\mathrm{B}} = 10^5\,\mathrm{M}_{\odot}$ (left two columns) and $m_{\mathrm{B}} = 8\times10^5\,\mathrm{M}_{\odot}$ (right two columns), and a star formation efficiency of $e_{\mathrm{sf}} = 0.32\,\mathrm{per\,cent}$. The combination of a larger gravitational softening scale ($\epsilon = 500\,\mathrm{pc}$) and the smaller minimum smoothing length ($h_{\mathrm{min}} = 7.75\,\mathrm{pc}$) result in a galaxy without artificial stellar clumps and without gas in the runaway collapse zone. The low star formation efficiency still results in a large amount of cold gas with $T<100\,\mathrm{K}$ but at better resolved gas densities.}
    \label{fig:gallerybest}
\end{figure*}

The number of clumps as well as their masses are not very sensitive to the mass resolution (increase $m_{\mathrm{B}}$; lower right). On the other hand, increasing the gravitational force softening length (increase $\epsilon$; upper right) or increasing the star formation efficiency (increase $e_{\mathrm{sf}}$; upper left) both reduce the number of stellar clumps drastically by reducing the amount of high-density gas and therefore the amount of gas within the runaway collapse zone. Interestingly, their gas distributions look very different\footnote{The colourscales are the same across all phase-space plots.}: while the simulation with the higher star formation efficiency ($e_{\mathrm{sf}}=1\,\mathrm{per\,cent}$) has very little cold ($< 1000\,\mathrm{K}$) gas, the galaxy with the larger softening length ($\epsilon = 500\,\mathrm{pc}$) has large amounts of cold gas at densities just outside the runaway collapse zone but its further collapse is delayed due to the larger softening scale. A tail towards very high densities it still noticeable but in this case it is limited to the central part of the galaxy and therefore does not result in clumps throughout the galactic disk. 

In contrast to the comparisons shown in Figs.~\ref{fig:GasPhaseOverview} and \ref{fig:gallerygasstars} for which only one parameter is varied at a time, we show in Fig.~\ref{fig:gallerybest} an alternative combination of values for $\epsilon$ and $h_{\mathrm{min}}$. For a particle mass of $m_{\mathrm{B}} = 10^5\,\mathrm{M}_{\odot}$ (first two columns in Fig.~\ref{fig:gallerybest}) the alternative parameter set fulfills the conditions in equation~(\ref{eq:hminmax}) for $h_{\mathrm{min}}$ and equation~(\ref{eq:epsilonD}) for $\epsilon$. Note that these parameters are not necessarily the best choice for every application but illustrate the impact of choosing values for $\epsilon$ and $h_{\mathrm{min}}$ that are informed by the conditions defined in sections~\ref{sec:runawaycollapse} and \ref{sec:adaptivesoft}.

Both simulations within each figure half in Fig.~\ref{fig:gallerybest} use the same number of particles (i.e. baryon particle mass, $m_{\mathrm{B}} = 10^5\,\mathrm{M}_{\odot}$, left two panels, and $m_{\mathrm{B}} = 8 \times 10^5\,\mathrm{M}_{\odot}$, right two panels) and all simulations use the same value for the star formation efficiency ($e_{\mathrm{sf}} = 0.32\,\mathrm{per\,cent}$). The gravitational softening length is increased from 250~pc (fiducial) to 500~pc (alternative). The minimum smoothing length is reduced from $h_{\mathrm{min}} = 77.5\,\mathrm{pc}$ (fiducial) to $h_{\mathrm{min}} = 7.75\,\mathrm{pc}$ (alternative). These changes move the runaway collapse zone to very high densities. In addition, the zone of induced collapse (zone~D in section~\ref{sec:adaptivesoft}, red shaded area at low densities in Fig.~\ref{fig:gallerybest}) is also slightly reduced. The slower gravitational collapse ensures that the majority of the cold gas is at more manageable densities ($<10^2\,\mathrm{cm}^{-3}$) and the density distribution does not extend into the runaway collapse zone. For both mass resolutions, the stellar mass surface density is smoother for the alternative resolution parameters and there are no dense star particle clumps, in contrast to the massive clumps present in the simulations with the fiducial resolution parameters. 

\begin{figure}
    \includegraphics[width=\linewidth]{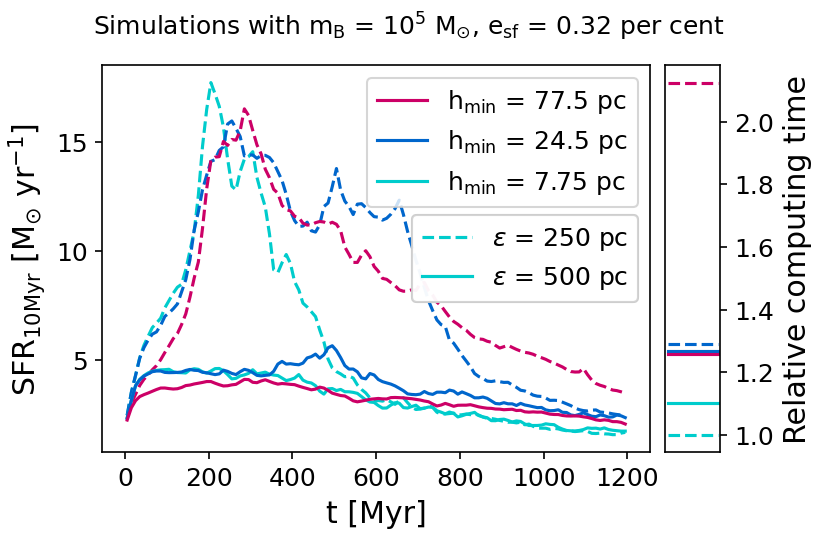}
    \caption{Main panel: Star formation histories (SFHs) for all simulations with a particle mass $m_{\mathrm{B}} = 10^5\,\mathrm{M}_{\odot}$ and a star formation efficiency of $e_{\mathrm{sf}} = 0.32\,\mathrm{per\,cent}$. Lines of different colours show the star formation rates (SFR, averaged over 10~Myr) of simulations with different minimum smoothing lengths, $h_{\mathrm{min}}$, and different line styles indicate different gravitational softening lengths, $\epsilon$. The red dashed line is the SFH for the simulation with the fiducial parameters (centre of Fig.~\ref{fig:gallerygasstars}, first column in Fig.~\ref{fig:gallerybest}) and the green solid line shows the SFH for the simulation with the alternative set of parameters in the second column of Fig.~\ref{fig:gallerybest}.  The small panel on the right gives an overview of the relative computing times normalized to the fastest simulation with the same linestyles as in the main panel.}
    \label{fig:SFH}
\end{figure}

Fig.~\ref{fig:SFH} shows both the star formation rates averaged over 10~Myr (large panel) and the relative computing time (small panel on the right) for simulations with a particle mass of $m_{\mathrm{B}} = 10^5\,\mathrm{M}_{\odot}$ and a star formation efficiency of $e_{\mathrm{sf}} = 0.32\,\mathrm{per\,cent}$. Variations in $h_{\mathrm{min}}$ (line colours) for isolated galaxies with a softening length of $\epsilon=500\,\mathrm{pc}$ (solid lines) do not have a large impact on neither the star formation history nor the computing time, because the large softening reduces the amount of high-density gas. For simulations with a softening length of $\epsilon=250\,\mathrm{pc}$ (dashed lines), the computing time is reduced by a factor of $\approx 2$ when reducing $h_{\mathrm{min}}$ by a factor of 10 (from $h_{\mathrm{min}} = 77.5\,\mathrm{pc}$, red dashed line, to $h_{\mathrm{min}} = 7.75\,\mathrm{pc}$, green dashed line). 

The increased computing time for simulations with larger values of $h_{\mathrm{min}}$ and therefore more clumping from runaway collapse is caused by substantial amounts of shock-heated gas at high densities $\log n_{\mathrm{H}} [\mathrm{cm}^{-3}]\gtrsim 2$) and temperatures of a few hundred to a few thousand K (see first and third column of Fig.~\ref{fig:gallerybest}). The temperature increase of a factor of 100 compared to simulations with smaller values of $h_{\mathrm{min}}$ (second and fourth column of Fig.~\ref{fig:gallerybest}) reduces the timestep size $\Delta t \propto m_{\mathrm{B}}^{1/3} n_{\mathrm{H}}^{-1/3} T^{1/2}$ (see e.g. \citealp{sphenix}) by a factor of 10. The higher ``real'' densities in artificially dense gas clumps in the fiducial run do not directly affect the timestep size because $\Delta t$ is calculated from the (underestimated) SPH densities. 

Summarizing, a too large value for $h_{\mathrm{min}}$ does not only produce artificially dense clumps of gas and star particles, but also increases the computing time by a factor of $\approx 2$ in our simulations.

\section{Discussion}\label{sec:discussion}

The runaway collapse zone, as defined in section~\ref{sec:runawaycollapse} and in particular equations~(\ref{eq:zonedens}) and (\ref{eq:zonetemp}), is an approximation because the clumping will depend on the exact particle configuration. Nevertheless, the isolated galaxy simulations show a clear correlation between the amount of gas in this zone and the presence of both artificial collapse (seen in the recalculated gas densities) as well as the number of dense stellar clumps, which is drastically reduced for smaller values of the minimum smoothing length (compare centre and bottom left panels in Fig.~\ref{fig:gallerygasstars}). 

The star formation efficiencies are varied in the presented simulations with values of $e_{\mathrm{sf}} = 0.32\,\mathrm{per\,cent}$ and $e_{\mathrm{sf}} = 1\,\mathrm{per\,cent}$ to illustrate the discussed issues. Higher values of $e_{\mathrm{sf}}$ mean that gas particles are converted into star particles on shorter timescales for a given density. The isolated galaxy simulations from this work do not produce artificial clumps for $e_{\mathrm{sf}}>1\,\mathrm{per\,cent}$, because dense gas is quickly converted into stars and therefore very little dense gas is present. Cosmological simulations of galaxy formation include the evolution of a large variety of galaxies with diverse properties and mass accretion histories and can occupy different density-temperature regions at different times. It is therefore plausible that the discussed issues are present in cosmological simulations also for  $e_{\mathrm{sf}}\ge 1\,\mathrm{per\,cent}$. 

One could expect that smaller values for the gravitational softening scale $\epsilon$ always lead to more accurate simulations than larger values because gravitational forces would be calculated correctly up to higher densities, but we discussed in section~\ref{sec:adaptivesoft} that the density - temperature zone within which the gravitational instabilities are modelled correctly ($\lambda_{\mathrm{J,N}}>l_{\mathrm{smooth}}$) only depends on the particle mass (i.e. the mass resolution). Gas with densities and temperatures such that $\lambda_{\mathrm{J,N}}<l_{\mathrm{smooth}}$, is formally unresolved independently of the softening length, which for a baryon particle mass of $m_{\mathrm{B}} = 10^5\,\mathrm{M}_{\odot}$ includes most of the cold neutral and molecular gas in galaxies (see the left panel of Fig.~\ref{fig:zonesatlsmooth}). A simulation with adaptive softening might report using $\epsilon_{\mathrm{min}} = 2\,\mathrm{pc}$ and a simulation with the same mass resolution of $m_{\mathrm{B}} = 10^5\,\mathrm{M}_{\odot}$ but constant softening might use $\epsilon = 100\,\mathrm{pc}$. Despite their very different values for $\epsilon$ and $\epsilon_{\mathrm{min}}$, neither simulation models gravitational instabilities correctly in the CNM nor in the MCs but both need to choose values for $\epsilon$ or $\epsilon_{\mathrm{min}}$ that avoid numerical issues. As described in section~\ref{sec:runawaycollapse}, this means a very small value for $\epsilon_{\mathrm{min}}$ to avoid numerical runaway collapse for adaptive softening (if $\epsilon_{\mathrm{min}} = h_{\mathrm{min}}$) and potentially a larger value for $\epsilon$ for constant softening if one wishes to suppress artificial sub-kernel gravitational instabilities (see section~\ref{sec:adaptivesoft}).

In very high (mass and force) resolution simulations, clusters of stars can form for physical reasons, but for the mass resolution in the isolated galaxies examples ($m_{\mathrm{B}} = 10^5\,\mathrm{M}_{\odot}$, $m_{\mathrm{B}} = 8 \times 10^5\,\mathrm{M}_{\odot}$), we showed that the high-density gas clumps and resulting dense star clusters are mainly numerical artefacts because their numbers and masses are drastically reduced in simulations with better hydrodynamic force resolution (compare the centre and bottom left panels in Fig.~\ref{fig:gallerygasstars}).

For cosmological simulations the choice of the resolution parameters (mass, gravity, hydro) also need to take into account the effect of artificial heating of the stellar disk by dark matter and stellar halo particles (see e.g.~\citealp{Ludlow2019stars, Ludlow2021, Ludlow2023, Wilkinson2022}), which is reduced for larger values of $\epsilon$ (appendix D in \citealp{Ludlow2021}). This effect is not included here because we do not use a live dark matter halo for these idealised simulations. 

The star formation rate for a gas particle in our simulation depends on the Newtonian free-fall time (see equation~\ref{eq:sfr}) but the collapse of gas clumps in the simulation follows the longer softened free-fall time (equation~\ref{eq:tffC2fit}). While for numerical stability considerations the softened properties (free-fall time, Jeans mass and length) are more relevant, we use the Newtonian free-fall time for the star formation rate because we want to model this subgrid process physically rather than numerically (see also the discussion in \citealp{Folkert2023arXiv230913750N}). 

Which resolution and star formation criteria to choose depends on the application of the simulation, but the softened Jeans criteria and the temperature - density zones defined in section~\ref{sec:application}, in particular the runaway collapse zone (section~\ref{sec:runawaycollapse}) provide a useful guideline to avoid undesired numerical behaviour that may introduce artefacts and can furthermore slow down the simulation.

To date most simulations of cosmological volumes that reach $z=0$ limit the pressure (or temperature) of the ISM by an effective pressure floor $P\propto \rho^{\gamma_{\mathrm{eff}}}$ which is related to the gas density $\rho$ through the effective polytropic index $\gamma_{\mathrm{eff}}$. It has been argued that for $\gamma_{\mathrm{eff}} > 4/3$, the Jeans mass does not increase with density and therefore prevents artificial fragmentation (e.g. \citealp{SchayeDallaVecchia2008}). In \textsc{eagle} \citep{EAGLE} $\gamma_{\mathrm{eff}} = 4/3$ for $n_{\mathrm{H}}>0.1\,\mathrm{cm}^{-3}$ and in the widely used \citet{SpringelHernquist} model, the polytropic index varies with density between $\gamma_{\mathrm{eff}} \approx 2.5$ at $0.1\,\mathrm{cm}^{-3}$ to $\gamma_{\mathrm{eff}} \approx 4/3$ at $100\,\mathrm{cm}^{-3}$ (for $z=0$; see figure~1 of \citealp{SpringelHernquist} for details). We argue here that such pressure (or entropy) floors are unnecessary in modern Lagrangian simulations, even for relatively low resolutions, provided that appropriate values for the softening and minimum smoothing lengths are used to avoid artificially inducing gravitational instabilities and a numerical runaway collapse.

The free-fall time in simulations with a constant softening length ($t_{\mathrm{ff,s}}$, equation~\ref{eq:tffsintext}) is longer than the Newtonian free-fall time ($t_{\mathrm{ff,N}}$, equation~\ref{eq:tffNintext}) in simulations with an adaptive softening length. In simulations of galaxy formation, the stellar feedback processes might have to start earlier to efficiently counteract the gravitational collapse when the gravitational softening is adaptive. Early feedback processes, such as radiation and stellar winds from young stars, or a small delay between star formation and core-collapse supernova explosions, might therefore be more important in simulations with adaptive softening than in simulations with a constant softening length.

\section{Summary}\label{sec:summary}

The Jeans stability criteria based on the analysis by \citet{Jeans1902} define a length scale, $\lambda_{\mathrm{J,N}}$, the so called Jeans length, by comparing the free-fall timescale to the sound crossing timescale (see the derivation in appendix~\ref{sec:derivation}). In a homogeneous medium with gas density $n_{\mathrm{H}}$ and temperature $T$, density perturbations on scales $>\lambda_{\mathrm{J,N}}$ are gravitationally unstable (i.e. the density perturbations grow) while density perturbations on scales $<\lambda_{\mathrm{J,N}}$ are gravitationally stable (i.e. the density perturbations decay with time until a new equilibrium is reached).

In this work we introduce ``softened Jeans criteria'' (section~\ref{sec:Jeans}) for which we re-derive the Jeans length in softened gravity, as used in Lagrangian simulations to suppress 2-body scattering. In parallel to the Newtonian Jeans criteria, the softened Jeans length (mass), $\lambda_{\mathrm{J,s}}$ ($M_{\mathrm{J,s}}$) describes the minimum length (mass) scale above which density perturbations grow and become gravitationally unstable in simulations with softened gravity.

For gas with densities and temperatures for which the Newtonian Jeans length is smaller that the gravitational softening length (densities above or temperatures below the red dashed line, $\lambda_{\mathrm{J,N}} = \epsilon$, in Fig.~\ref{fig:MJ100pc}), gravitational fragmentation is described by the softened Jeans criteria instead of the Newtonian Jeans criteria. The further gravitational collapse is slowed down in softened gravity and better described by the softened free-fall time (equation~\ref{eq:tffsintext}) than the Newtonian free-fall time (equation~\ref{eq:tffNintext}). 

Depending on the gravitational softening length, $\epsilon$, the softened Jeans mass can exceed the Newtonian Jeans mass $M_{\mathrm{J,N}}$ by several orders of magnitude for gas densities and temperatures typical of the cold interstellar medium. For example, at a gas temperature of a few tens of K and a gas density of $\approx 100\,\mathrm{cm}^{-3}$, the Newtonian Jeans mass is $M_{\mathrm{J,N}}\approx 10^2 \,\mathrm{M}_{\odot}$, and the softened Jeans mass for a constant softening length of $\epsilon=100\,\mathrm{pc}$ is $M_{\mathrm{J,s}}\approx 10^5 \,\mathrm{M}_{\odot}$ (Fig.~\ref{fig:MJ100pc}). 

In simulations with particle masses of $\approx 10^5\,\mathrm{M}_{\odot}$ that do not impose an artificial pressure or entropy floor in the form of an effective equation of state, cold neutral and molecular ISM gas is formally unresolved because gravitational instabilities should form within an individual smoothing kernel ($\lambda_{\mathrm{J,N}}<l_{\mathrm{smooth}}$). Modelling a multi-phase interstellar medium at these mass resolutions therefore requires an understanding of how instabilities behave at and below the resolution limit (see section~\ref{sec:application}). 

If the hydrodynamic forces are resolved by at least one smoothing kernel, gravitational instabilities would be modelled correctly (if $\lambda_{\mathrm{J,s}} = \lambda_{\mathrm{J,N}}$), or be suppressed ($\lambda_{\mathrm{J,s}} > \lambda_{\mathrm{J,N}}$) for all gas densities and temperatures in simulations, because softened gravity never exceeds Newtonian gravity and therefore $\lambda_{\mathrm{J,s}} \ge \lambda_{\mathrm{J,N}}$. Yet, we show in section~\ref{sec:application} that perturbations can grow under specific conditions within a hydrodynamic smoothing kernel. Two distinct pathways are identified for numerically induced instabilities, both related to an inaccurate calculation of the hydrodynamic properties below the resolution limit: instabilities caused by (i) an underestimated gas density (subsection~\ref{sec:runawaycollapse}) and (ii) pressure that is smoothed on length scales larger than those on which gravity is softened (subsection~\ref{sec:adaptivesoft}). The former instability is relevant for simulations with both adaptive and constant softening lengths, while the latter only applies to simulations with a constant softening length (see discussion in section~\ref{sec:application}).

The effects of sub-kernel instabilities is demonstrated in section~\ref{sec:simulations} in simulations of isolated galaxies using the \textsc{Swift} code with a constant softening length, $\epsilon$. As outlined in subsection~\ref{sec:runawaycollapse}, the density of gas clumps that are smaller than the minimum allowed smoothing length, $h_{\mathrm{min}}$, are under-estimated. Re-calculating the densities of all gas particles with a vanishing value for $h_{\mathrm{min}}$ reveals gas clumps with several orders of magnitude higher gas densities (compare top and bottom row of Fig.~\ref{fig:GasPhaseOverview}). These gas clumps result in dense star clusters that largely disappear for smaller minimum smoothing length values and are therefore artificial (Fig.~\ref{fig:gallerygasstars}).

Based on the analysis in section~\ref{sec:runawaycollapse}, we recommend for both simulations with constant and adaptive softening lengths to set the minimum smoothing lengths to a value small enough value that the smoothing lengths are not limited to a constant value in a simulation with particle mass, $m_{\mathrm{B}}$ and a maximum expected density, $n_{\mathrm{sim,max}}$:

\begin{equation}\label{eq:hminA}
    h_{\mathrm{min}} < 8.3\,\mathrm{pc} \left ( \frac{m_{\mathrm{B}}}{10^5\,\mathrm{M}_{\odot}} \right )^{1/3} \left ( \frac{n_{\mathrm{sim,max}}}{10^4\,\mathrm{cm}^{-3}} \right )^{-1/3}\,.
\end{equation}

\noindent
We recommend to generously fulfill this condition and could not identify a benefit of a larger value for $h_{\mathrm{min}}$. If the minimum values for softening and smoothing lengths are identical ($\epsilon_{\mathrm{min}} = h_{\mathrm{min}}$) in simulations with adaptive softening, a small value for $\epsilon_{\mathrm{min}}$ is also required.

Gravitational instabilities that would form within a smoothing kernel are unresolved by definition, even if both the minimum smoothing length and the softening length are approaching zero\footnote{A simulation with a given particle mass $m_{\mathrm{B}}$ does not have a higher resolution just because it uses a smaller value for $\epsilon$, see also discussion in section~\ref{sec:discussion}.}. Assuming $h_{\mathrm{min}}\rightarrow 0$, the fundamental differences that remain between simulations with adaptive and constant softening lengths for instabilities within a kernel are described in section~\ref{sec:adaptivesoft}. Neither method gives the right solution for sub-kernel gravitational instabilities: an adaptive softening length that follows the smoothing length of the kernel leads to artificial suppression of sub-kernel instabilities while a constant softening length can result in artificially inducing sub-kernel instabilities. Because these length scales are by definition below the resolution limit of the simulation, their detailed impact on galaxy properties needs further investigation and is beyond the scope of this work. 

Suppressing sub-kernel instabilities in simulations with constant softening lengths requires a softening length that exceeds the value given by equation~\ref{eq:epsilonD}. However, increasing the softening length artificially stabilizes gravitational instabilities at the hydrodynamical spatial resolution limit: the size of a smoothing kernel, $l_{\mathrm{smooth}}$, for larger regions in density-temperature space (see the middle panel in Fig.~\ref{fig:zonesatlsmooth}). The impact of sub-kernel instabilities on galaxy properties remains to be tested, because cooling rates, star formation rates and other density- or pressure-dependent sub-grid models use the smoother SPH estimates. The optimal value of a constant softening length, $\epsilon$, will therefore depend on the application.

Finally, we argue that SPH simulations with relatively low baryon mass resolution (shown here up to $m_{\mathrm{B}} = 8\times 10^5\,\mathrm{M}_{\odot}$ but the dependence on $m_{\mathrm{B}}$ is weak) do not depend on an effective pressure floor for numerical stability\footnote{Simulations such as \textsc{eagle} \citep{EAGLE} needed the implemented pressure floor to suppress the runaway collapse that could otherwise occur due to their large adopted value for $h_{\mathrm{min}}$ ($h_{\mathrm{min}} = 108.5\,\mathrm{pc}$ for their $(100\,\mathrm{Mpc})^{3}$ simulation).}. Simulations with comparable mass resolutions and without an artificial pressure floor do not suffer from the numerical issue discussed in section~\ref{sec:runawaycollapse} if $h_{\mathrm{min}}$ satisfies equation~(\ref{eq:hminA}). While the collapse of individual gas clouds is suppressed or slowed down when gravity is softened, this can be taken into account by subgrid prescriptions, e.g. for star formation, and is generally not a bottleneck if the aim of a simulation is to reproduce general galaxy properties.

The ideal simulation has both the mass and force resolution to accurately model the gravitational fragmentation and collapse of individual molecular clouds but in this work we showed a numerically stable alternative, both for adaptive and constant softening, for projects for which this is computationally too expensive.

\section*{Acknowledgements}

The authors would like to thank the anonymous referee for a very constructive referee report. SP acknowledges helpful discussions, especially with Vadim Semenov, at the Computational Galaxy Formation meeting, organized by Thorsten Naab and Volker Springel in Ringberg, Germany in April 2022. This research was funded in part by the Austrian Science Fund (FWF) grant number V~982-N.
This paper made use of the following python packages: astropy \citep{astropy2022}, numpy \citep{numpy}, scipy \citep{scipy}, matplotlib \citep{matplotlib}, unyt \citep{unyt} and swiftsimio \citep{swiftsimio,swiftsimio2021}. This work used the DiRAC@Durham facility managed by the Institute for Computational Cosmology on behalf of the STFC DiRAC HPC Facility (www.dirac.ac.uk). The equipment was funded by BEIS capital funding via STFC capital grants ST/K00042X/1, ST/P002293/1, ST/R002371/1 and ST/S002502/1, Durham University and STFC operations grant ST/R000832/1. DiRAC is part of the National e-Infrastructure. The computational results presented have been achieved in part using the Vienna Scientific Cluster (VSC).

%%%%%%%%%%%%%%%%%%%%%%%%%%%%%%%%%%%%%%%%%%%%%%%%%%
\section*{Data Availability}

The simulations were run with the public SPH code \textsc{Swift} \citep{swift2023} (\href{https://swift.dur.ac.uk/}{www.swiftsim.com}) version: 0.9.0, revision v0.9.0-1182-g423e9dd8. The initial conditions are discussed in \citet{Folkert2023arXiv230913750N} and public within the ``IsolatedGalaxy'' example within \textsc{Swift} (here the ``M5\_disk.hdf5'' and ``M6\_disk.hdf5'' files are used). The routines to set up, run and analyse the simulations, including the routines to create each figure, are public on gitlab (\href{https://gitlab.phaidra.org/softenedjeanscriteria}{https://gitlab.phaidra.org/softenedjeanscriteria}). The results are therefore fully reproducible, and only require limited computational resources.



\bibliographystyle{mnras}
\bibliography{SoftenedJeansCriteria} % if your bibtex file is called example.bib

%%%%%%%%%%%%%%%%%%%%%%%%%%%%%%%%%%%%%%%%%%%%%%%%%%

%%%%%%%%%%%%%%%%% APPENDICES %%%%%%%%%%%%%%%%%%%%%

\appendix

\section{Derivation of softened Jeans criteria}\label{sec:derivation}

In the following, we re-derive the Jeans length for softened gravity. For simplicity we focus on comparing the free-fall time, $t_{\mathrm{ff}}$, to the sound crossing time, $t_{\mathrm{sc}}$, of a self-gravitating structure with an initial radius, $R$. This derivation can easily be extended to general gravitational accelerations.

A system is in equilibrium if it has a radius $R$ for which the free-fall time equals the sound-crossing time ($R \equiv \lambda_{\mathrm{J}}$, the Jeans length). If the structure (here: a gas cloud) has a radius larger than the Jeans length, the cloud collapses or fragments, otherwise it expands. The sound crossing time is defined as

\begin{equation}\label{eq:tsc}
    t_{\mathrm{sc}} = \frac{R}{c_{\mathrm{s}}}
\end{equation}

\noindent
with the sound speed

\begin{equation}\label{eq:soundspeed}
    c_{\mathrm{s}} = \left (\frac{\gamma k_{\mathrm{B}} T}{\mu m_{\mathrm{H}}} \right )^{1/2}
\end{equation}

\noindent
and $t_{\mathrm{sc}}$ is independent of the shape of the gravitational potential. Here, $\gamma$ is ratio of specific heats, $k_{\mathrm{B}}$ is the Boltzmann constant, $T$ is the gas temperature, $\mu$ the mean particle mass, and $m_{\mathrm{H}}$ the hydrogen particle mass.

For a general definition of the free-fall time, we integrate the time it takes a test mass to fall from rest at a distance $R$ to the centre of the potential:

\begin{equation}\label{eq:ff1}
    t_{\mathrm{ff}} = \int_{r=R}^0 \mathrm{d}t =\int_{r=R}^0  \frac{\mathrm{d}t}{\mathrm{d}r}\mathrm{d}r = \int_{r=R}^0 \frac{1}{v(r)} \mathrm{d}r \, .
\end{equation}

\noindent
Integrating the equation of motion
\begin{align}
    v \frac{\mathrm{d}v}{\mathrm{d}r} &= - |a(r)| 
\end{align}

\noindent 
from $R$ to $r$ leads to 

\begin{equation}\label{eq:velocityff}
    v(r) = \left (2  \int_{r'=R}^{r} - |a(r')| \mathrm{d}r' \right )^{1/2} \, .
\end{equation}

\noindent
which is used in equation~\ref{eq:ff1} to calculate the free-fall time

\begin{equation}\label{eq:tffgeneral}
    t_{\mathrm{ff}}(R) = \int_{r=R}^{0} v(r)^{-1} \mathrm{d}r = \int_{r=R}^{0} \left ( 2 \int_{r' = R}^{r} -| a(r') | \; \mathrm{d}r'  \right )^{-1/2} \mathrm{d}r \, .
\end{equation}

\noindent
A pressure-less collapse and therefore no shell-crossing is assumed, which means that the mass inside the radius of the test mass remains constant throughout the free-fall: $M(<r)=M= 4\pi R^3 /3 \rho$. Finally, the Jeans length is calculated by solving the equation $ t_{\mathrm{ff}}(R) = t_{\mathrm{sc}} (R, c_{\mathrm{s}})$ for $R$.

\subsection{Newtonian (unsoftened) gravity}

The Jeans length for Newtonian gravity can be found in many textbooks but depending on the derivation the prefactors are slightly different. We repeat one of the textbook derivations here briefly to allow for a direct comparison of the individual steps with the derivation of the softened Jeans length and to demonstrate which prefactors are included in this derivation\footnote{A pre-factor of order unity is not important due to the approximate nature of the Jeans criterion in realistic applications but for comparisons to the softenened Jeans criteria and for the sake of completeness, we keep all pre-factors.}.  

For Newtonian gravity, 

\begin{equation}
    |a(r)| = |a_{\mathrm{N}}(r)| = \frac{GM}{r^2} 
\end{equation}
\noindent
and the velocity for the free-fall is

\begin{equation}\label{eq:velffN}
    v_{\mathrm{N}}(r) = \left (2GM\right )^{1/2}\left ( \frac{1}{r} - \frac{1}{R}\right )^{1/2} \quad .
\end{equation}

\noindent
The free-fall time is

\begin{equation}
    t_{\mathrm{ff,N}} = \left ( \frac{\pi^2 R^3}{8GM}\right )^{1/2} = \left ( \frac{3 \pi}{32 G \rho} \right )^{1/2}
\end{equation}

\noindent
or
\begin{align}\label{eq:tffNewton}
    t_{\mathrm{ff,N}}   &= 4.4 \, \mathrm{Myr} \left ( \frac{n_{\mathrm{H}}}{100\, \mathrm{cm}^{-3}}\right )^{-1/2}
\end{align}

\noindent 
for the gas density $\rho = n_{\mathrm{H}} m_{\mathrm{H}} / X_{\mathrm{H}}$ with the hydrogen number density, $n_{\mathrm{H}}$, the hydrogen mass fraction $X_{\mathrm{H}}$ (here we use $X_{\mathrm{H}} = 0.74$ for solar chemical composition, \citealp{Asplund2009}). The Jeans length for Newtonian gravity is  

\begin{align}
    \lambda_{\mathrm{J,N}} &= t_{\mathrm{ff,N}} \cdot c_{\mathrm{s}} \\
                           &= \left (\frac{3 \pi \gamma X_{\mathrm{H}} k_{\mathrm{B}}T}{32 G \mu m_{\mathrm{H}}^2 n_{\mathrm{H}}}\right )^{1/2} \\
                           &= 1.5\,\mathrm{pc}\,\left( \frac{T}{10\,\mathrm{K}} \right)^{1/2} \left ( \frac{n_{\mathrm{H}}}{100\,\mathrm{cm}^{-3}}\right )^{-1/2} \; , \label{eq:ljeansNewt}
\end{align}

\noindent
which corresponds to a Jeans mass of

\begin{align} \label{eq:MJ0}
    M_{\mathrm{J,N}} &= \frac{4\pi\rho \lambda_{\mathrm{J,N}}^3 }{3}  \nonumber \\
                        &= 46 \,\mathrm{M}_{\odot} \,\left( \frac{T}{10\,\mathrm{K}} \right)^{3/2} \left ( \frac{n_{\mathrm{H}}}{100\,\mathrm{cm}^{-3}}\right )^{-1/2} \; ,
\end{align}

\noindent
where we use $\gamma = 5/3$ and $\mu = 1.28$, values typical for a neutral atomic ideas gas with solar metallicity. For diatomic gas, such as molecular hydrogen, the ratio of specific heats would be $\gamma = 1.4$ and $\mu = 2$, but these are comparable to other order of unity effects, such as the concrete particle distribution. 

\begin{figure}
	\includegraphics[width=\linewidth]{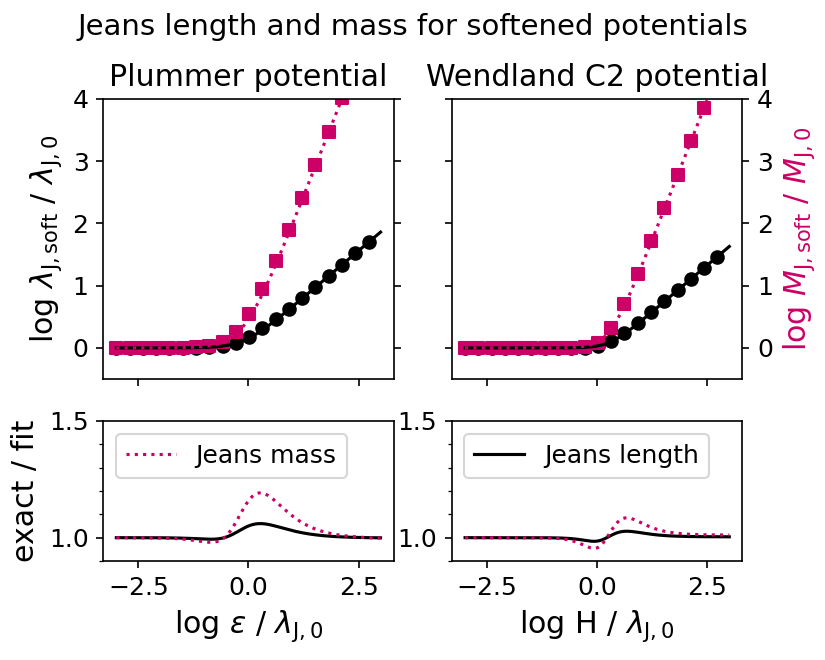}
    \caption{Top panels: Ratios of gravitationally softened Jeans length $\lambda_{\mathrm{J,s}}$ to Newtonian Jeans length $\lambda_{\mathrm{J,N}}$ (black solid lines and filled circles, left y-axis) and softened Jeans mass to Newtonian Jeans mass (red dotted lines and squares, right y-axis) for a Plummer potential with a softening length $\epsilon$ (left panels) and a Wendland C2 potential with a softening length $H=3\epsilon$ (right panels).
    The lines in the top panels are the fits (equations~\ref{eq:lJfitPlummer}, ~\ref{eq:MJP} for the Plummer potential and equations~\ref{eq:lJfitWendland}, ~\ref{eq:MJC2} for the Wendland C2 potential) to the numerical root finding (symbols, only selected points shown) and the bottom panels illustrate the accuracy of the fit.}
    \label{fig:lJ}
\end{figure}

\subsection{Plummer softening}

For a \citet{Plummer1911} softened potential with

\begin{equation}
    |a(r)| = |a_{\mathrm{P}}(r)| = \frac{GMr}{\left (r^2 + \epsilon^2\right )^{3/2}}  \, ,
\end{equation}

\noindent
and a softening scale $\epsilon$, the free-fall velocity, equation (\ref{eq:velocityff}), becomes

\begin{equation}
    v_{\mathrm{P}}(r) = \left (2GM\right )^{1/2}\left ( \frac{1}{\left (r^2 + \epsilon^2 \right )^{1/2}} - \frac{1}{\left (R^2 + \epsilon^2 \right )^{1/2}}\right )^{1/2} \, ,
\end{equation}

\noindent 
and the free-fall time (equation~\ref{eq:tffgeneral}) is 

\begin{align}
    t_{\mathrm{ff,P}} = &\left ( 2GM \right )^{-1/2} \cdot \nonumber\\
    &\int_{r=R}^0 \left ( \frac{1}{\left (r^2 + \epsilon^2 \right )^{1/2}} - \frac{1}{\left (R^2 + \epsilon^2 \right )^{1/2}} \right )^{-1/2} \mathrm{d}r \; .
\end{align}

\noindent

\noindent
This expression can be re-written as

\begin{equation}
    t_{\mathrm{ff,P}} = t_{\mathrm{ff,N}} \int_{x=0}^{1} \frac{2}{\pi} \left ( \left [ x^2+f_{\mathrm{P}}^2\right ]^{-1/2} - \left [ 1+f_{\mathrm{P}}^2\right ]^{-1/2} \right )^{-1/2}\mathrm{d}x \; ,
\end{equation}

\noindent
with the Newtonian free fall time $t_{\mathrm{ff,N}}$ and the dimensionless parameters $x=r/R$ and $f_{\mathrm{P}}=\epsilon/R$. This integral cannot be solved analytically but it can be numerically evaluated for different values of $f_{\mathrm{P}}$. The following fit matches these data points to within a few percent: 

\begin{equation} \label{eq:tffPlummerfit}
    t_{\mathrm{ff,P,fit}} = t_{\mathrm{ff,N}} \left ( 1 + 2^{2/3} f_{\mathrm{P}}^2 \right )^{3/4} \; .
\end{equation}

\noindent 
The free-fall in a Plummer softened potential is therefore slowed down by a factor of $\approx 2.0$ if the softening scale equals the size of the clump ($f_{\mathrm{P}} = 1$) and by a factor $\approx 45$ for a softening scale that is 10 times the size of the clump ($f_{\mathrm{P}} = 10$).

For the limiting case $f_{\mathrm{P}} \gg 1$ (i.e. softening is important) the Plummer free-fall time can be calculated analytically (see \citealp{Folkert2023arXiv230913750N}) and for $f_{\mathrm{P}} \ll 1$ the Plummer free-fall time approaches the Newtonian free-fall time. 
For the full expression of the Jeans length (solving $t_{\mathrm{ff,P,fit}} = R/c_{\mathrm{s}}$ for $R$ with $\lambda_{\mathrm{J,N}} = t_{\mathrm{ff,N}} c_{\mathrm{s}}$, the Newtonian Jeans length) we use a root finding algorithm for 

\begin{equation}
    \left ( \frac{\lambda_{\mathrm{J,P}}}{\lambda_{\mathrm{J,N}}} \right)^{10/3} - \left ( \frac{\lambda_{\mathrm{J,P}}}{\lambda_{\mathrm{J,N}}} \right)^{2} - 2^{2/3}\left ( \frac{\epsilon}{\lambda_{\mathrm{J,N}}} \right)^{2} = 0 \; .
\end{equation}

\noindent
This dimensionless equation is solved numerically for values of $\epsilon/\lambda_{\mathrm{J,N}}$ between $0.01$ and $10^{5}$. The results are fit by 

\begin{equation}\label{eq:lJfitPlummer}
    \lambda_{\mathrm{J,P,fit}}= \lambda_{\mathrm{J,N}}  \left ( 1 + 1.42 \left ( \frac{\epsilon}{\lambda_{\mathrm{J,N}}}\right )^{3/2}\right )^{2/5} \; ,
\end{equation}

\noindent
and the softened Jeans for the Plummer potential is fit by 

\begin{equation}\label{eq:MJP}
    M_{\mathrm{J,P,fit}} = M_{\mathrm{J,N}} \left ( 1 + 1.42 \left ( \frac{\epsilon}{\lambda_{\mathrm{J,N}}}\right )^{3/2}\right )^{6/5} \; .
\end{equation}

\noindent
The top left panel in Fig.~\ref{fig:lJ} illustrates the Jeans length (black, left y-axis) and Jeans mass (red, right y-axis) as a function of the ratio between the gravitational softening length, $\epsilon$, and the Newtonian Jeans length $\lambda_{\mathrm{J,N}}$. For small values of $\epsilon/\lambda_{\mathrm{J,N}}$ (i.e.~the gas cloud is much larger than the softening scale), the softened Jeans length approaches the Newtonian Jeans length ($\log \lambda_{\mathrm{J,P}} / \lambda_{\mathrm{J,N}}\rightarrow 0$) but for large values of $\epsilon/\lambda_{\mathrm{J,N}}$ the softened Jeans length and the softened Jeans mass are much larger than their Newtonian counterparts. 

The bottom left panel shows the ratio between the exact (numerical) solution and the fits from equations~(\ref{eq:lJfitPlummer}) and (\ref{eq:MJP}) which differ by up to 20~per~cent for the Jeans mass. This is small compared to other order of unity effects (i.e. different derivations, non-spherical particle distribution). 

\subsection{Wendland C2 softening}

For the \citet{Wendland1995} C2 kernel the acceleration $|a| = |a_{\mathrm{W}}|$ is defined in two parts, $a_{\mathrm{W}}(r<H,H)$ and $a_{\mathrm{W}}(r\ge H) = a_{\mathrm{N}}(r)$. This means that the free-fall velocity has distinct formulations depending on the relative size of the cloud, $R$, and the softening scale $H=3\epsilon$, where $\epsilon$ is the Plummer equivalent softening length. 

If the size of the cloud is smaller than $H$ ($f_{\mathrm{W}} \equiv H/R \ge 1$), the gravitational acceleration is softened throughout the free-fall and the velocity is calculated from

\begin{equation}
    v(r) = \left [ -2 \int_{r' = R}^r  \frac{GMr'}{H^3} V(u) \mathrm{d}r' \right ]^{1/2}
\end{equation}

\noindent
with the Wendland C2 kernel

\begin{align}
    W(u) &= -3u^7 + 15u^6-28u^5 + 21u^4 - 7u^2 + 3 \\
    V(u) &= -W'(u)/u = 21u^5 - 90u^4 + 140u^3 -84u^2 + 14
\end{align}

\noindent 
where $u=r/H$. If the size of the cloud is larger than the softening length, the free-fall velocity is Newtonian for $r > H$ and softened for $r \le H$ (or: $u\equiv r/H \le 1$). The full integral for the velocity for $r<H$ consists therefore of two parts:

\begin{align}
    v(r)& = \nonumber \\
    & \sqrt{2} \left [ \int_{r' = R}^H -|a_{\mathrm{N}}(r)|\, \mathrm{d}r' + \int_{r' = H}^0 - |a_{\mathrm{W}}(r<H,H)|\, \mathrm{d}r' \right ]^{1/2}\;,
\end{align}

\noindent
and the free-fall velocity is Newtonian (equation~\ref{eq:velffN}) for $u>1$. Combining all these cases results in the following expression for the free-fall velocity in a Wendland C2 softened gravitational potential:

\begin{align}
  &v_{\mathrm{W}}(r) = \nonumber \\ 
  &\sqrt {2GM }
    \begin{cases}
      H^{-1/2} \left [ W(u) - W\left (\frac{1}{f_{\mathrm{W}} }\right )\right ]^{1/2} & \text{if $f_{\mathrm{W}}  \ge 1$}\\[2ex]
      H^{-1/2} \left [ W(u) - f_{\mathrm{W}}\right ]^{1/2} & \text{if $f_{\mathrm{W}}  < 1 \wedge u \le 1$}\\[2ex]
      \left [ \frac{1}{r} - \frac{1}{R} \right ]^{1/2} & \text{if $f_{\mathrm{W}}  < 1 \wedge u > 1$}
    \end{cases}       
\end{align}

\noindent
These expressions have been numerically verified by an explicit leapfrog integration with a small, constant timestep of 0.1~Myr.

\paragraph*{Case 1: free-fall always softened ($f_{\mathrm{W}} \ge 1$ or $R \le H$):}

The free-fall time is 

\begin{equation}
    t_{\mathrm{ff,W}}(f_{\mathrm{W}}\ge 1) = \int_{r=R}^0 \left[ v_{\mathrm{W}} (r, f_{\mathrm{W}} \ge 1)\right]^{-1}\,\mathrm{d}r \; ,
\end{equation}

\noindent
which can be expressed relative to the Newtonian free-fall time $t_{\mathrm{ff,N}}$ and the dimensionless variables, $x=r/R$ and $f_{\mathrm{W}}=H/R$ as

\begin{align}
    t_{\mathrm{ff,W}}&(f_{\mathrm{W}}\ge 1) = t_{\mathrm{ff,N}} \frac{2}{\pi} f_{\mathrm{W}}^4 \, \cdot \nonumber \\
    \int_{x=1}^0  & \left (-3x^7 + 15x^6f_{\mathrm{W}} -28x^5f_{\mathrm{W}}^2 + 21x^4f_{\mathrm{W}}^3 - 7x^2f_{\mathrm{W}}^5 \right . \nonumber \\
    & \left . +3 -15f_{\mathrm{W}} +28f_{\mathrm{W}}^2-21f_{\mathrm{W}}^3+7f_{\mathrm{W}}^5 \right )^{-1/2} \mathrm{d}x \; .
\end{align}

\noindent 
The dimensionless integral and therefore $t_{\mathrm{ff,W}}/f_{\mathrm{ff,N}}$ is solved numerically for values of $f_{\mathrm{W}}$ between $1$ and $10^3$.  

\paragraph*{Case 2: free-fall partially softened ($f_{\mathrm{W}} < 1$ or $R > H$):}

If the size, $R$, of the self-gravitating structure is larger than the softening length, $H$, the free-fall time consists of two parts and is defined as

\begin{equation}
\begin{split}  
    t_{\mathrm{ff,W}}(f_{\mathrm{W}}< 1) &= \int_{r=R}^{H} v_{N}(r)^{-1} \mathrm{d}r \\& + \int_{r=H}^{0} v_{\mathrm{W}}(r, f_{\mathrm{W}} < 1 \wedge u \le 1)^{-1}\mathrm{d}r
\end{split}
\end{equation}

\noindent
with the Newtonian free-fall velocity $v_{N}$. The first integral can be solved analytically

\begin{equation}
    \int_{r=R}^{H} v_{\mathrm{N}}(r)^{-1} \mathrm{d}r = t_{\mathrm{ff,N}} \frac{2}{\pi} \left [ f_{\mathrm{W}} C_{\mathrm{W}} + \mathrm{tan}^{-1} \left (C_{\mathrm{W}}\right) \right]
\end{equation}

\noindent
with $C_{\mathrm{W}} = (1/f_{\mathrm{W}} -1)^{1/2}$. The second integral is rewritten as

\begin{align}
    \int_{r=H}^{0} v_{\mathrm{W}}(r, f_{\mathrm{W}} < 1 \wedge u \le 1)^{-1}\mathrm{d}r &= t_{\mathrm{ff,N}} \frac{2}{\pi} f_{\mathrm{W}}^{3/2} \cdot \nonumber \\
    &\int_{u = 1}^{0} (W(u) -f_{\mathrm{W}})^{-1/2}\mathrm{d}u \; .
\end{align}

\noindent
 The free-fall time for a partially softened trajectory in a Wendland C2 potential is 
 
 \begin{equation}
 \begin{split}
     t_{\mathrm{ff,W}}(f_{\mathrm{W}}< 1) = t_{\mathrm{ff,N}}\frac{2}{\pi} & \left [ f_{\mathrm{W}} C_{\mathrm{W}} + \mathrm{tan}^{-1} \left ( C_{\mathrm{W}}\right) \right .\\
     &\left .+ f_{\mathrm{W}}^{3/2} \int_{u = 1}^{0} (W(u) -f_{\mathrm{W}})^{-1/2}\mathrm{d}u \right] 
\end{split}
\end{equation}
 
\noindent
and is solved numerically for $f_{\mathrm{W}}$ between $10^{-3}$ and $1$.

\paragraph*{Combined:} The numerically calculated values for $t_{\mathrm{ff,W}}/t_{\mathrm{ff,N}}$ are fit with 

\begin{equation}\label{eq:tffC2fit}
    t_{\mathrm{ff,W,fit}} = t_{\mathrm{ff,N}} \left (1 + \frac{1}{7} f_{\mathrm{W}}^3 \right )^{1/2}\; .
\end{equation}

\noindent
The Jeans length for a Wendland C2 softened gravitational potential is calculated by solving $t_{\mathrm{ff,W,fit}} = R/c_{\mathrm{s}}$ for $R$ with $\lambda_{\mathrm{J,N}} = t_{\mathrm{ff,N}} c_{\mathrm{s}}$, the Newtonian Jeans length. We use a root finding algorithm for

\begin{equation}
    \left ( \frac{\lambda_{\mathrm{J,W}}}{\lambda_{\mathrm{J,N}}} \right )^5 - \left ( \frac{\lambda_{\mathrm{J,W}}}{\lambda_{\mathrm{J,N}}} \right )^3 - \frac{1}{7} \left ( \frac{H}{\lambda_{\mathrm{J,N}}} \right )^3  = 0
\end{equation}

\noindent
and this dimensionless equation is fit with 

\begin{equation}\label{eq:lJfitWendland}
   \lambda_{\mathrm{J,W,fit}}= \lambda_{\mathrm{J,N}} \cdot \left ( 1 + 0.27 \left ( \frac{H}{\lambda_{\mathrm{J,N}}}\right )^2\right )^{3/10} \; ,
\end{equation}

\noindent
which leads us the softened Jeans mass

\begin{equation}\label{eq:MJC2}
    M_{\mathrm{J,W,fit}} = M_{\mathrm{J,N}} \left ( 1 + 0.27 \left ( \frac{H}{\lambda_{\mathrm{J,N}}}\right )^2\right )^{9/10} \; .
\end{equation}

\noindent
The expressions from equations~(\ref{eq:lJfitWendland}) and (\ref{eq:MJC2}) as well as the accuracy of the fits are shown in the right panels of Fig.~\ref{fig:lJ}. The behaviour is very similar to the softened Jeans criteria for the Plummer potential (left panels of Fig.~\ref{fig:lJ}) with a slight offset. 

\section{Application to higher-resolution simulations}\label{sec:highres}

Throughout this work, we have discussed the derived instability criteria for large-scale simulations with baryon particle masses of $\gtrsim 10^{5}\,\mathrm{M}_{\odot}$. Here, we briefly demonstrate how the discussion in section~\ref{sec:application} applies to simulations with much higher mass resolution. 

\begin{figure}
    \centering
    \includegraphics[width=\linewidth]{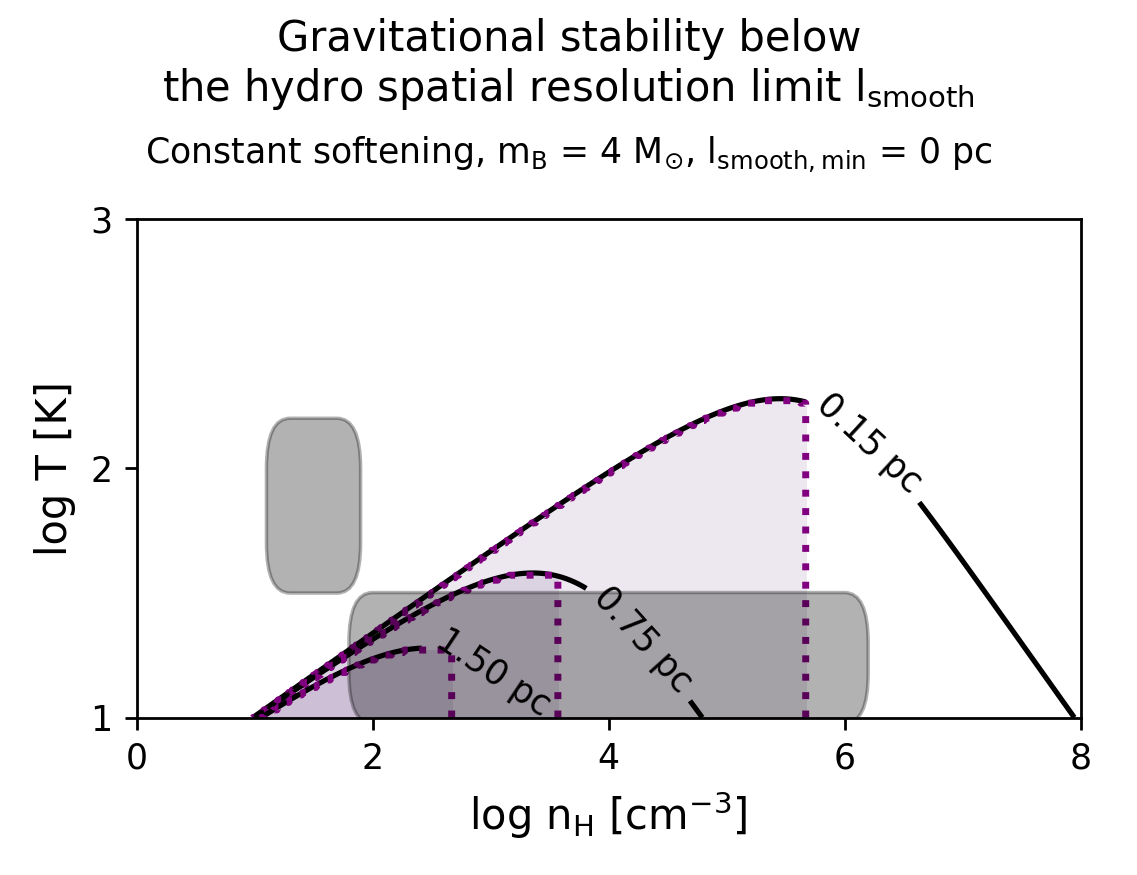}
    \caption{Lines and colours are as in the right panel of Fig.~\ref{fig:zonesbelowlsmooth} but for simulations with much higher mass resolution ($m_{\mathrm{B}} = 4\,\mathrm{M}_{\odot}$) and constant softening lengths of $l_{\mathrm{soft}} = 0.15$, $0.75$, and $1.5\,\mathrm{pc}$ ($\epsilon = 0.1$, $0.5$, and $1\,\mathrm{pc}$). Typical densities and temperatures for the CNM and MCs are indicated with dark patches, as in Fig.~\ref{fig:zoneexamples}.}
    \label{fig:highres}
\end{figure}

Assuming a vanishing value for the minimum smoothing length, sub-kernel clumping may be of concern in simulations with a constant softening length. Fig.~\ref{fig:highres} repeats the analysis from Fig.~\ref{fig:zonesbelowlsmooth} for simulations with a particle mass of $m_{\mathrm{B}} = 4\,\mathrm{M}_{\odot}$ and constant softening lengths of $l_{\mathrm{soft}} = 0.15$, $0.75$, and $1.5\,\mathrm{pc}$ ($\epsilon = 0.1$, $0.5$, and $1\,\mathrm{pc}$). Gas with densities and temperatures below the black lines is gravitationally unstable in the respective simulation to perturbations with a length scale of $l_{\mathrm{smooth}}$, the size of an individual kernel and the spatial resolution limit of the hydrodynamic solver. In the low-density part (purple shaded area, defined as $l_{\mathrm{soft}} < l_{\mathrm{smooth}}$) instabilities within the kernel may grow (see sections~\ref{sec:gravinstatlsmooth} and \ref{sec:gravinstbelowlsmooth} for a detailed discussion). For $m_{\mathrm{B}} = 4\,\mathrm{M}_{\odot}$ and $l_{\mathrm{soft}} = 0.15\,\mathrm{pc}$ (e.g. \citealp{Hislop2022,Steinwandel2023}), sub-kernel instabilities may form over the full density and temperature range typical for molecular gas in the Milky Way (indicated as dark patch at $10^2 \lesssim n_{\mathrm{H}} [\mathrm{cm}^{-3}] \lesssim 10^6$, $T\approx 20\,\mathrm{K}$). 

\citet{Steinwandel2023} mention that the global galactic properties are not very sensitive to variations in the softening length but that reducing $\epsilon$ to below $0.1\,\mathrm{pc}$, leads to numerical issues caused by runaway stars with very high velocities. For the same particle mass ($m_{\mathrm{B}} = 4\,\mathrm{M}_{\odot}$), \citet{Hu2017} varied the softening length between $\epsilon = 0.5\,\mathrm{pc}$ and $\epsilon = 2\,\mathrm{pc}$ in their appendix B2. 
They find that the gas density distribution in their simulations extends to higher densities for smaller values of $\epsilon$, but that integrated galaxy properties, such as the total star formation rate only has a weak dependence on $\epsilon$. It would be interesting to see if the discussed sub-kernel clumping is visible upon closer inspection in these simulations with very small values for $\epsilon$, but such an analysis is beyond the scope of this work. 

Based on Fig.~\ref{fig:highres}, sub-kernel clumping can be avoided for simulations with $m_{\mathrm{B}} = 4\,\mathrm{M}_{\odot}$, for softening lengths of $\epsilon \ge 2\,\mathrm{pc}$, as e.g. used in \citet{Hu2016, Hu2017}.

\section{Robustness of results}

The simulations presented in section~\ref{sec:simulations} illustrate the process of a numerical runaway collapse as outlined in section~\ref{sec:runawaycollapse}. In this section we vary the energy per supernova (\ref{sec:ESNvariation}) and the thermal state of the interstellar medium (\ref{sec:ISRFvariation}) to demonstrate that our conclusions are robust to these changes. 

\subsection{Supernova energy}\label{sec:ESNvariation}

The simulations in section~\ref{sec:simulations} use an energy per supernova of $E_{\mathrm{SN}} = 10^{51}\,\mathrm{erg}$. This energy is stochastically injected into the nearest gas particle by increasing its temperature by $\Delta T_{\mathrm{heat}} = 10^{7.5}\,\mathrm{K}$. \citet{DallaVecchiaSchaye2012} estimate the maximum gas density up to which thermal feedback is efficient by comparing the gas cooling time, $t_{\mathrm{c}}$, to the sound-crossing time, $t_{\mathrm{sc}}$, across a heated resolution element. If $t_{\mathrm{c}} \gg t_{\mathrm{sc}}$ (e.g. by a factor of $f_{\mathrm{t}}\equiv t_{\mathrm{c}}/f_{\mathrm{sc}} = 10$), the gas expands adiabatically before the thermal energy is lost through radiative cooling. 

\begin{figure}
    \centering
    \includegraphics[width=\linewidth,trim={0.8cm 0.8cm 1.5cm 7.0cm},clip]{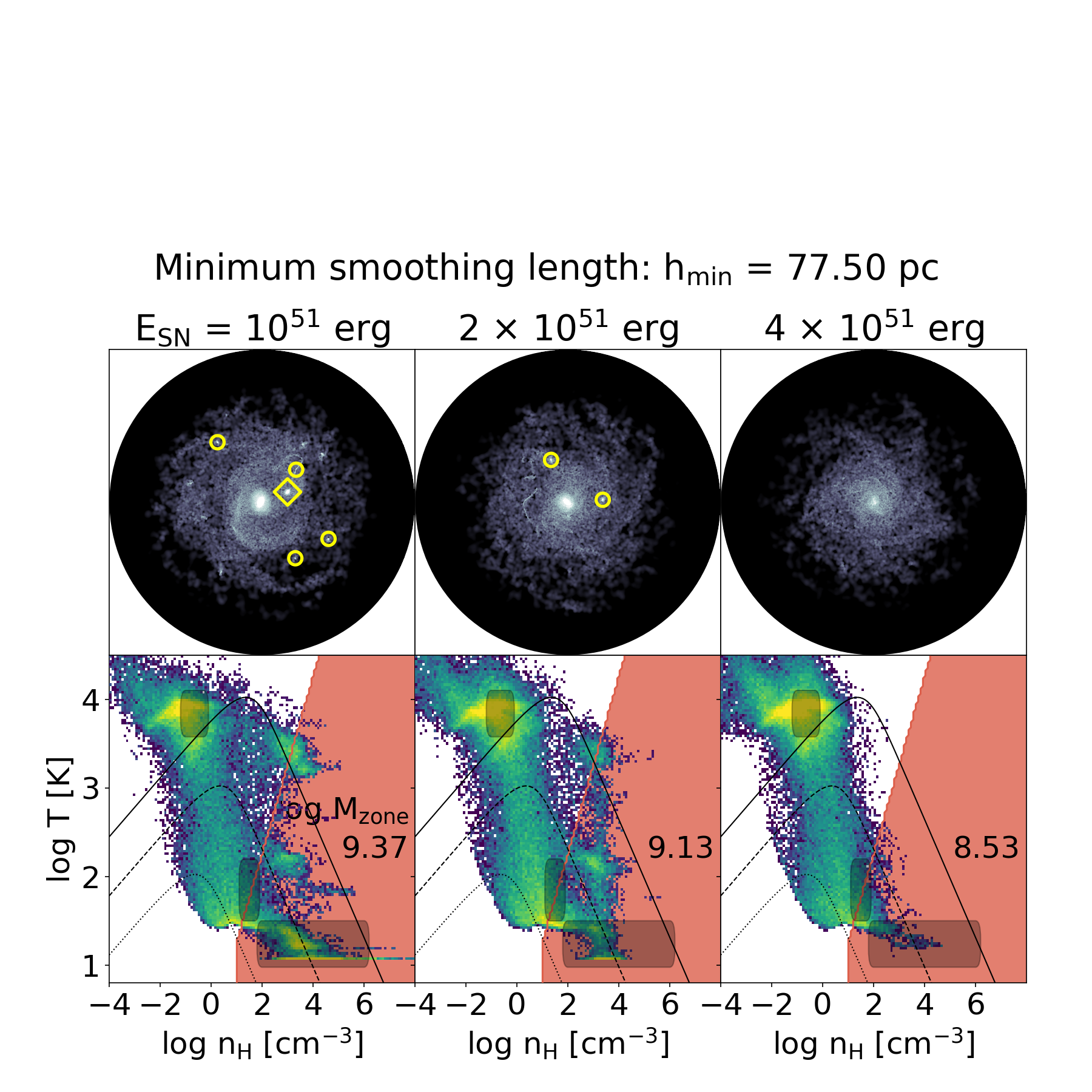}
    \caption{Stellar mass surface density maps (top row) and gas density-temperature histograms for $h_{\mathrm{min}} = 77.5\,\mathrm{pc}$, as in Fig.~\ref{fig:UVBvariations}. Increasing the energy deposited per supernova from the canonical value of $E_{\mathrm{SN}} = 10^{51}\,\mathrm{erg}$ (1st column) by a factor of 2 (2nd column) and by a factor of 4 (3rd column) reduces the amount of gas in the runaway collapse zone (red shaded area), resulting in reduced stellar clumping. Typical densities and temperatures for the WNM, CNM, and MCs are indicated with dark patches, as in Fig.~\ref{fig:zoneexamples}.}
    \label{fig:ESNvariations}
\end{figure}

As discussed in section~\ref{sec:simulations}, the maximum density below which thermal feedback is efficient is $\approx 10\,\mathrm{cm^{-3}}$ (equation 18 from \citealp{DallaVecchiaSchaye2012}) for the fiducial simulations. In Fig.~\ref{fig:ESNvariations}, we show simulations with a 2 and 4 times higher energy per supernova, while keeping the heating temperature constant at $\Delta T_{\mathrm{heat}} = 10^{7.5}\,\mathrm{K}$. In the stochastic model, this means more frequent thermal energy injections. For reference, \citet{EAGLE} used an energy per supernova that varies between $E_{\mathrm{SN}} = 0.3\times 10^{51}\,\mathrm{erg}$ and $3 \times 10^{51}\,\mathrm{erg}$ depending on metallicity (higher $E_{\mathrm{SN}}$ for lower metallicity) and gas density at the time of star formation (higher $E_{\mathrm{SN}}$ for higher gas density).  

Increasing $E_{\mathrm{SN}}$ ($10^{51}\,\mathrm{erg}$, $2 \times 10^{51}\,\mathrm{erg}$, $4 \times 10^{51}\,\mathrm{erg}$, columns 1 to 3) indeed reduces the clumping of the stars as well as the amount of shock-heated gas, even for large values of the minimum smoothing length (here: $h_{\mathrm{min}} = 77.5\,\mathrm{pc}$). This agrees with the prediction from section~\ref{sec:runawaycollapse} that the artificial clumping depends on the amount of gas in the runaway collapse zone. A larger value for $E_{\mathrm{SN}}$ reduces the highest density gas within the galaxy and for the largest energy per supernova ($E_{\mathrm{SN}} = 4\times10^{51}\,\mathrm{erg}$, 3rd column), barely any molecular gas (with $n_{\mathrm{H}} > 100\,\mathrm{cm}^{-3}$) is left within the galaxy. On the other hand, a smaller value of $h_{\mathrm{min}}$ reduces the artificial clumping of stars also in the presence of cold gas.

\subsection{Photoheating and cosmic rays}\label{sec:ISRFvariation}

The simulations presented in section~\ref{sec:simulations} use the radiative cooling and heating rates from \citet{PS20}. The rates are pre-calculated, assuming a pressure-dependent interstellar radiation field (ISRF) and cosmic ray (CR) rate (see \citealp{PS20} for details) and a redshift-dependent radiation background from distant galaxies. In addition to the fiducial table (``UVB\_dust1\_CR1\_G1\_shield1"), \citet{PS20} also provide tables without an ISRF or CRs (``UVB\_dust1\_CR0\_G0\_shield1"), and with a 10 times higher normalization for the ISRF and CR rates (``UVB\_dust1\_CR2\_G2\_shield1"). 

\begin{figure*}
    \centering
    \includegraphics[width=0.47\linewidth,trim={0.8cm 0.8cm 1.5cm 7.0cm},clip]{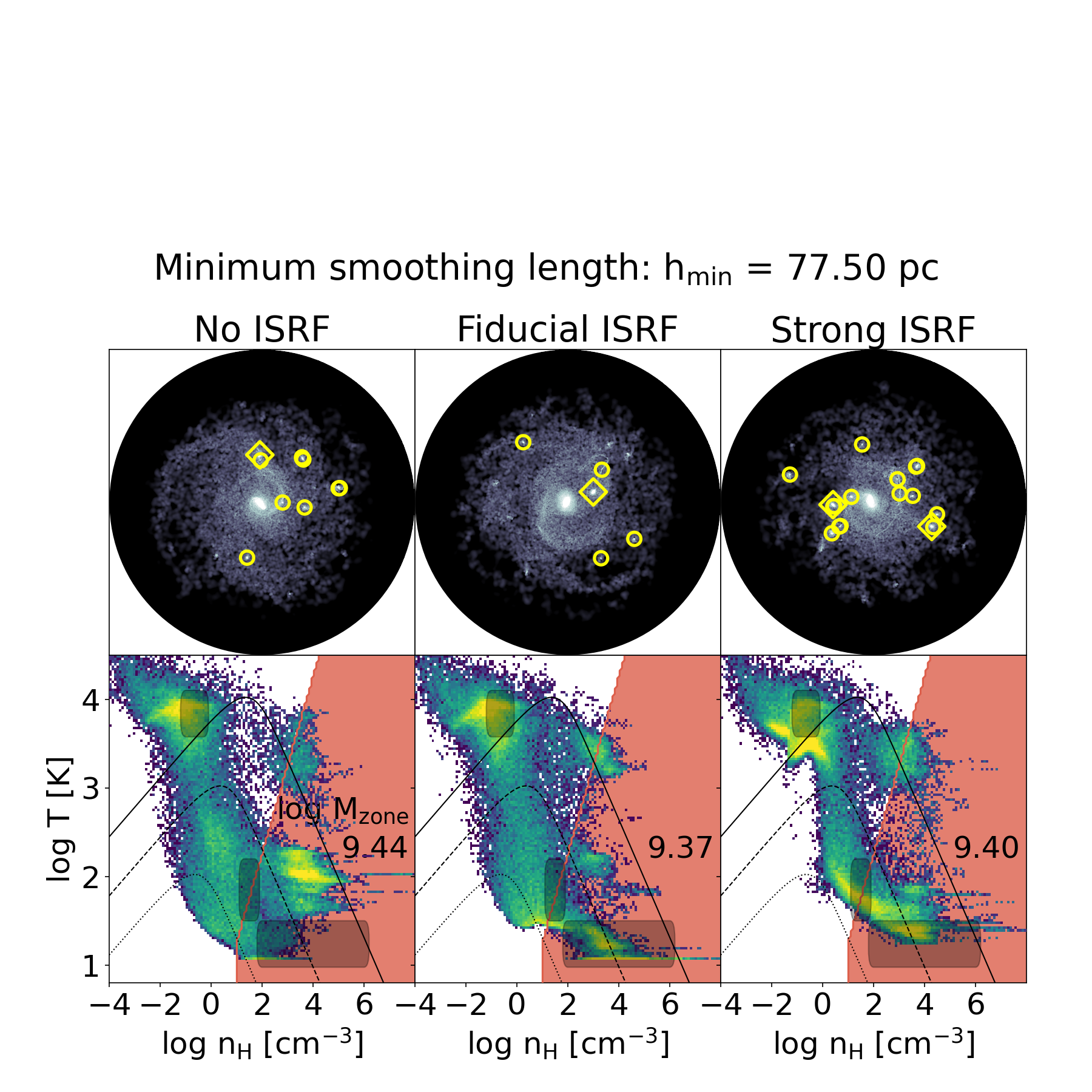}
    \includegraphics[width=0.47\linewidth,trim={0.8cm 0.8cm 1.5cm 7.0cm},clip]{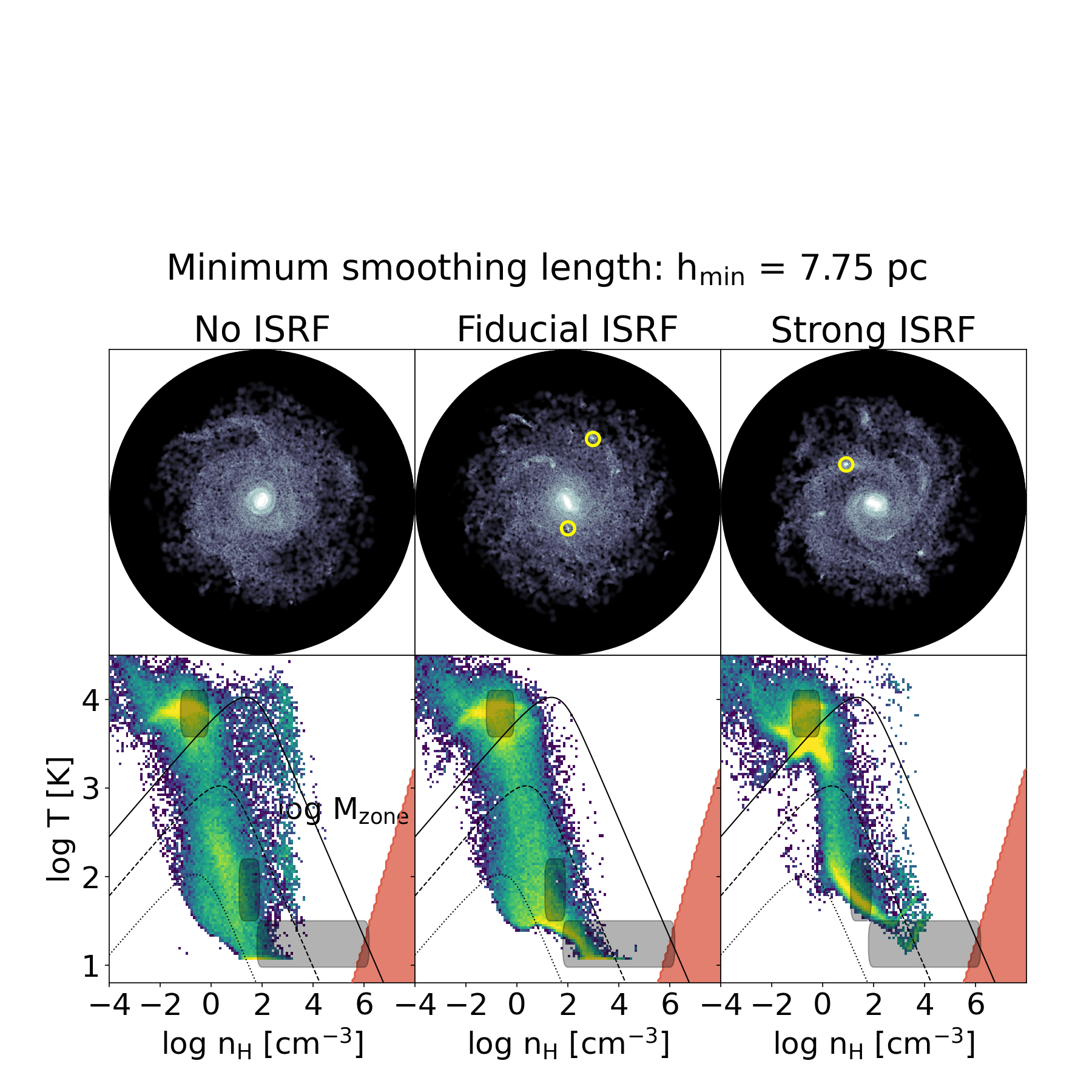}
    \caption{Stellar mass surface density maps (top row) and gas density-temperature histograms, as in Fig.~\ref{fig:gallerybest}. The drastically reduced clumping in stars, as well as the reduced amount of dense shock-heated gas is apparent for simulations with different cooling functions (no ISRF: 1st and 4th column, fiducial ISRF: 2nd and 5th column, strong ISRF: 3rd and 6th column) when reducing the minimum smoothing length, $h_{\mathrm{min}}$ from $77.5\,\mathrm{pc}$ (columns 1 to 3) to $7.75\,\mathrm{pc}$ (columns 4 to 6).  Typical densities and temperatures for the WNM, CNM, and MCs are indicated with dark patches, as in Fig.~\ref{fig:zoneexamples}.}
    \label{fig:UVBvariations}
\end{figure*}

The fiducial simulation ($m_{\mathrm{B}} = 10^5\,\mathrm{M}_{\odot}$, $\epsilon = 250\,\mathrm{pc}$) was rerun without an ISRF (``No ISRF") which also does not include CRs, and with the cooling and heating rates that correspond to the higher ISRF and CR normalization (``Strong ISRF"). Fig.~\ref{fig:UVBvariations} demonstrates that the features that we have identified for large values of the minimum smoothing length, $h_{\mathrm{min}}$: (i) stellar clumps and (ii) shock-heated, high-density ($n_{\mathrm{H}}\gtrsim100\,\mathrm{cm}^{-3}$) gas with temperatures of a few hundred to a few thousand K, are insensitive to variations in the radiation field, if the value of $h_{\mathrm{min}}$ is large ($77.5\,\mathrm{pc}$, columns 1 to 3) and many gas particles are located within the runaway collapse zone. In turn, the presence of both features is drastically reduced, when reducing the value of $h_{\mathrm{min}}$ to $7.75\,\mathrm{pc}$ (columns 4 to 6), confirming the findings from section~\ref{sec:simulations}.

%%%%%%%%%%%%%%%%%%%%%%%%%%%%%%%%%%%%%%%%%%%%%%%%%%

% Don't change these lines
\bsp	% typesetting comment
\label{lastpage}
\end{document}